\documentclass[oneside,reqno]{amsart}
\usepackage{setspace}
\usepackage[latin1]{inputenc}
\usepackage[swedish,english]{babel} 
\usepackage{amssymb,mathrsfs}
\usepackage{ifpdf} 
\ifpdf
\usepackage[pdftex]{graphicx} 
\usepackage{epstopdf} 
\else
\usepackage[dvips]{graphicx} 
\fi 
\usepackage{caption} 
\usepackage{multirow}
\usepackage[pdftitle = {pipEMCres},
  pdfauthor = {Liu Engblom Nettelblad},
  pdffitwindow = true,
  breaklinks = true,
  colorlinks = true,
  urlcolor = blue,
  linkcolor = red,
  citecolor = red,
  anchorcolor = red]{hyperref}
\usepackage[numbers,sort]{natbib} 

\usepackage[subrefformat=parens,labelformat=parens,labelfont={rm}]{subfig}
\usepackage{listings}
\usepackage{color}
\usepackage{cleveref}
\usepackage[flushleft]{threeparttable}

\usepackage{hyperref}
\hypersetup{
    colorlinks=true,
    linkcolor=black,
    citecolor=black,
    filecolor=black,
    urlcolor=black
}

\newcommand*{\refsec}[1]{%
  \begingroup
    \hypersetup{
      linkcolor=blue,
    }%
    \S\ref{#1}%
  \endgroup
}

\usepackage[inline]{enumitem}

\newcommand{\Mpix}{M_{\mbox{{\tiny pix}}}}
\newcommand{\Mrot}{M_{\mbox{{\tiny rot}}}} 
\newcommand{\Mdata}{M_{\mbox{{\tiny data}}}}

\newcommand{\Prob}{\mathbf{P}} 

\newcommand{\Wgrid}{\mathbb{W}}

\newcommand{\Fourier} {\mathscr{F}}

\begin{document}

\title[FXI 3D analysis pipeline]{Flash X-ray diffraction imaging in
  3D: a proposed analysis pipeline}
 
\author[J.~Liu]{Jing Liu}
\author[S.~Engblom]{Stefan Engblom}
\author[C.~Nettelblad]{Carl Nettelblad}

\address[J.~Liu]{Laboratory of Molecular Biophysics,
  Department of Cell and Molecular Biology, Uppsala university, SE-751
  24 Uppsala, Sweden.}
\email{jing.liu@icm.uu.se}

\address[S.~Engblom \and J.~Liu \and C.~Nettelblad]{Division of Scientific Computing,
  Department of Information Technology, Uppsala university, SE-751 05
  Uppsala, Sweden.}
\urladdr[S. Engblom]{\url{http://user.it.uu.se/~stefane}}
\email{stefane, jing.liu, carl.nettelblad@it.uu.se}

\address[C.~Nettelblad]{
   Division of Scientific Computing, Science for Life Laboratory, Department of Information Technology, Uppsala
    university, SE-751 05 Uppsala, Sweden.}	

  \date{\today}
	


\begin{abstract}
  Modern Flash X-ray diffraction Imaging (FXI) acquires diffraction
  signals from single biomolecules at a high repetition rate from X-ray  
  Free Electron Lasers (XFELs), easily obtaining millions of 2D
  diffraction patterns from a single experiment. Due to the stochastic
  nature of FXI experiments and the massive volumes of data,
  retrieving 3D electron densities from raw 2D diffraction
  patterns is a challenging and time-consuming task.

  We propose a semi-automatic data analysis pipeline for FXI
  experiments, which includes four steps: hit finding and preliminary
  filtering, pattern classification, 3D Fourier reconstruction, and
  post analysis. We also include a recently developed bootstrap
  methodology in the post-analysis step for uncertainty analysis and
  quality control. To achieve the best possible resolution, we further
  suggest using background subtraction, signal windowing, and convex
  optimization techniques when retrieving the Fourier phases in the
  post-analysis step.
  
  As an application example, we quantified the 3D electron structure
  of the PR772 virus using the proposed data-analysis pipeline. The
  retrieved structure was above the detector-edge resolution and
  clearly showed the pseudo-icosahedral capsid of the PR772.
\end{abstract}


\maketitle


\section{Introduction}
\label{sec:intro}

Flash X-ray diffraction Imaging (FXI) for single-particle imaging
(SPI) is a nascent technology for exploring the structure of
individual biological molecules in their natural states without
crystallization or cryofixation.  Relying on X-ray Free Electron
Lasers (XFELs), FXI can outrun the radiation damage to biomolecules
and capture interpretable 2D diffraction signals at the high
repetition rate available at XFELs, thus allowing the study of
single-particle structures at femtosecond timescales. This idea, using
XFELs to probe atomic resolution imaging of non-crystallized samples,
was first suggested in \cite{dad}, and demonstrated using artificial
planar object in \cite{cowbox}. In 2011, the first FXI experiment on
single mimiviruses was performed \cite{mimivirus_Xrays} at the Linac
Coherent Light Source (LCLS). Mimivirus is an icosahedral virus with
double-stranded DNAs, and its capsid is around 500~nm in diameter.
The first FXI mimiviruse experiments produced 2D mimivirus diffraction
patterns at the resolution of 32~nm, and since then, FXI has get
attentions in determining structures of biomolecules in varying shapes
and sizes \cite{mimivirus_Xrays,Hantke2014,Schot2015, Daurer2017,
  3d_mimi, Hosseinizadeh2017,Kurta2017,pr772_ros,Rice}.

In FXI experiments, a stream of sample particles is intersected with
the X-ray focus, with individual interactions between sample and X-ray
pulses occurring stochastically. The dataset therefore consists of
diffraction signals of varying quality, including, e.g., misses,
single hits, double hits, impurities, and background
noise. Furthermore, the data volume of the FXI experiment follows that
of the XFEL repetition rate. With LCLS \cite{LCLS}, the FXI
experiments operate at 120 Hz and can produce over 400,000 diffraction
patterns per hour, i.e., more than 1.6 TB per hour or 38 TB per
day. The newest facility, the European XFEL \cite{EXFEL} operates at
up to 27,000 Hz, i.e., 225 times more than the LCLS, and can produce
more than 12.6 million images per hour \cite{AGIPD1,AGIPD2}. The
massive volumes of data and the data complexity \cite{Hantke2014,
  Daurer2017} make a manual analysis impossible, and hence we seek an
automatic, robust, high-quality, and fast FXI analysis pipeline which
enables the efficient determination of 3D-structures.

\begin{figure}[!htbp]
  \centering
  \includegraphics[width = .8\textwidth]{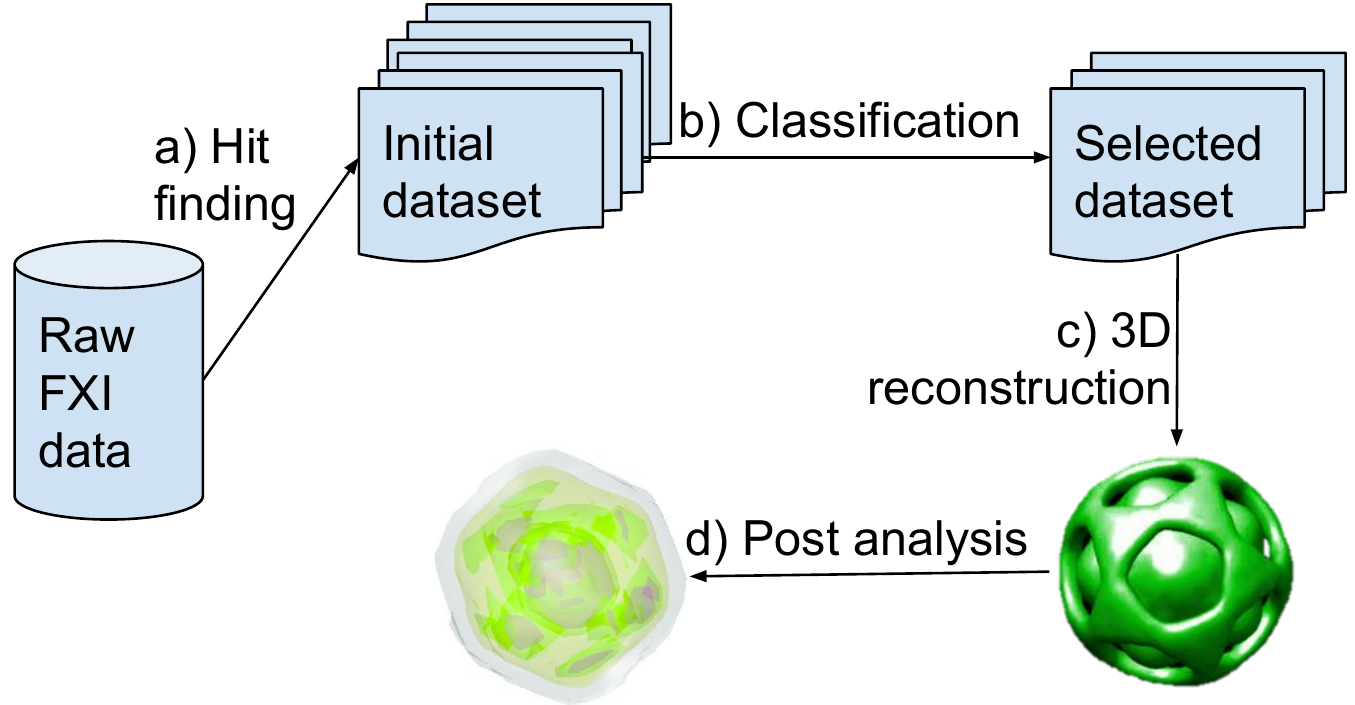}
  \caption{The proposed FXI data-analysis pipeline. a) The hit finding
    procedure selects an initial dataset from the raw FXI data.  b)
    From the initial dataset, the classification procedure selects
    high-quality single-particle diffraction patterns. c) The 3D
    reconstruction procedure produces a Maximum-Likelihood estimate of
    the 3D Fourier intensity. d) In the post-analysis step, the 3D
    real-space structure is estimated together with the overall
    resolution uncertainty.}
  \label{fig:pipeline}
\end{figure}

In this paper, we design such a data-analysis pipeline and we
illustrate its use for a PR772 virus dataset
\cite{pr772_sd,cxidb_entry58}, which can be downloaded from the
Coherent X-ray Imaging Data Bank (CXIDB) \cite{cxidb}. These results
are then complementary to recent work using the same experimental data
\cite{Hosseinizadeh2017, Kurta2017, pr772_ros}. Our proposed pipeline,
illustrated in Figure~\ref{fig:pipeline}, is designed to select a
limited amount of data frames for determining the 3D Fourier
intensities, putting considerable effort in the post-analysis, i.e.,
phase retrieval, background handling, bootstrap analysis, pattern
adjustment, and so on. Requiring a reasonable amount of computing
power, the overall design goal is to be able to use our compute
pipeline for handling FXI data on site, possibly even while the
experiment is still producing new data. Further, our pipeline allows
accessing uncertainties in the reconstructed 3D structures, which
provides another view of resolution other than the phase retrieval
transfer function (PRTF) or the Fourier-shell correlation (FSC).

Our proposed pipeline consists of four steps:
\begin {enumerate}
\item[a)] The hit finding/preliminary filtering procedure finds
  single-particle hits from the raw FXI data. Previous studies on
  hit-finding by thresholding can be found in \cite{cheetah}, or via
  methods such as spectral clustering techniques
  \cite{PCAclassification,spcl}, manifold embedding \cite{me}, 
  and diffusion map embedding \cite{novelsort}.

\item[b)] The classification procedure developed by us \cite{psort}
  classifies the initial dataset and selects high-quality diffraction
  patterns with specific properties, such as icosahedral shape or in a
  correct size range, and so on. It is possible to combine this
  procedure with the hit finding step to reduce the time for selecting
  high-quality patterns.

\item[c)] The 3D reconstruction procedure has been refined by us
  \cite{hpcEMC,algEMC} and assembles a 3D Fourier intensity from the
  selected 2D diffraction patterns using the
  Expansion-Maximization-Compression (EMC) algorithm, originally
  developed in \cite{EMC}.

\item[d)] The post-analysis step transfers the 3D Fourier intensity
  into a real-space structure \cite{Fienup:82,phase_recent,
    marchesini2005phase,rra,Li2017}. Before retrieving phases, we may
  subtract background noise, heal \cite{COACS} and window the 3D
  Fourier intensity. The reconstruction uncertainty may be estimated
  via a bootstrap procedure \cite{algEMC}.
\end{enumerate}

An introduction to FXI Methodology and the beam description for the
PR772 experiment are found in \refsec{sec:Methodology}, and our
classification procedure is summarized in \refsec{sec:class}. In
\refsec{sec:EMC}, we outline the procedure to retrieve the 3D Fourier
intensity from the 2D diffraction patterns. In \refsec{sec:post}, we
discuss the phasing problem and how to validate and improve on the
results. A concluding discussion is found in \refsec{sec:conclusion}.

\section{Methodology}
\label{sec:Methodology}

\begin{figure}[!htbp]
  \includegraphics[width=.8\textwidth]{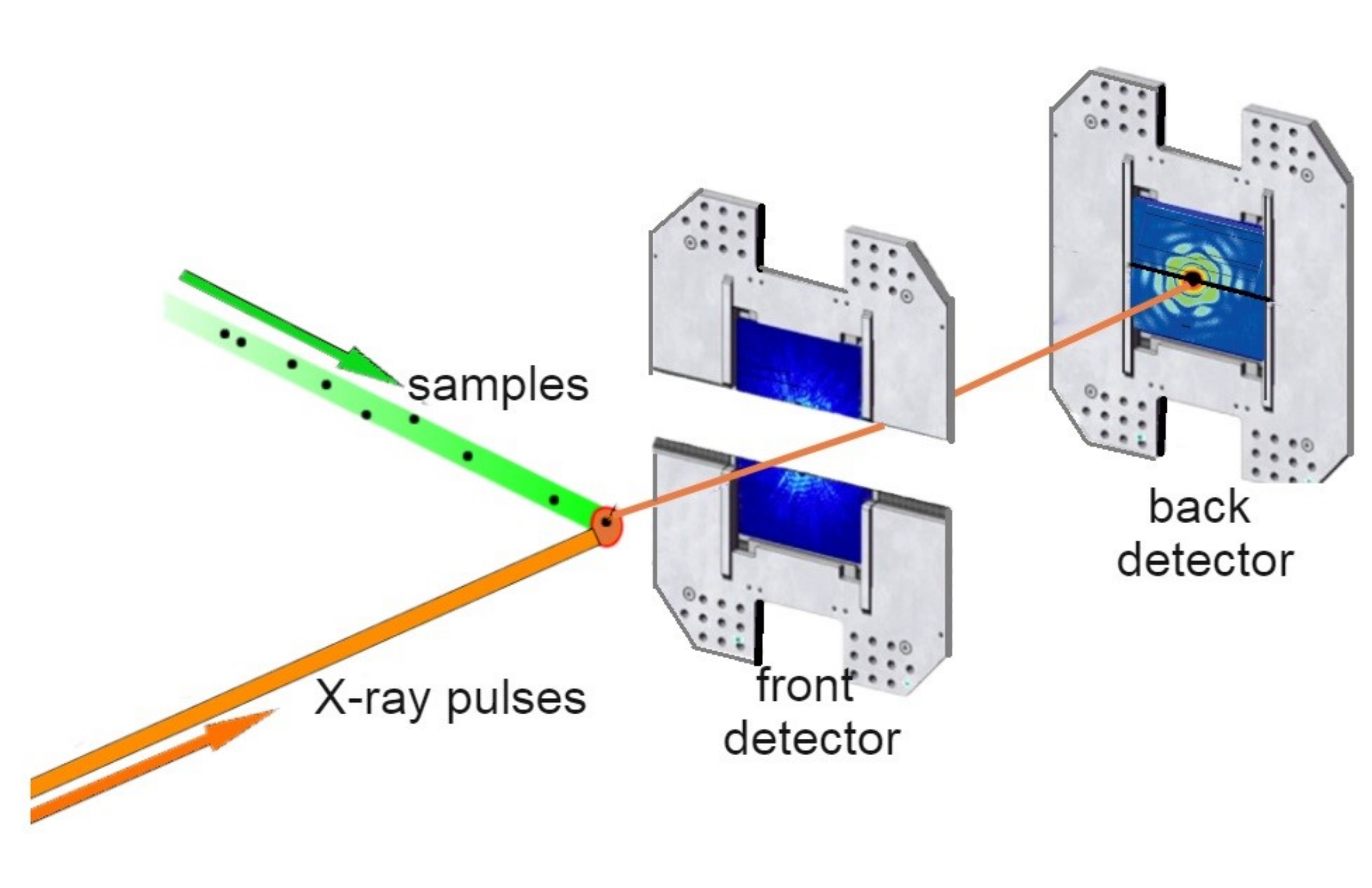}
  \caption{An illustration of the typical FXI experimental setup. The
    front detector captures signals at higher diffraction angles and
    the back detector captures signals at lower angles. The sample
    particles are injected into the X-ray beam at random and hence we
    get diffraction patterns of various types and qualitites, i.e.,
    from single particles, multiple particles, contaminants, and blank
    frames, etc. The particle orientations are unknown, and so is the
    relative strengths (photon fluence $\phi$) of the diffraction
    signals.}
  \label{fig:setup}
\end{figure}

XFEL pulses are short and intense enough to create a perceptible
diffraction signal from one single particle. FXI, which relies on the
``diffract and destroy''~\cite{dad} scheme, uses XFEL pulses to probe
single particles and digital detectors to capture the diffraction
signals depicting the sample just before it turns into an exploding
plasma due to the energy deposited. In a typical FXI experiment setup
(see Figure~\ref{fig:setup}), an effective injector, such as a Gas
Dynamic Virtual Nozzle (GDVN) injector \cite{gdvn, es_uppsala},
creates an evaporated aerosol from particle samples originally
dissolved in a liquid buffer, and injects the aerosol stream into the
X-ray interaction region, where particles interact with the X-ray
pulses randomly. Digital detectors, such as pnCCD \cite{pnccd}, AGIPD
\cite{AGIPD1, AGIPD2}, and other area detectors \cite{CSPAD, EPIC},
are used to record the intensities of the scattered waves.

According to classical diffraction theory \cite{CoherentXrayOptics},
we may consider a sample particle as a collection of infinitesimal
point scatterers. Let $\textbf{x}'$ be a position of the sample
particle, and $\textbf{x}$ a position on the detector. We may then
write the propagated signal of a coherent X-ray plane wave
$\Psi(\textbf{x})$ from sample position $\textbf{x}'$ to detector
position $\textbf{x}$ as follows:
\begin{align}
  \nonumber
  \Psi(\textbf{x}) &= \Psi^{(0)}(\textbf{x}) + \Psi^{(1)}(\textbf{x}) \\
  &=  \Psi_{0} \text{exp} (\textit{i} \, \textbf{k}_0 \textbf{x}) +
  \Psi_{0} \iiint  \frac{\textit{k}^2}{4\pi} [1-\textit{n}^2(\textbf{x}')] \text{exp} (\textit{i} \textbf{k}_0 \textbf{x}') \dfrac{\text{exp} (-\textit{i} \textit{k}|\textbf{x} -\textbf{x}' | )}{|\textbf{x}-\textbf{x}'|}d\textbf{x}',
  \label{eq:scattering_theory}
\end{align}
where $\Psi^{(0)}(\textbf{x})$ represents the unscattered wave of
$\Psi(\textbf{x})$ at $\textbf{x}$ with amplitude $\Psi_0$. Further,
$\Psi^{(1)}(\textbf{x})$ is the scattered component, which accounts
for the superposition of the many spherical waves that emits from
point scatterers $\textbf{x}'$ to detector position
$\textbf{x}$. Further, $\textit{n}$ is the refractive index, $
\textbf{k}_0$ is the wave vector of the incoming wave, and
$\textit{k}$ is the wave number.

Considering the following characteristics of FXI experiments: i)
detectors are in the far-field capturing 2D intensities of the
scattered wave, ii) the captured signal contains only a small part of
the diffraction angles, taken together this means that we can
approximate the intensity of the scattered wave on the detector as
\begin{align}
I \propto I_0 | \Fourier(\textbf{q}_{\perp}) | ^2,
\label{eq:I_p}
\end{align}
with $I_0$ the X-ray pulse intensity at the interaction region, and
$\Fourier$ the Fourier transformation. Further, $\textbf{q}$ is a
scattering vector that lies on a sphere in diffraction space, i.e.,
the so-called Ewald sphere, and $\textbf{q}_{\perp}$ denotes the
subset of scattering vectors which are perpendicular to the beam
direction.

In fact, to obtain interpretable signals containing structural
information of single particles (or non-periodic objects), FXI
requires X-ray pulses of a few femtosecond duration of more than
$10^{12}$ photons, and the beam focus is much larger than the
particles \cite{XFELbook2018}. The ultra-short duration can outrun
conventional radiation damage processes, allowing the capture of the
\emph{continuous} diffraction patterns on the detectors.  Note that
FXI experiments also have many stochastic features.  First, the number
of particles interacting with the X-ray pulse can vary, i.e., the raw
dataset from an experiment will contain single-particle diffraction
patterns, multiple-particles patterns, blank frames with only
background noise, and frames with signals from contaminants, etc. We
therefore need to find the high-quality single-particle diffraction
patterns in steps a) and b) of our proposed pipeline. Second,
particles are observed in random orientations and hence the inversion
procedure from 2D single-particle diffraction patterns into 3D
structures is critical. Third, the strength of the diffraction signals
will vary a lot, mainly due to the location of the particle within the
X-ray focus. We refer the relative strength of the diffraction signal
to the \emph{photon fluence} and denote it by $\phi$. In our pipeline,
we determine diffraction-pattern orientations and photon-fluence
distribution together in step c). Last but not least, the diffraction
patterns and the reconstructed 3D objects are Fourier intensities
without phase information. We need to handle noise, aliasing effects,
etc., together with measuring uncertainties of phase retrieval and 3D
reconstruction process in our post-analysis step d).

\subsection{Properties of the PR772 FXI Experiment at LCLS Beamtime AMO86615 }
\label{sec:data}

The bacteriophage PR772 virus is a member of the Tectiviridae family
and has an approximate 70-nm-diameter icosahedral protein capsid,
which encapsulates dsDNA and various internal
proteins~\cite{772_cyroEM}. The PR772 virus is also known to have high
structural homogeneity and uniform size distribution. Given these
merits, we choose to use the bacteriophage PR772 virus AMO88615
experimental data \cite{pr772_sd} conducted at the Atomic Molecular
Optics (AMO) instruments at the LCLS to illustrate our proposed
pipeline.

The Beamtime AMO86615 used a photon energy of 1.6 keV and a 70-fs
pulse duration \cite{pr772_sd}. The beam focus was around 1.5 $\mu
m^2$, and the power density was approximately $10^{13}$ photons per
pulse without beam losses. The sample viruses were aerosolized by a
GDVN~\cite{gdvn} injector before being injected into an X-ray beam at
random orientations. The diffraction signals were captured by two
pairs of pnCCD detectors \cite{pnccd} in the far-field. In this paper,
we ignore the data from the front detector (placed at 100 mm
downstream of the interaction region), and only use diffraction
signals from the back detector (located 581 mm downstream of the
interaction region). The back detector had a 75 $\mu m^2$ pixel size,
and $1024 \times 1024$ pixels in total. Further, the resolution of the
back detector was 11.6 nm at the edge or 8.3 nm in the corner.

The full PR772 dataset \cite{pr772_sd} of AMO86615 consisted of
$2,976,568$ raw diffraction patterns, which contained patterns with
varying quality, including, e.g., misses, single hits, multiple hits,
impurities, etc. Further, \cite{pr772_sd, cxidb_entry58} reported a
selected subset of $N_{14k} = 14,772$ patterns using diffusion map
embedding~\cite{chi,novelsort}.  In $N_{14k}$ only 12,678 patterns
were identified as single particle hits.  It also reported a subset of
$N_{7k} = 7,992$ patterns from the principal component
analysis~\cite{PCAclassification}, and a subset of $N_{37k}=37,550$
patterns from manifold embedding techniques~\cite{me}.

In the following sections, we use the first subset
$N_{14k}$~\cite{pr772_sd}, which was downloaded from the Coherent
X-ray Imaging Data Bank (CXIDB)~\cite{cxidb} entry
58~\cite{cxidb_entry58}. The downloaded diffraction patterns in
$N_{14k}$ are in units of photon counts and are binned into a
$256\times 256$ grid.


\section{Classification}
\label{sec:class}

With the EigenImage classifier proposed in \cite{psort}, the aim is to
push a selected and limited number of high quality diffraction
patterns to the next step of the analysis pipeline such that the 3D
reconstruction step becomes as efficient as possible. To select
suitable patterns from $N_{14k}$, we used a synthetic training dataset
constructed by Condor \cite{Condor}, and consisting of 1,000 randomly
oriented uniform-density icosahedra, 70~nm in size.  The EigenImage
classifier trained from only icosahedral patterns performs badly if
presented with non-icosahedral patterns. Unfortunately, the dataset
$N_{14k}$ contained a small portion of unwanted particles, such as
spherical patterns and multiple-hits patterns. To find those unwanted
patterns, we added one 70~nm spherical pattern into the training
dataset, and we eliminated the patterns matched with the spherical
pattern from the final results. All synthetic patterns used for
training were masked with a central $36$-pixel in diameter circular
mask and a central 5-pixel-width vertical strip mask.

\begin{figure*}[!htbp]
  \centering
  \subfloat[]{\includegraphics{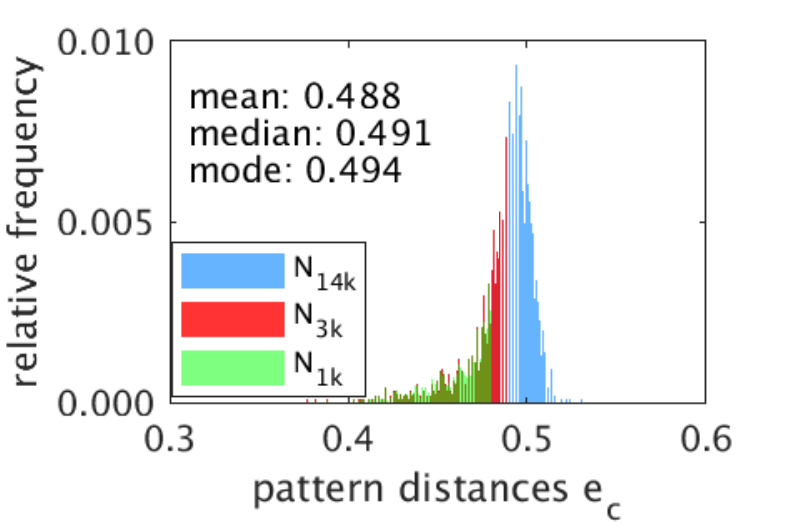} \label{fig:DCerror}}
  \subfloat[]{\includegraphics{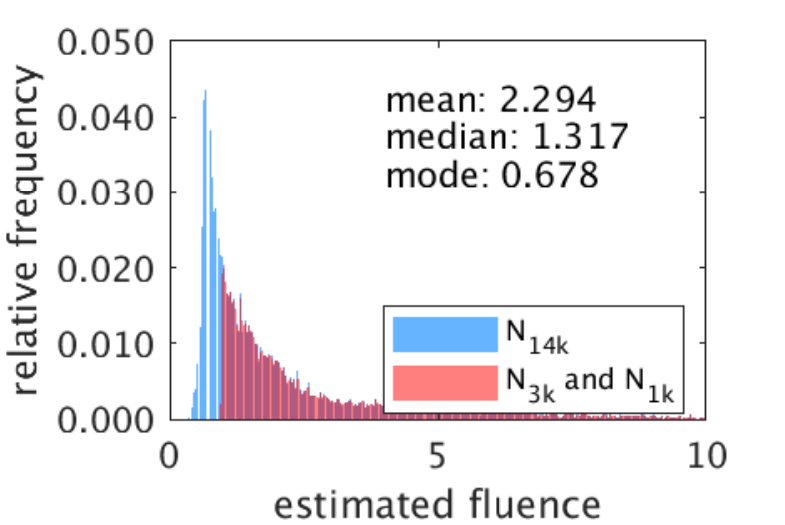}  \label{fig:Dfluence}} \\
  \subfloat[]{\includegraphics{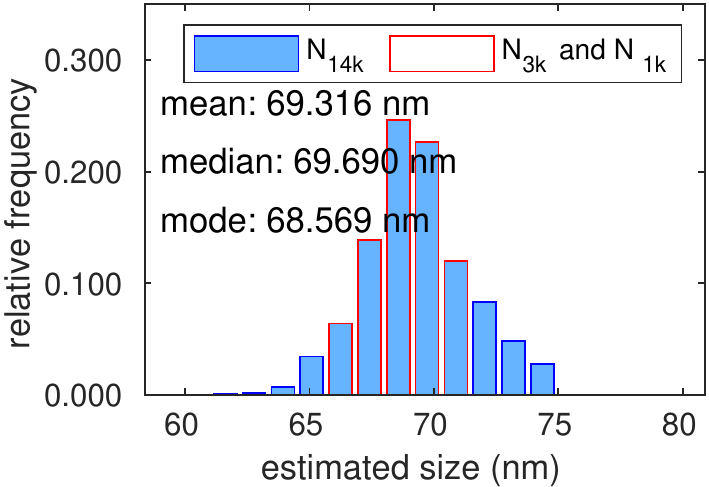} \label{fig:DPsize}}
  \caption{Statistical measures from the EigenImage
    classifier. \protect\subref{fig:DCerror}: pattern distances (the
    relative Euclidean distance between the testing pattern and the
    best matched template) \protect\subref{fig:Dfluence}: estimated
    fluence (the ratio of the total photon counts in the test pattern
    to the best-matched template), and \protect\subref{fig:DPsize}:
    estimated particle sizes.  In \protect\subref{fig:DCerror} and
    \protect\subref{fig:Dfluence}, we used 1000 equal sized bins, and
    the bin sizes were $2.062 \times 10^{-4}$ and 0.0381 for
    \protect\subref{fig:DCerror} and \protect\subref{fig:Dfluence},
    respectively. In \protect\subref{fig:DPsize}, we used 12 equal
    sized bins of size 1.159~nm. Consult \cite{psort} for the notation
    and definitions of pattern distance and estimated fluence.}
  \label{fig:classification_res}
\end{figure*}

With the EigenImage classifier, we selected two smaller subsets from
$N_{14k}$ based on the pattern distance (the relative Euclidean
distance $e_c$ between the testing pattern and the best matched
template \cite{psort}), see Figure~\ref{fig:classification_res}. When
selecting, we simultaneously excluded spherical patterns, and weak
patterns for which the estimated fluence (the ratio of total photons
in the testing pattern to the best matched template) was smaller than
1 (Figure~\subref*{fig:Dfluence}), and the estimated particle size was
either larger than 72~nm or smaller than 67~nm
(Figure~\subref*{fig:DPsize}). By this procedure we selected in total
$N_{1k} = 1,084$ patterns with pattern distance $e_c < 0.48$ and
$N_{3k}=3,140$ frames with $e_c < 0.49$
(Figure~\subref*{fig:DCerror}). On average, the selected particle size
was 68.9~nm for the dataset $N_{1k}$ and 69.4~nm for the dataset
$N_{3k}$.

\section{Maximum Likelihood Imaging}
\label{sec:EMC}

To reconstruct 3D Fourier intensities, we applied the
Expansion-Maximization-Compression (EMC) algorithm as detailed in
\cite{hpcEMC} using the scaled Poissonian probability model
\cite{algEMC}. The scaled Poissonian assumes that the $i$th pixel of
the $k$th measured diffraction pattern $K_{ik}$ is Poissonian around
the unknown Fourier intensity $W_{ij}$ with the relative fluence
$\phi_{jk}$:
\begin{equation}
  \Prob(K_{ik} = \kappa | W_{ij},R_j,\phi_{jk}) = \prod_{j=1}^{\Mrot} \dfrac{(W_{ij} \phi_{jk})^\kappa e^{-W_{ij} \phi_{jk}}}{\kappa!},
  \label{eq:scaledPoisson}
\end{equation}
where $i$, $j$, and $k$ are indices to image pixels, rotations (or
slices), and images, respectively. $R_j$ is the $j$th sample rotation,
and $\Mrot$ is the total number of rotation samples, where
$\Mrot = 50,100$ is a typical resolution. With both efficiency and
robustness in mind we binned the selected diffraction patterns
$4\times$, and those binned images were used in the computations until
EMC met its stopping criterion \cite{algEMC}. Further, to reduce the
smearing effect in the final 3D Fourier intensity, we reran the final
compression step using slightly more detailed patterns ($3\times$
binning) with 5-pixels-wide zero paddings around the patterns,
yielding a final 3D Fourier intensity of size $96 \times 96 \times 96$
voxels.

\begin{figure}[!htbp]
  \centering
  \subfloat[]{\includegraphics{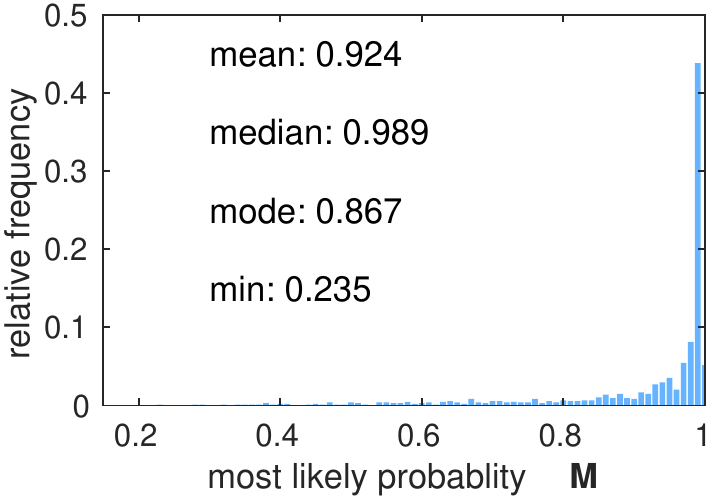} \label{fig:1kprob}}
  \subfloat[]{\includegraphics{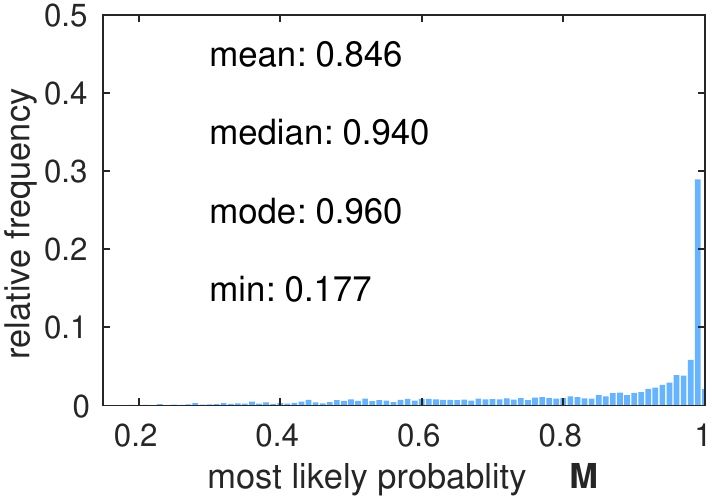}\label{fig:3kprob}}  \\
  \subfloat[]{\includegraphics[width=.5\textwidth]{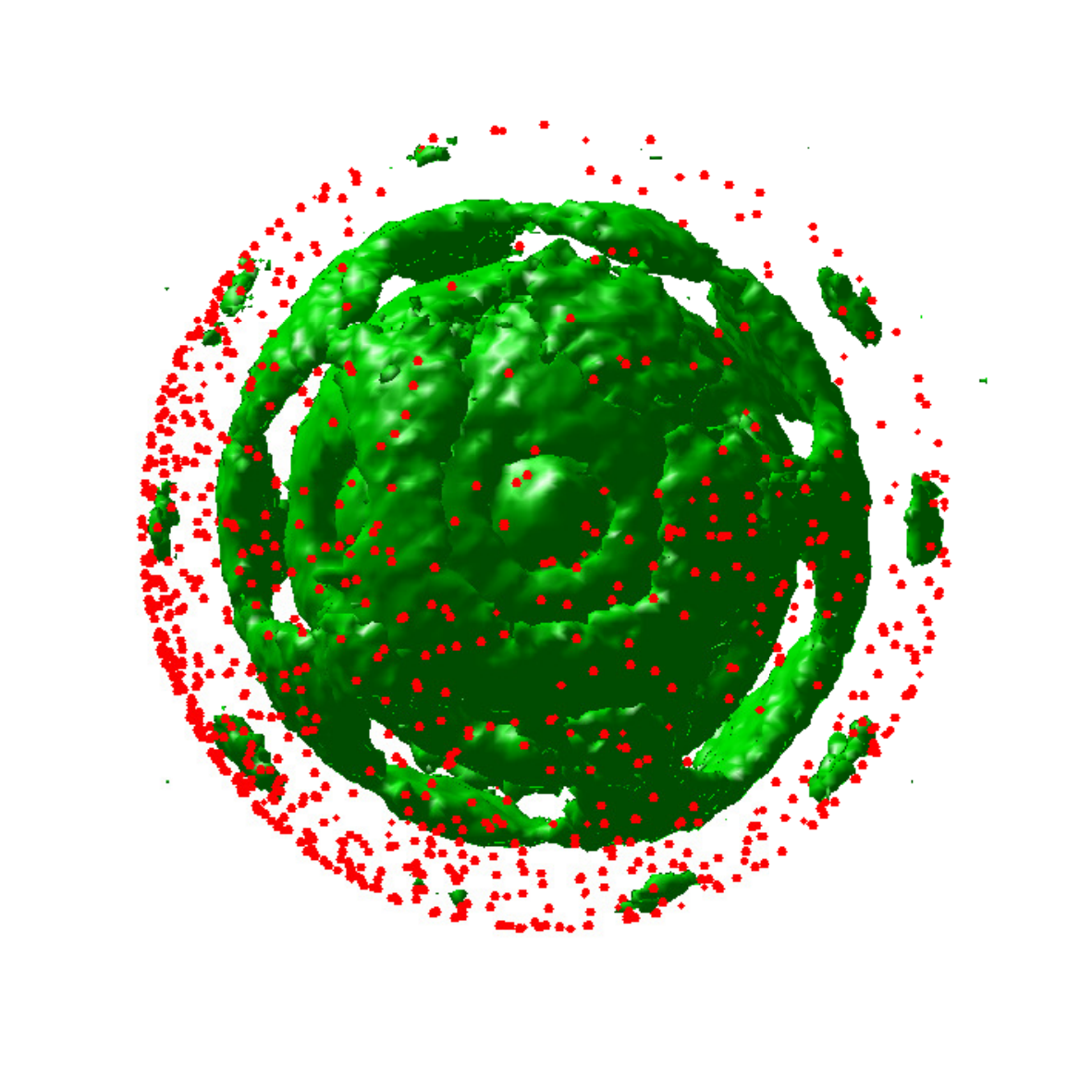} \label{fig:1kEMC}}
  \subfloat[]{\includegraphics[width=.5\textwidth]{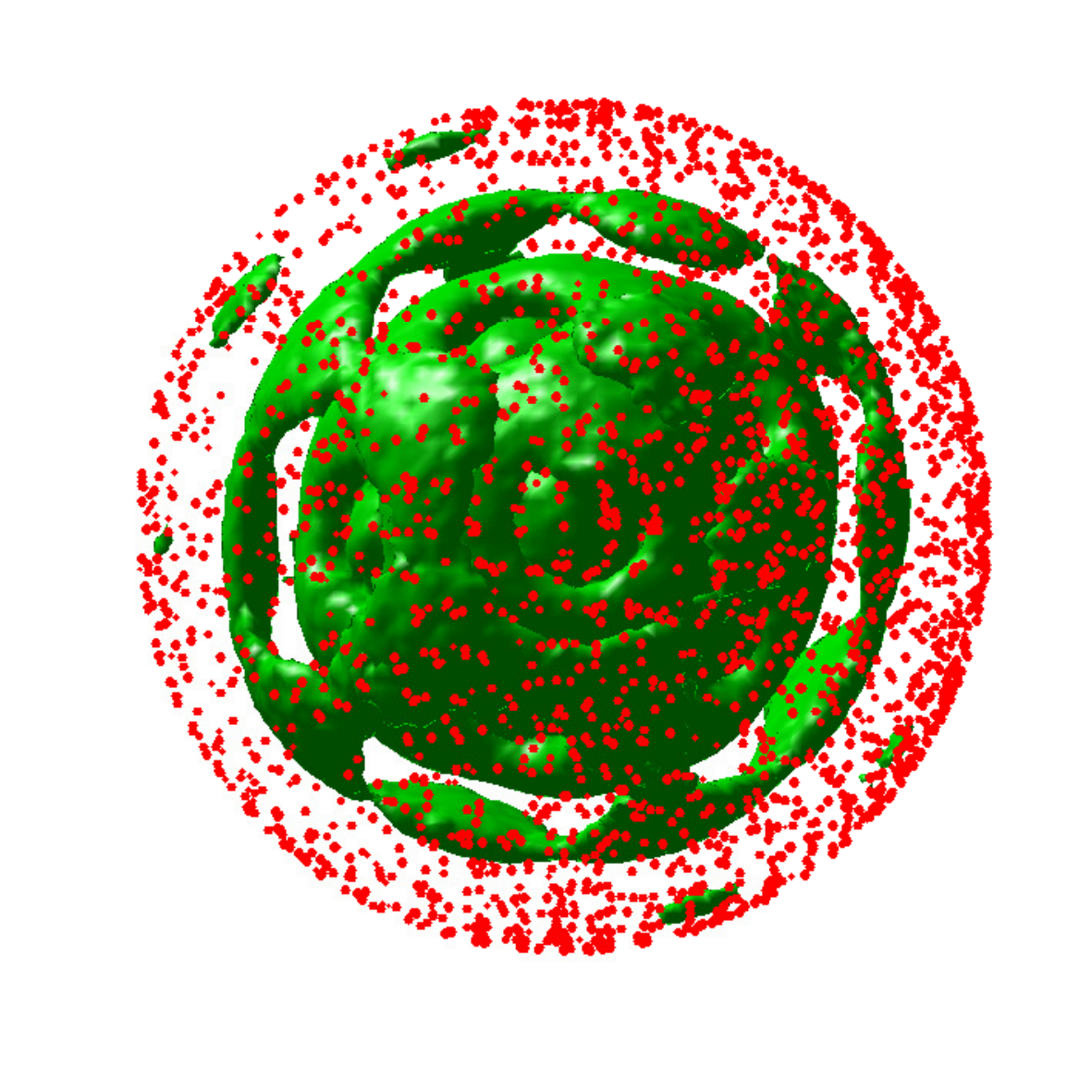}\label{fig:3kEMC}}\\
  \caption{The most likely probability $\mathbf{M}$ at the final EMC
    iteration for datasets $N_{1k}$ \protect\subref{fig:1kprob} and
    $N_{3k}$ \protect\subref{fig:3kprob}, cf.~\eqref{eq:maxP}. The
    corresponding isosurface plots of Fourier intensities are
    \protect\subref{fig:1kEMC} and \protect\subref{fig:3kEMC},
    respectively, and the red dots show the corresponding rotations of
    the most likely probabilities.}
  \label{fig:emc_prob}
\end{figure}

To quantify the fitness of the datasets to the scaled Poissonian
probability model, we studied the most likely rotations for each
diffraction pattern, see Figure~\ref{fig:emc_prob}. Let $P =(P_{jk})$
for $j = 1,\ldots,\Mrot$ and $k=1,\ldots,\Mdata$ be the normalized
rotational probabilities calculated at the Expectation step of the EMC
algorithm, for $\Mrot$ the number of sampled rations, and $\Mdata$ the
number of input diffraction patterns. The most likely probabilities
are then
\begin{align}
 \mathbf{M}_k = \max_j P_{jk},
 \label{eq:maxP}
\end{align}
Since $P$ is normalized ($\sum_j P_{jk} = 1 $), the most likely
probabilities $\mathbf{M}$ range from $\Mrot^{-1}$ to one. For a
pattern composed of random noise, the expected probability is
approximately $1/\Mrot \approx 2\times10^{-5}$ with $\Mrot =
50,100$. For a pattern fitted into one specified rotation perfectly,
the most likely probability of that pattern will be one. For our
selected datasets, the smallest most likely probability (the smallest
value in $\mathbf{M}$) of $N_{1k}$ and $N_{3k}$ were 0.235 and 0.177,
respectively, indicating that the selected patterns fitted well into a
3D intensity using the scaled Poissonian model. Further, more than
43\% and 28\% patterns from $N_{1k}$ and $N_{3k}$ fitted into only one
definite rotation. Moreover, the locations of the most likely
rotations were not evenly distributed in 3D space, suggesting an
asymmetric structure in the particle electron density in
real space. We also observe that Figure~\subref*{fig:3kEMC} is much
denser and its isosurface is smoother than Figure~\subref*{fig:1kEMC},
as the number of frames in $N_{3k}$ are almost 3 times larger than in
$N_{1k}$.

We measured the contrast of local intensity extrema \cite{pr772_ros}
to quantify the quality of the obtained 3D Fourier intensity,
\begin{align}
  C = \dfrac{I_{max} - I_{min}}{I_{max} + I_{min}},
  \label{eq:contrast}
\end{align}
where $I_{max}$ and $I_{min}$ are neighbouring maxima and minima pairs
along a line that crossed the particle center.  For each
reconstruction, we averaged the contrast over all maxima and minima
pairs for 100 random lines and and we obtained the average contrasts
0.5792 for $N_{1k}$ and 0.5874 for $N_{3k}$. Notably both contrasts
were better than the contrast of the 3D Fourier intensity with
background noises in \cite{pr772_ros} (0.4955).

Our scaled Poisson EMC implementation took up to 40 iterations for
recovering the rotational probabilities of both datasets. The
computing time was around 10 minutes for $N_{1k}$ and 41 minutes for
$N_{3k}$ using 3 Nvidia GeForce GTX 680 GPUs in parallel  (see 
Appendix~A for more details of the reconstruction).


\section{Post Analysis}
\label{sec:post}


In this section, we detail our design of the post-analysis workflow
with respect to phase retrieval, background noise removal, and the
bootstrap analyses in both real and Fourier space. We also provide
phasing error statistics and the shape analysis of the retrieved
electron electron densities of the PR772 virus for both $N_{1k}$, and
$N_{3k}$.
	 
To phase an EMC intensity model and thus retrieve a 3D electron
density distributions, we used a combination of algorithms --- 10,000
iterations of the relaxed averaged alternating reflections (RAAR)
\cite{rra,Li2017} followed by 2,000 iterations of the Error Reduction
(ER) \cite{Fienup:82} algorithm. The final electron density
distribution was computed as an average of 100 phased objects. Note
that we reconstruct and phase our Fourier intensities without any
symmetry constraints. As suggested in \cite{marchesini2005phase}, we took
the average of the the absolute values of the phased objects to
further reduce the low-order phase errors.

\subsection{Original Object}
\label{subsec:phase_ori}

\begin{figure}[!htbp]
  \centering
  \subfloat[]{\includegraphics[width=.2\textwidth]{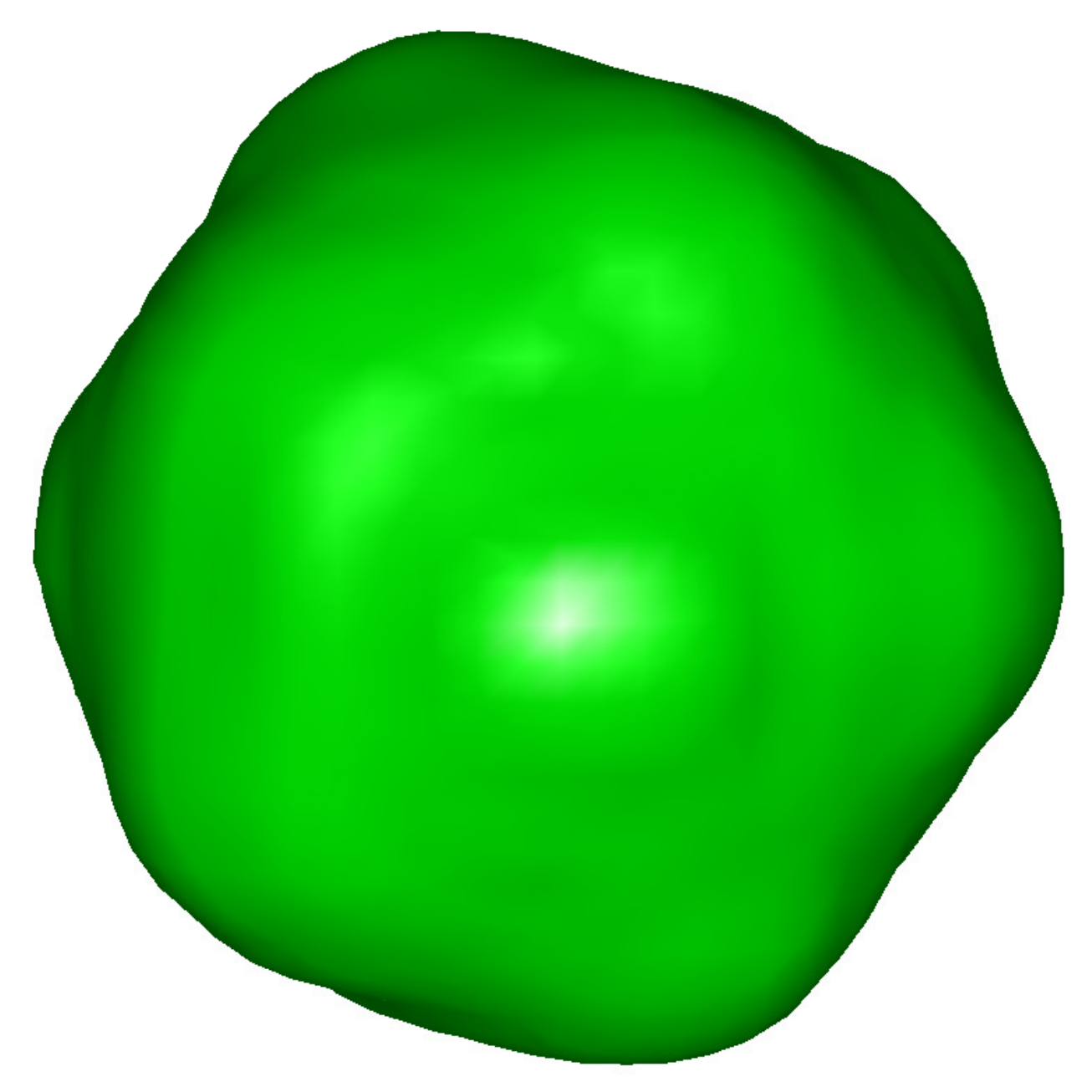} \label{fig:3D_1k_ori_iso}}
  \subfloat[]{\includegraphics[width=.2\textwidth]{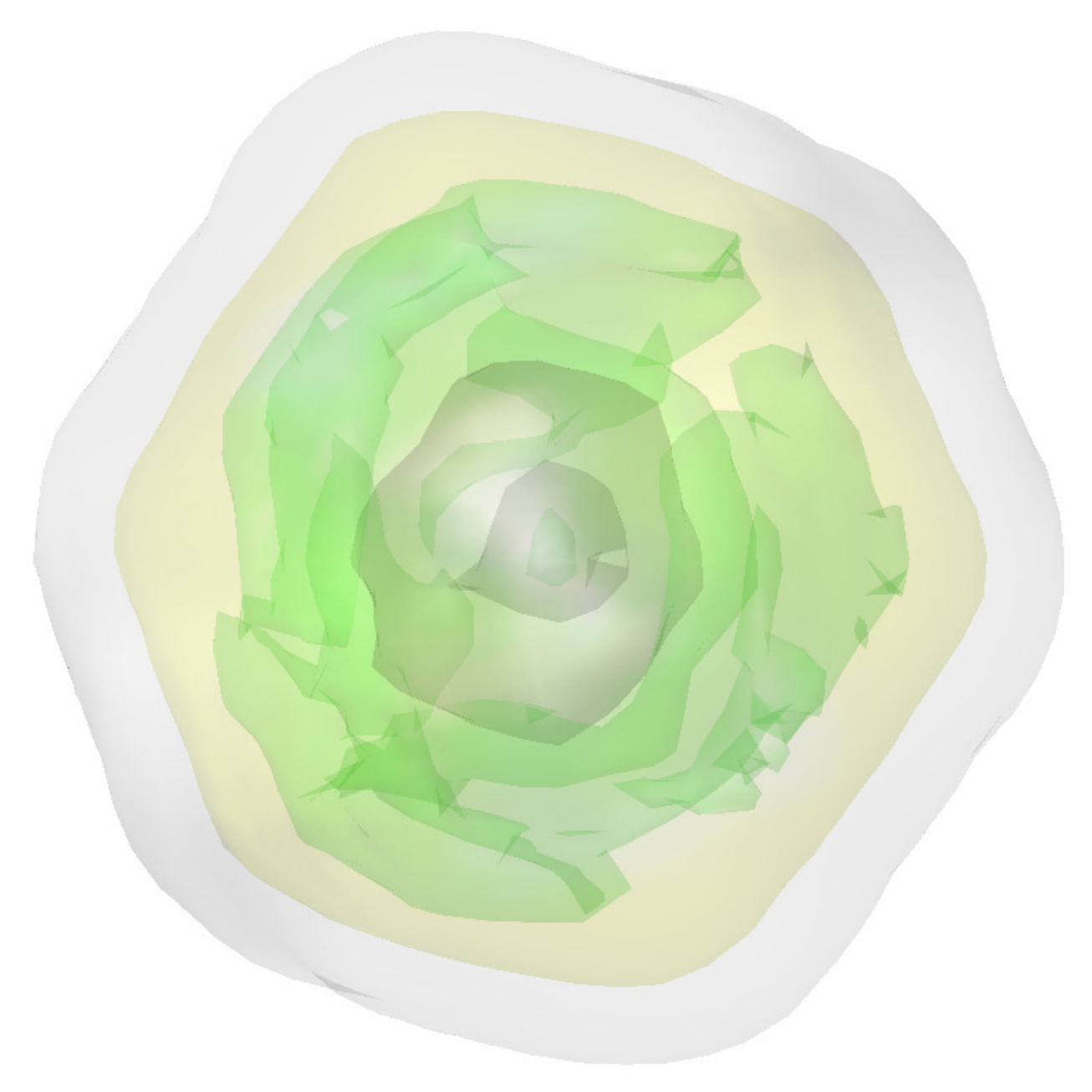} \label{fig:3D_1k_ori_real}}
  \subfloat[]{\includegraphics[width=.2\textwidth]{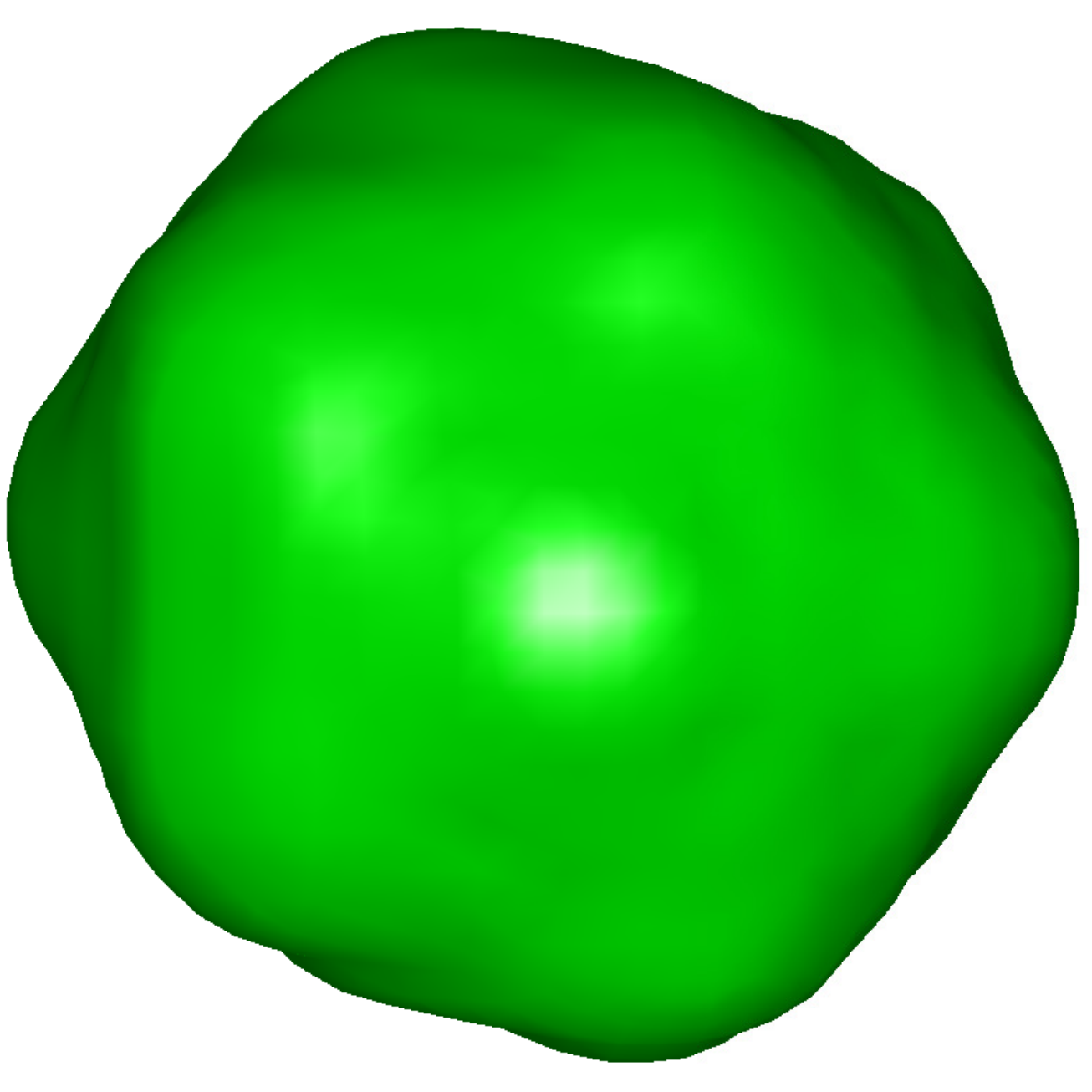} \label{fig:3D_3k_ori_iso}}
  \subfloat[]{\includegraphics[width=.2\textwidth]{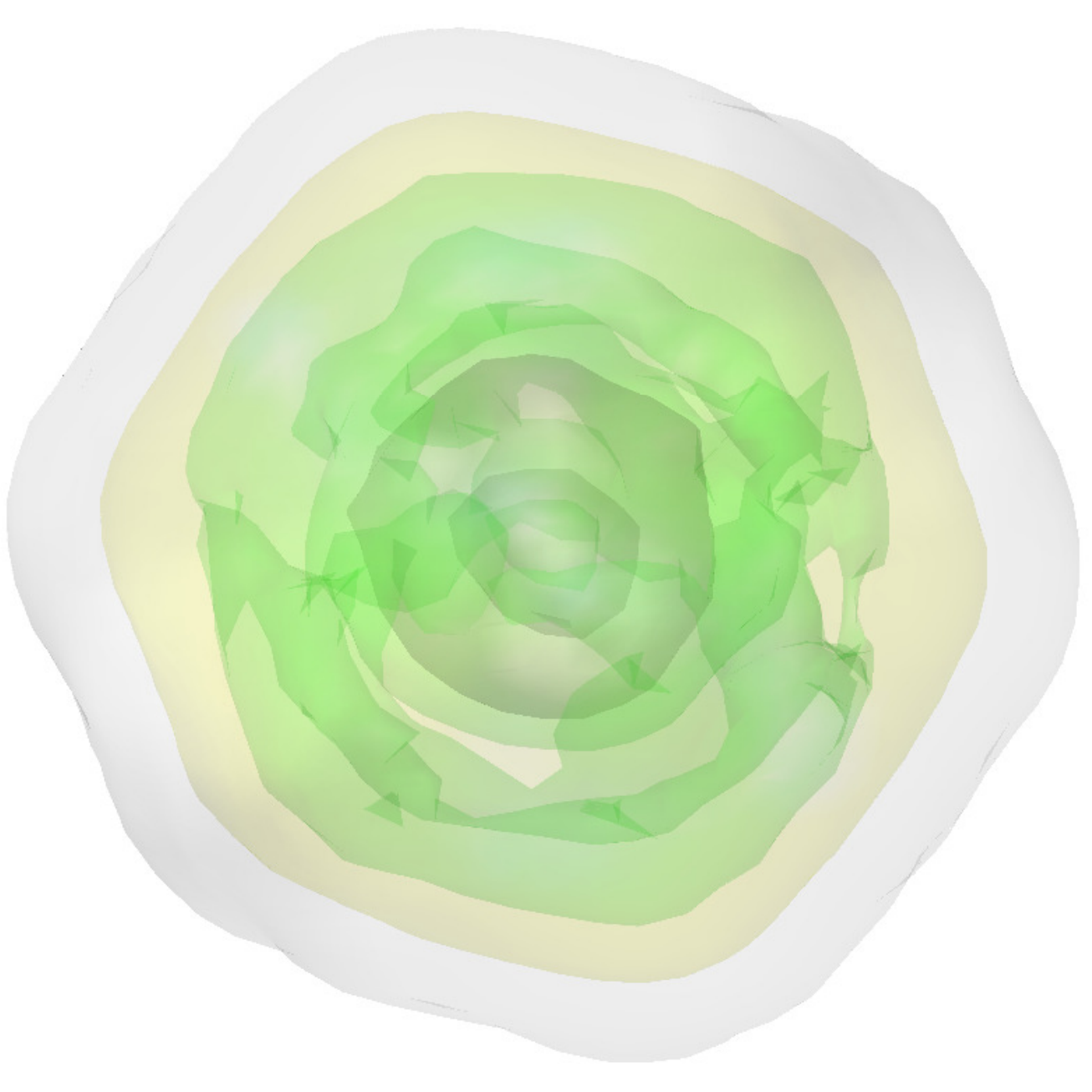} \label{fig:3D_3k_ori_real}}
  \hfill\null
  \\
  \subfloat[]{\includegraphics[width=.2\textwidth]{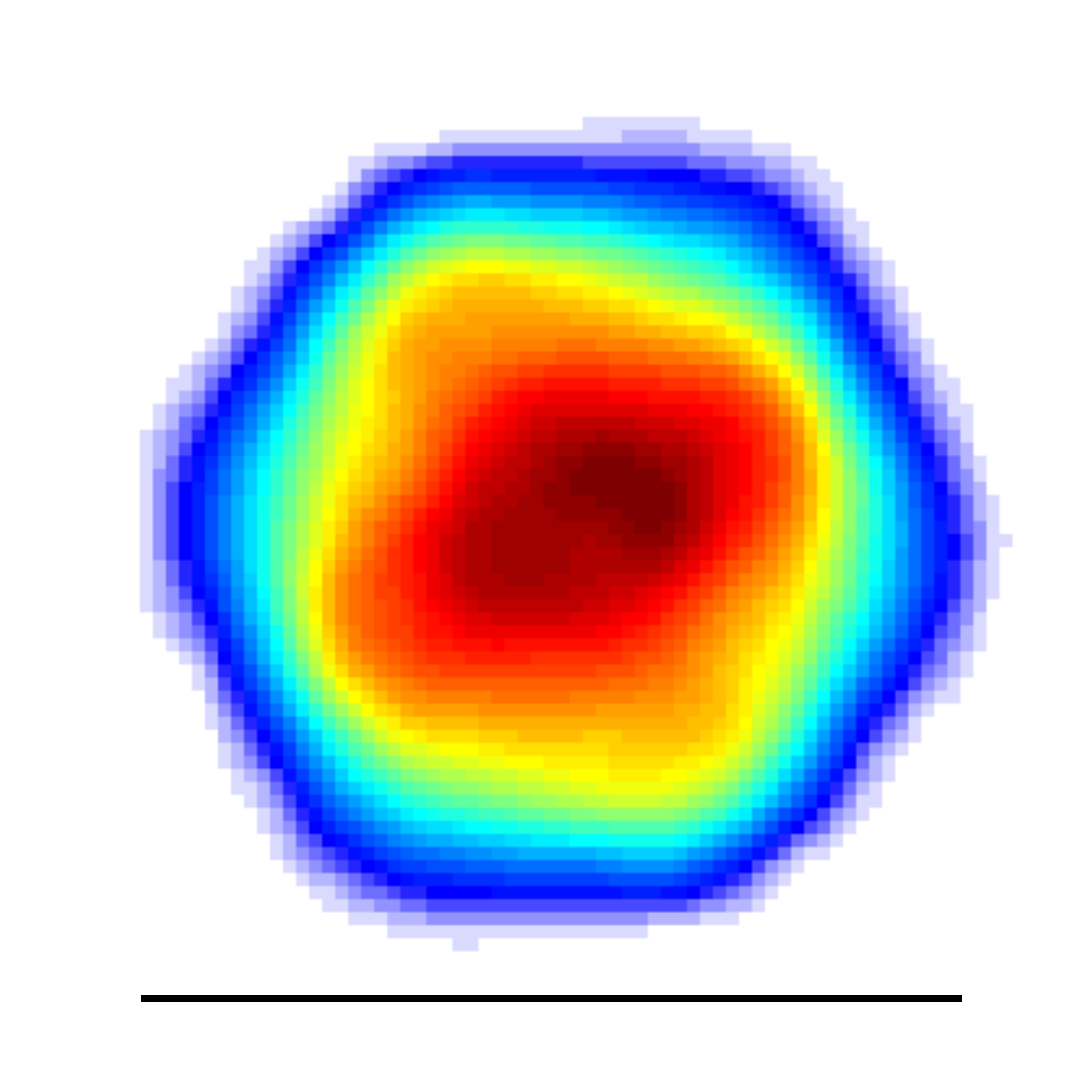} \label{fig:3D_1k_ori_proj}}
  \subfloat[]{\includegraphics[width=.2\textwidth]{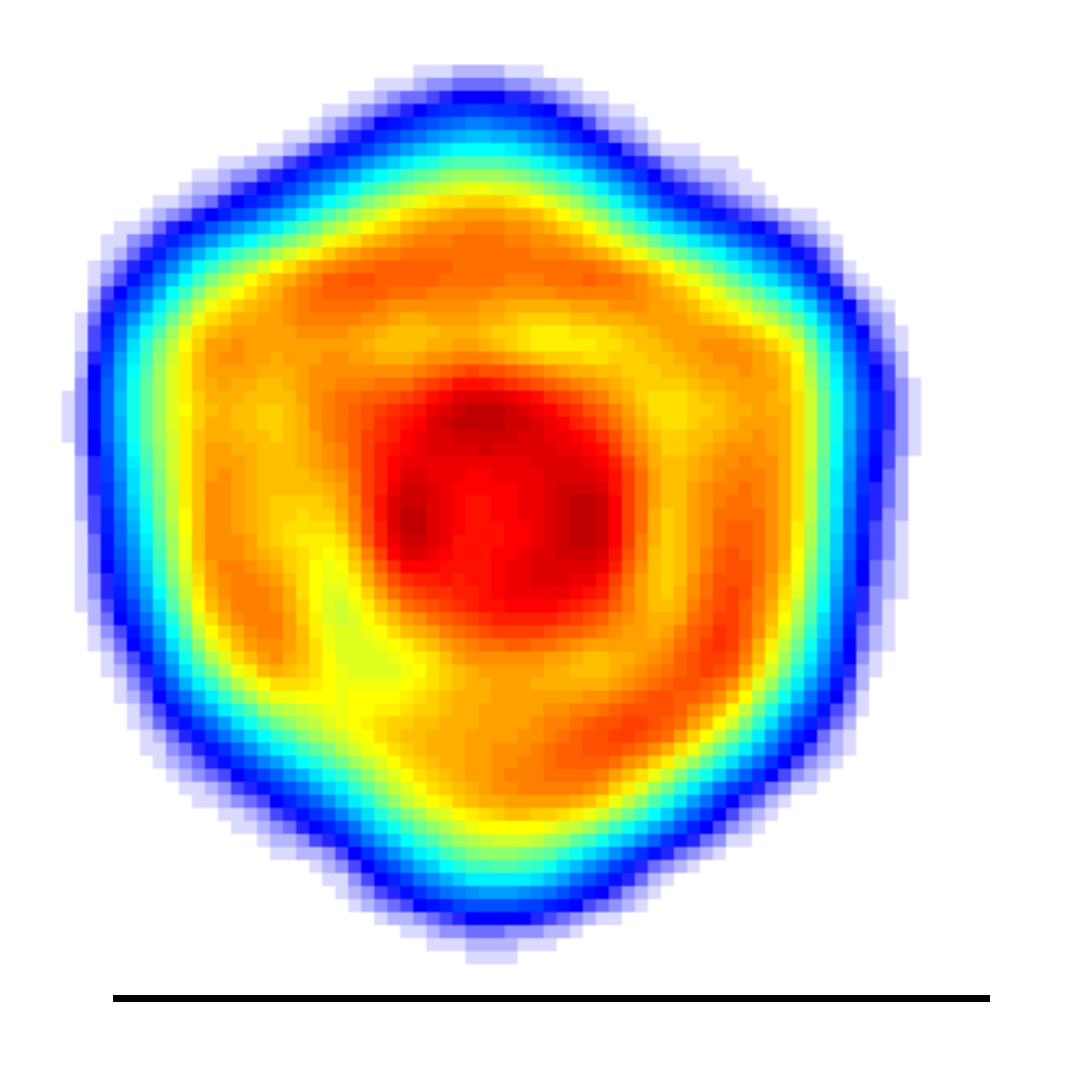} \label{fig:3D_1k_ori_slice}}
  \subfloat[]{\includegraphics[width=.2\textwidth]{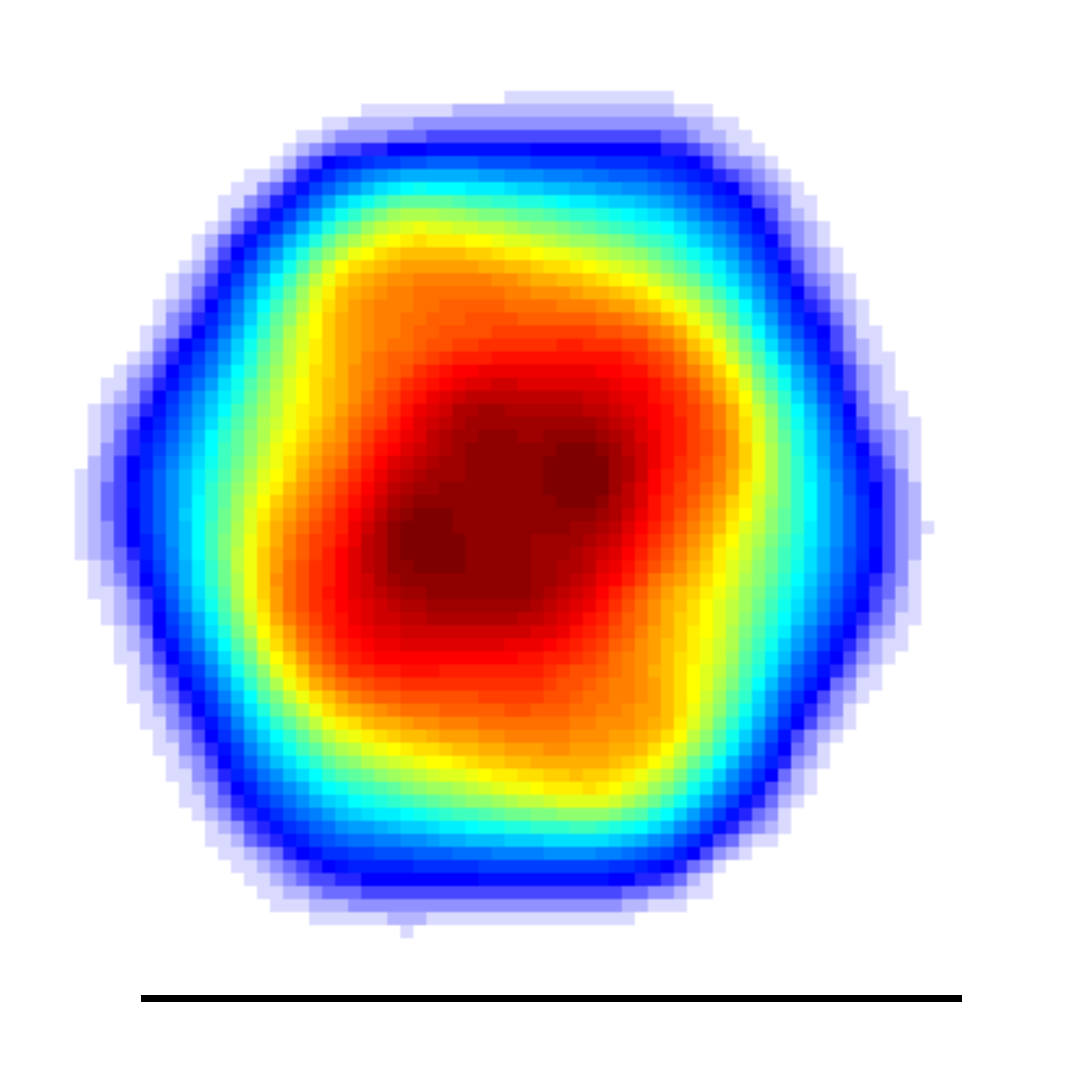} \label{fig:3D_3k_ori_proj}}
  \subfloat[]{\includegraphics[width=.2\textwidth]{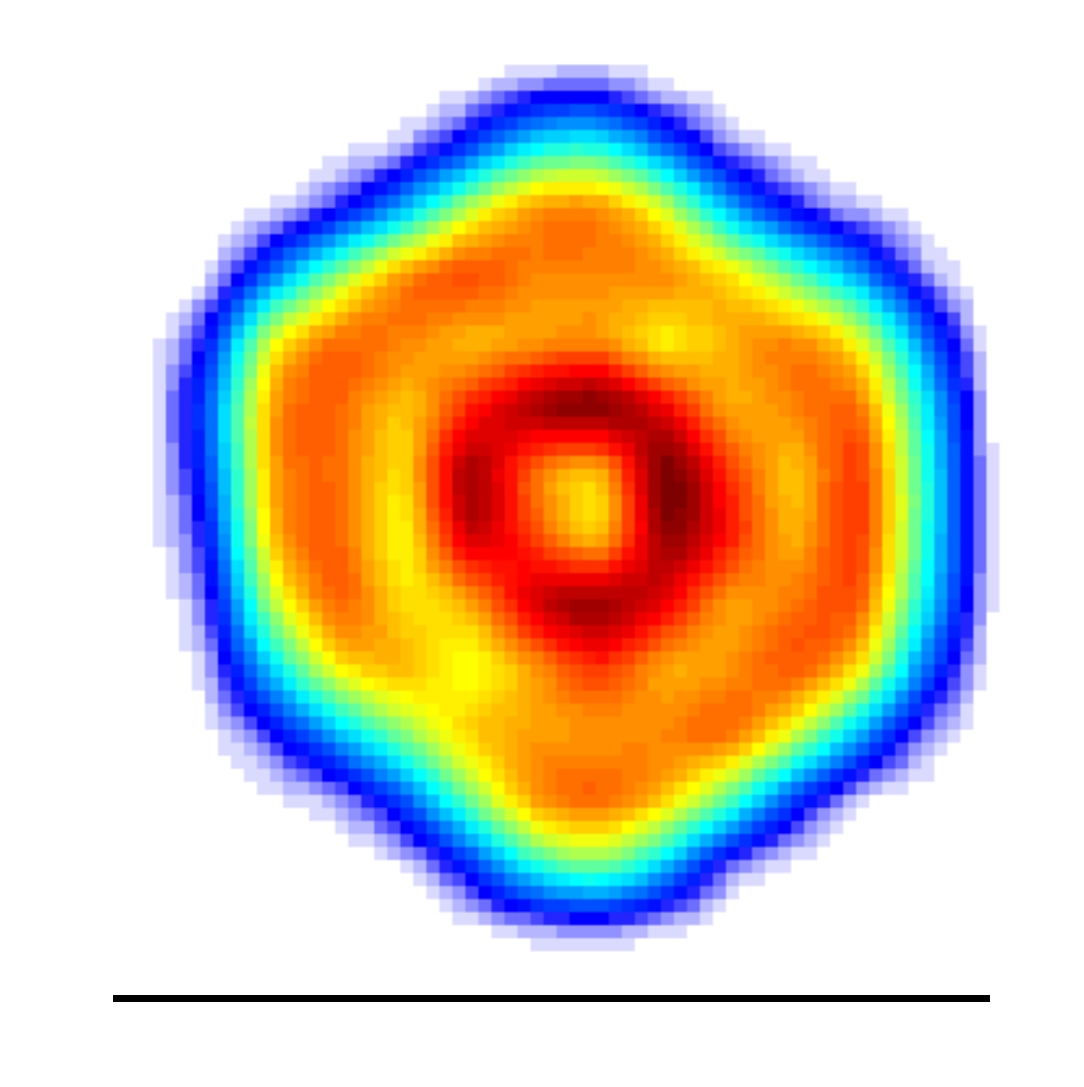} \label{fig:3D_3k_ori_slice}}
  \subfloat{\includegraphics[width=.2\textwidth]{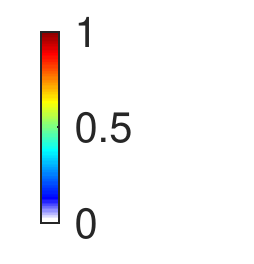}  \label{fig:ori_colorbar}}\\
  \caption{The 3D electron density distributions of the PR772 virus.
    \protect\subref{fig:3D_1k_ori_iso} and
    \protect\subref{fig:3D_3k_ori_iso} are the isosurface plots at
    10\% values of the maximum electron density for $N_{1k}$ and
    $N_{3k}$, respectively.  \protect\subref{fig:3D_1k_ori_real} and
    \protect\subref{fig:3D_3k_ori_real} are the corresponding interior
    structures as isosurface plots at 10\%, 50\%, 79\% and 89\% values
    of the maximum electron density.  In the second row,
    \protect\subref{fig:3D_1k_ori_proj} and
    \protect\subref{fig:3D_3k_ori_proj} are projection images from
    \protect\subref{fig:3D_1k_ori_iso} and
    \protect\subref{fig:3D_3k_ori_iso}, and
    \protect\subref{fig:3D_1k_ori_slice} and
    \protect\subref{fig:3D_3k_ori_slice} are cross-section slices
    through the particle center.  The black scale bars denote
    70~nm. The resolution was 10.7~nm according to a PRTF analysis,
    which was better than the detector edge resolution (11.6~nm), see
    also Figure~\ref{fig:prtfs}. The particle size was determined to
    be 68.9~nm and 69.4~nm, respectively, for $N_{1k}$ and $N_{3k}$.}
  \label{fig:phased_ori}
\end{figure}

The result of the straight-forward phase retrieval is shown in
Figure~\ref{fig:phased_ori}.
As can be seen, the recovered particle had pseudo-icosahedral capsids
with asymmetric interior structures. The particle shape deviated from
an ideal icosahedral symmetry as a low density hole existed closed to
one of the facets, see [Figure~\subref*{fig:3D_1k_ori_slice},
Figure~\subref*{fig:3D_3k_ori_slice}]. Moreover, we observed three
concentric layers, which might reflect the character of the
Tectiviridae family~\cite{Miyazaki}. However, given the detector edge
resolution of 11.6~nm, we judge that the concentric-layers structure
is an artefact due to aliasing and noise (see Appendix~B for more
details). Further, the recovered particle from $N_{3k}$ was smoother,
denser, and slightly larger than the one from $N_{1k}$. The
resolutions determined by the phase retrieval transfer function (PRTF)
\cite{Chapman:06} were around 10.7 nm for both reconstructions at the
$e^{-1}$ threshold, see Figure~\ref{fig:prtfs}.

\subsection{Background Noise}
\label{subsec:phase_bgrm}

\begin{figure}[!htbp]
  \centering
  \subfloat[]{\includegraphics[width=.2\textwidth]{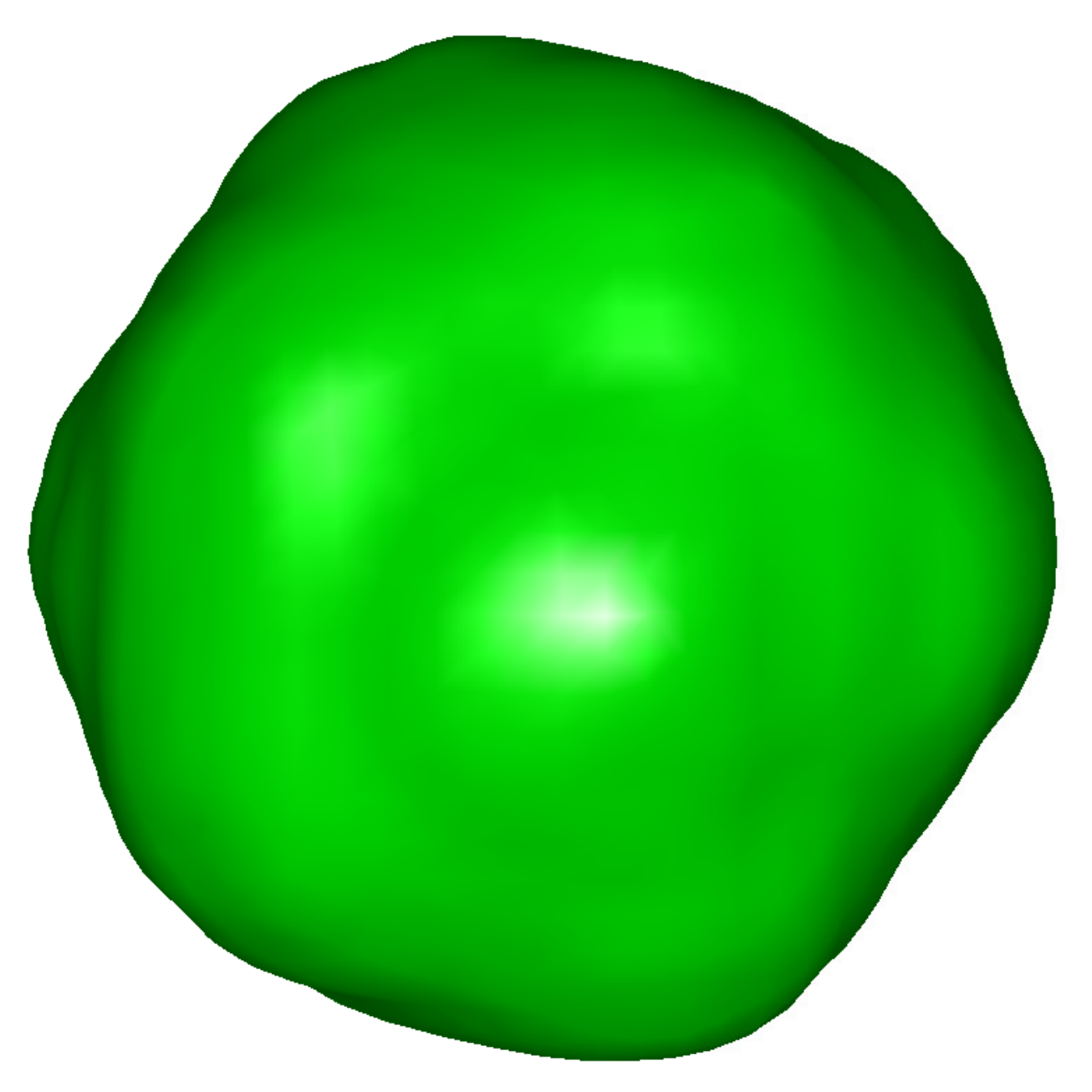} \label{fig:3D_1k_bgrm_iso}}
  \subfloat[]{\includegraphics[width=.2\textwidth]{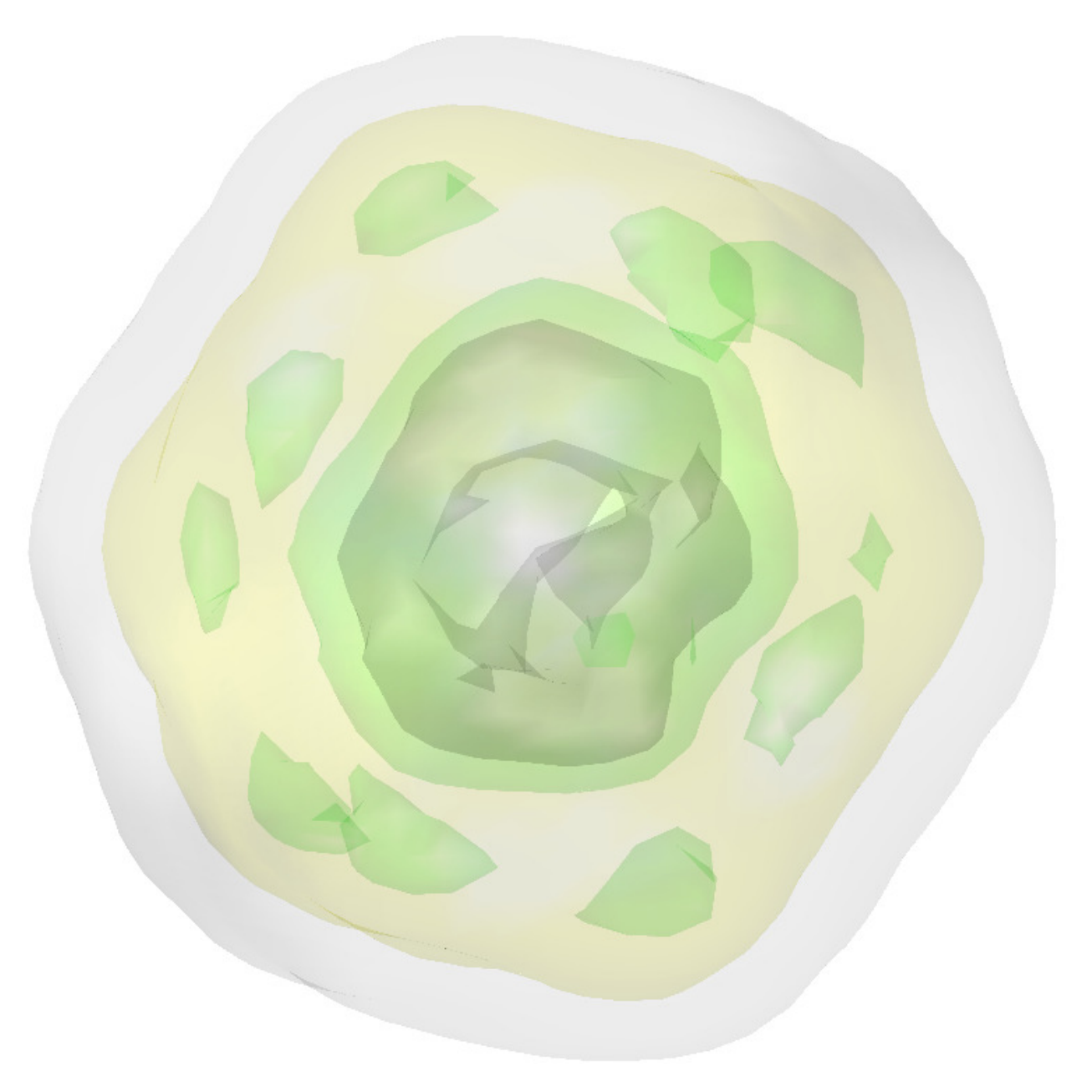} \label{fig:3D_1k_bgrm_real}}
  \subfloat[]{\includegraphics[width=.2\textwidth]{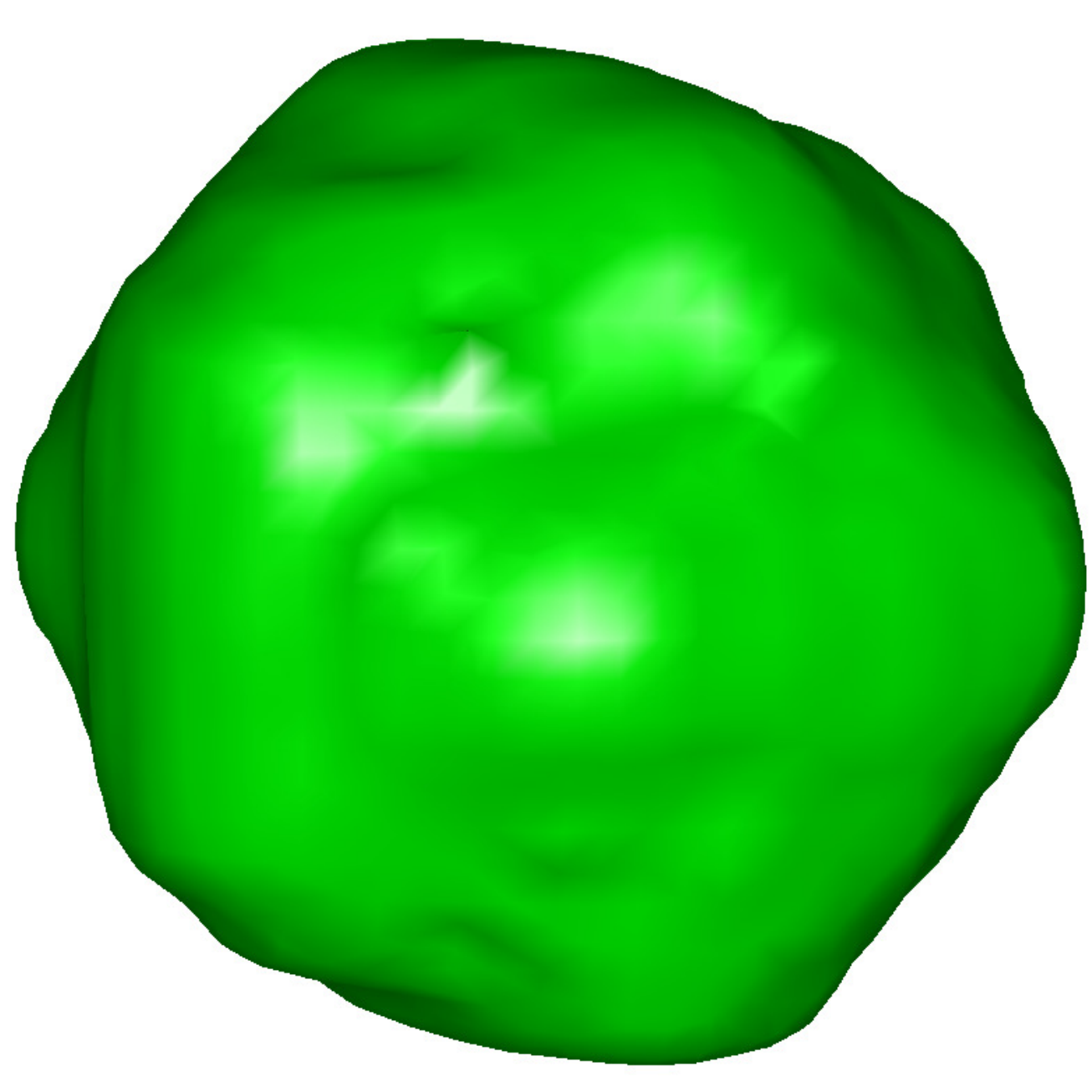} \label{fig:3D_3k_bgrm_iso}}
  \subfloat[]{\includegraphics[width=.2\textwidth]{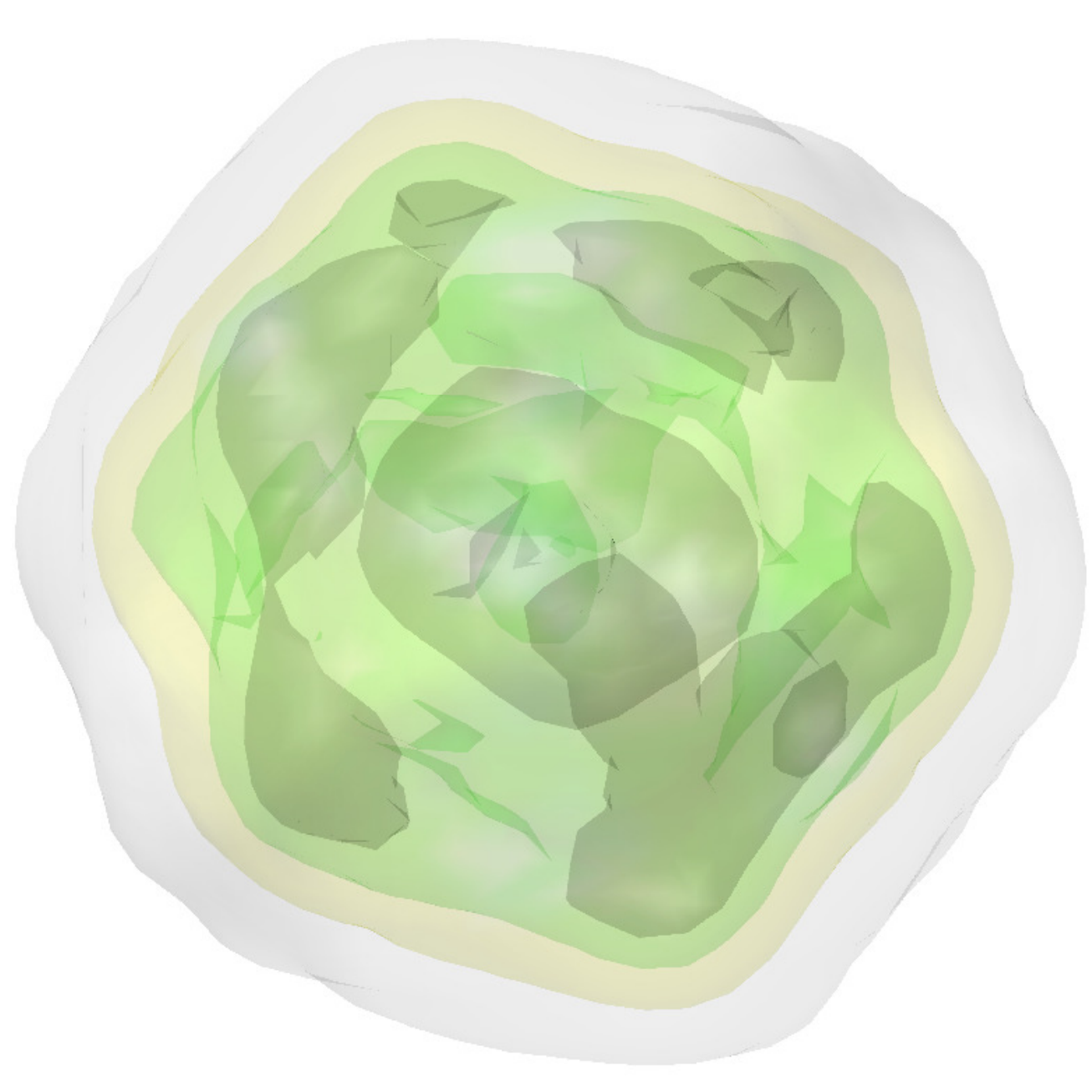} \label{fig:3D_3k_bgrm_real}}
  \hfill\null
  \\
  \subfloat[]{\includegraphics[width=.2\textwidth]{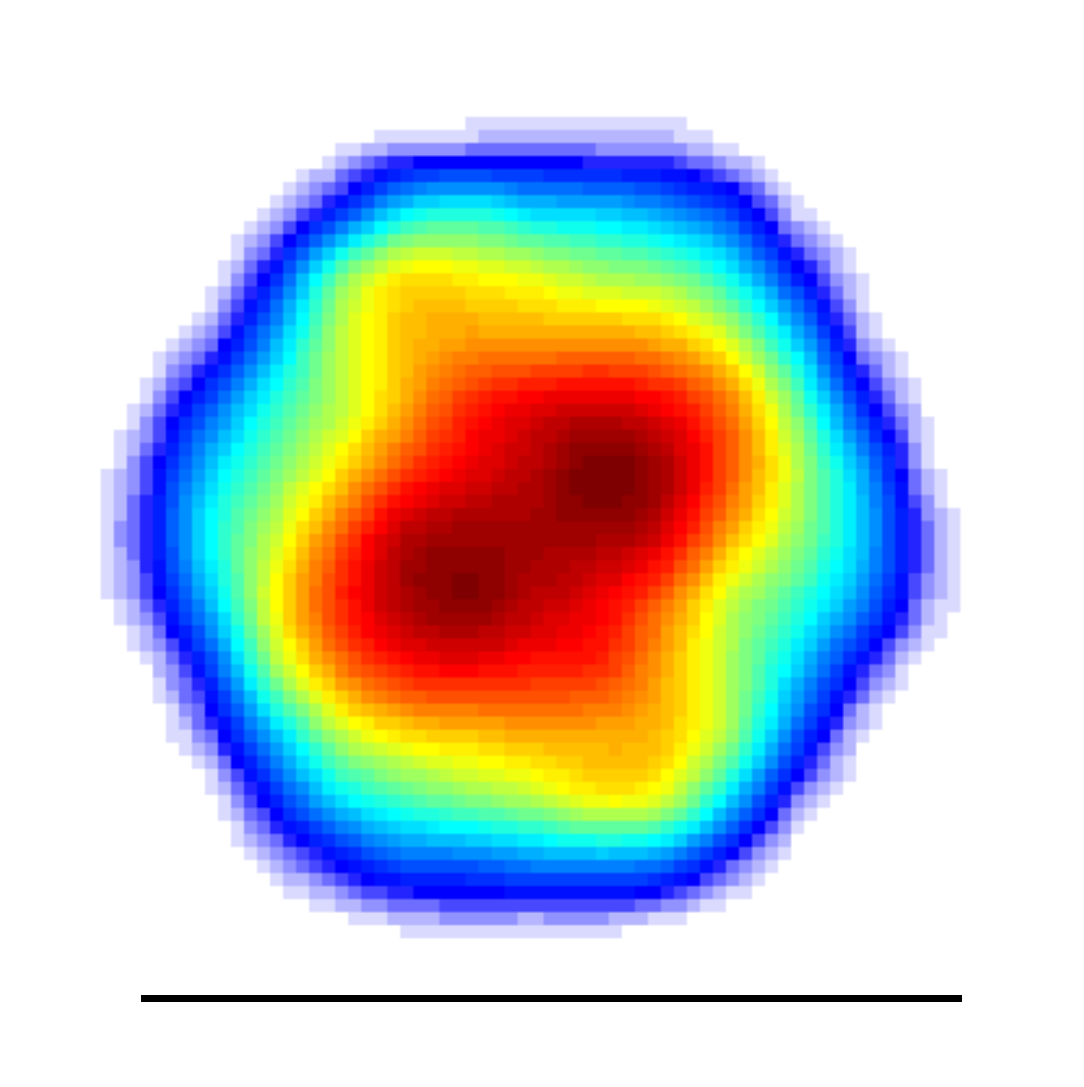} \label{fig:3D_1k_bgrm_proj}}
  \subfloat[]{\includegraphics[width=.2\textwidth]{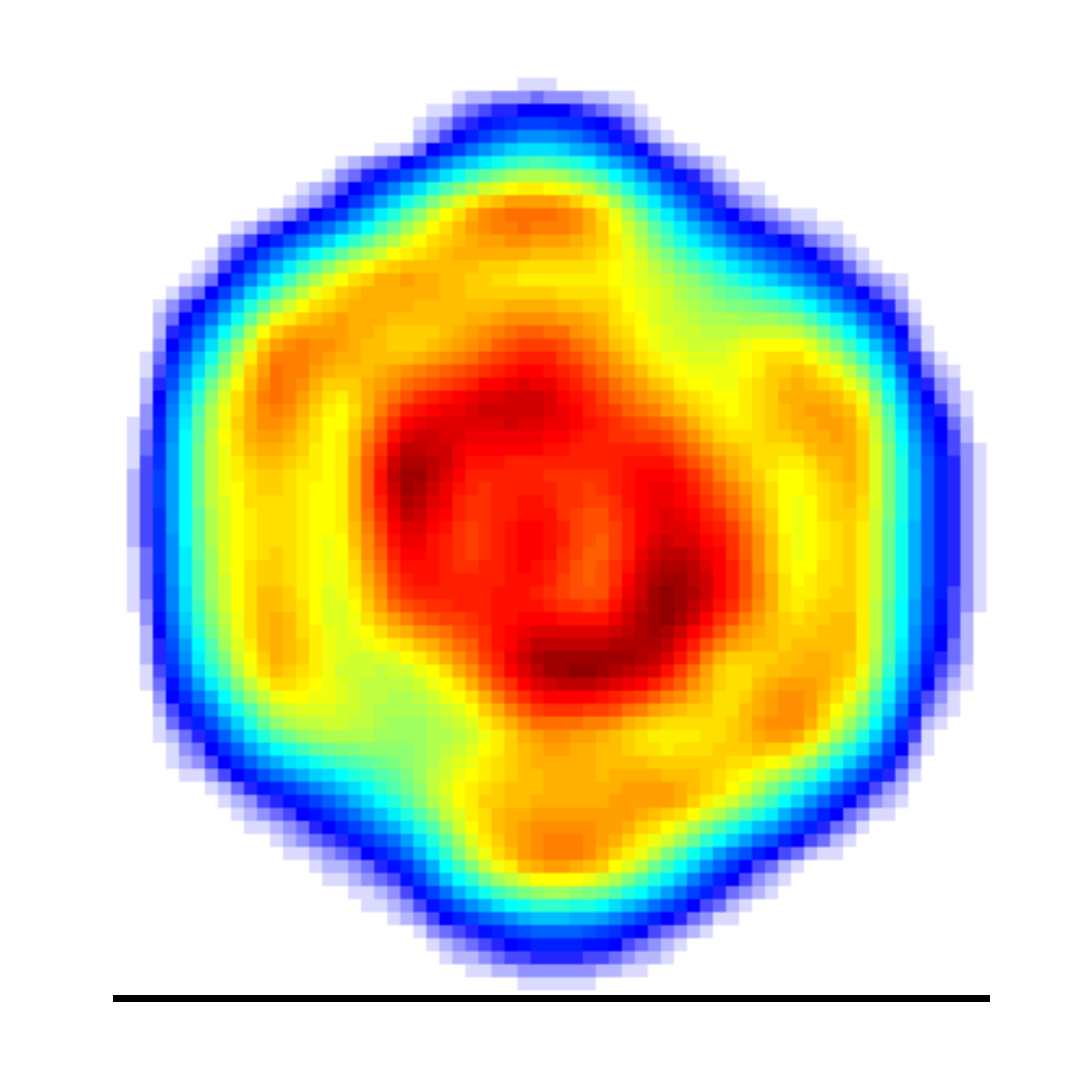} \label{fig:3D_1k_bgrm_slice}}
  \subfloat[]{\includegraphics[width=.2\textwidth]{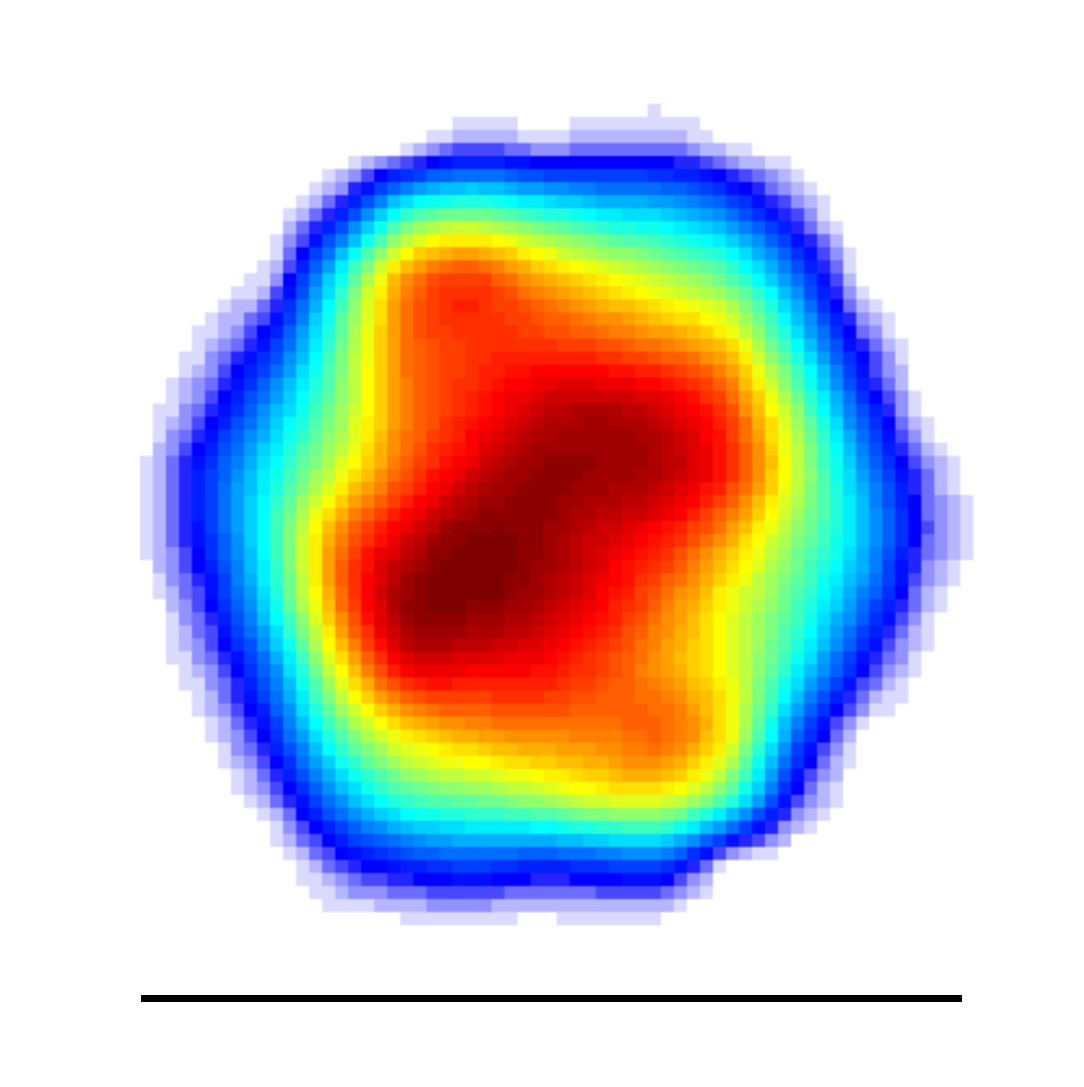} \label{fig:3D_3k_bgrm_proj}}
  \subfloat[]{\includegraphics[width=.2\textwidth]{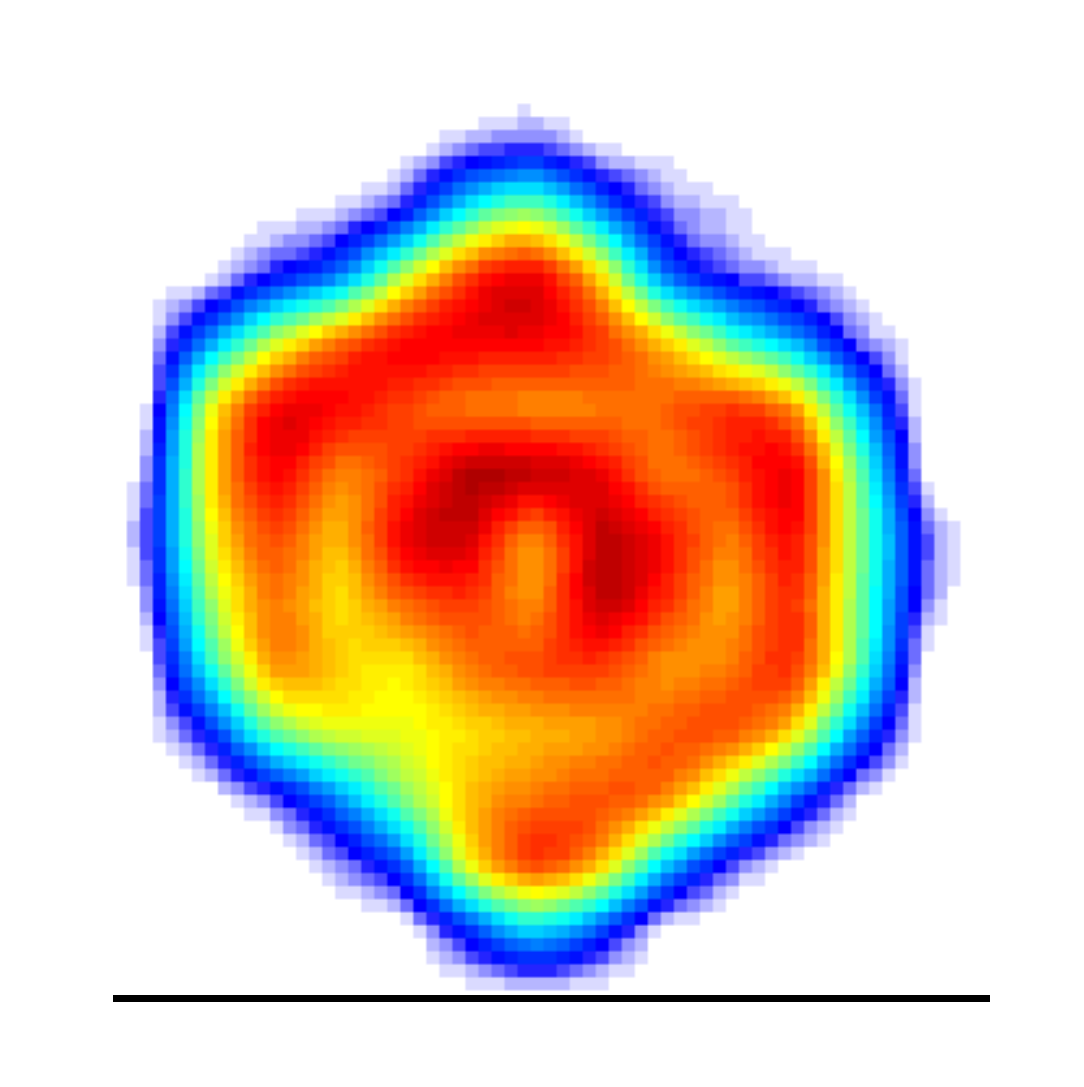} \label{fig:3D_3k_bgrm_slice}}
  \subfloat{\includegraphics[width=.2\textwidth]{fig/colorbar}}
  \caption{The 3D electron density distributions of the PR772 virus
    with background subtraction. See Figure~\ref{fig:phased_ori} for a
    detailed description. The resolution was 9.5~nm from a PRTF
    analysis for $N_{1k}$ and 8.4~nm for $N_{3k}$, and the particle
    size was determined to be 68.6~nm and 68.9~nm, respectively.}
  \label{fig:bgrm_ori}
\end{figure}

The recovered intensity in Figure~\ref{fig:phased_ori} is strongly
concentrated in the central rings, which might partially be due to
background noise. By subtracting 50\% of the minimum values from the
Fourier intensities at each frequency (see Appendix~C for details), we
improved the contrast of the Fourier intensities
(eq.~\eqref{eq:contrast}) to 0.67 and 0.71 (from 0.58 and 0.59 with
background) for $N_{1k}$ and $N_{3k}$, respectively, and the retrieved
density distributions were less concentrated, yet the
concentric layer structure was maintained, see Figure~\ref{fig:bgrm_ori}. The sizes
of the retrieved particle became slightly smaller than 69~nm after
background subtraction, and the PRTF analysis gave a resolution of
$9.5$~nm and $8.4$~nm, respectively, for $N_{1k}$ and $N_{3k}$.

\subsection{Windowed Signal}

\begin{figure*}[!htbp]
  \centering
  \subfloat[]{\includegraphics[width=.2\textwidth]{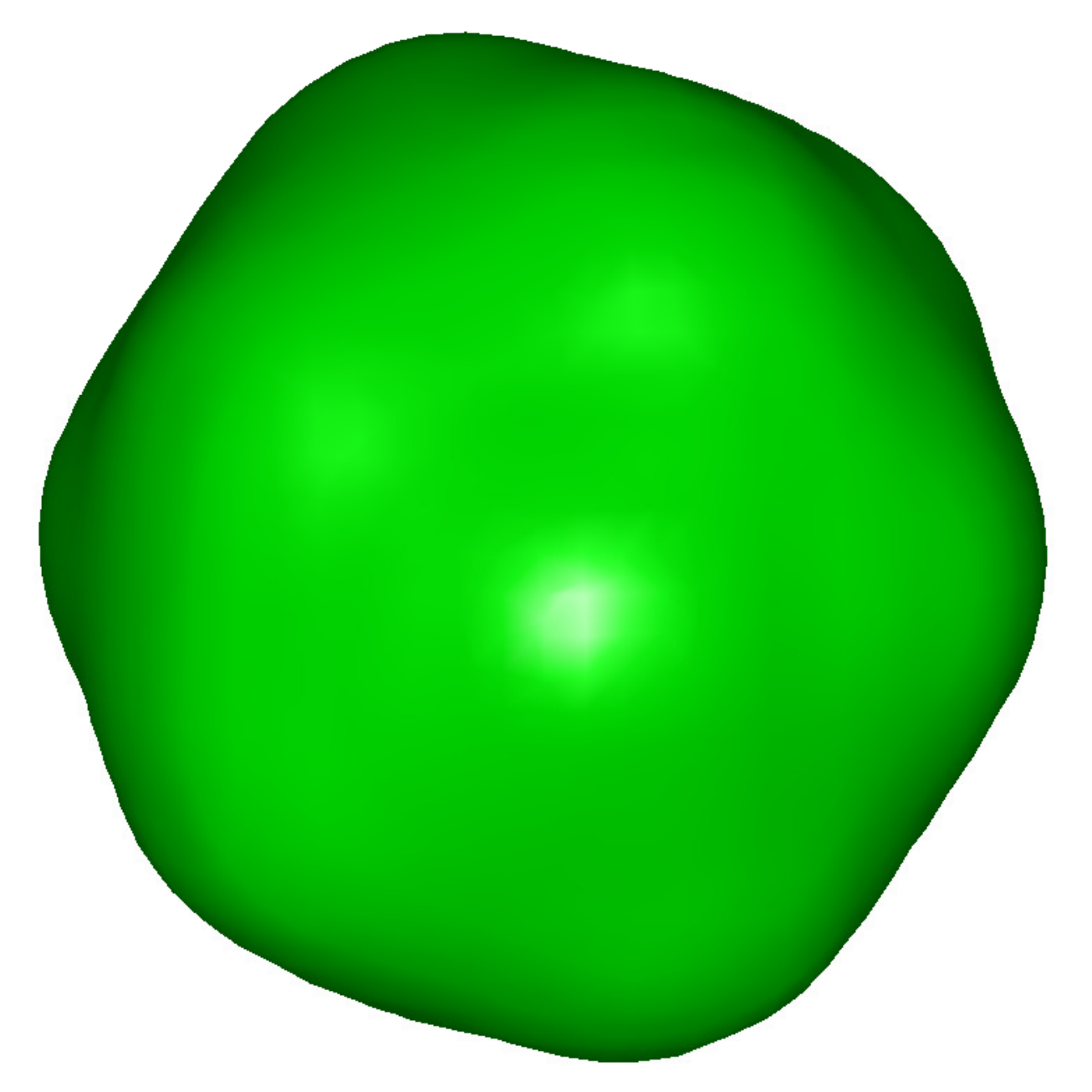} \label{fig:3D_1k_hann_iso}}
  \subfloat[]{\includegraphics[width=.2\textwidth]{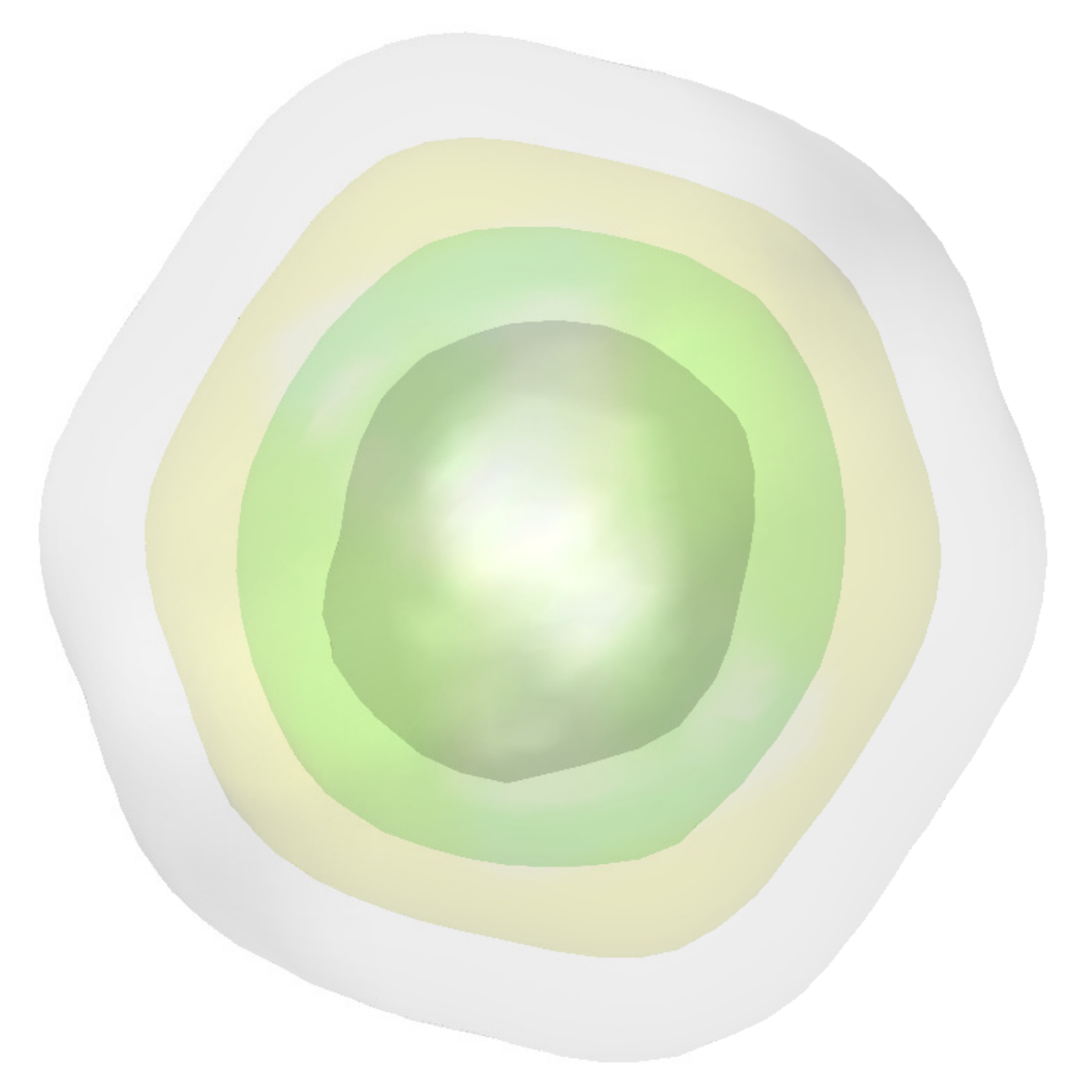} \label{fig:3D_1k_hann_real}}
  \subfloat[]{\includegraphics[width=.2\textwidth]{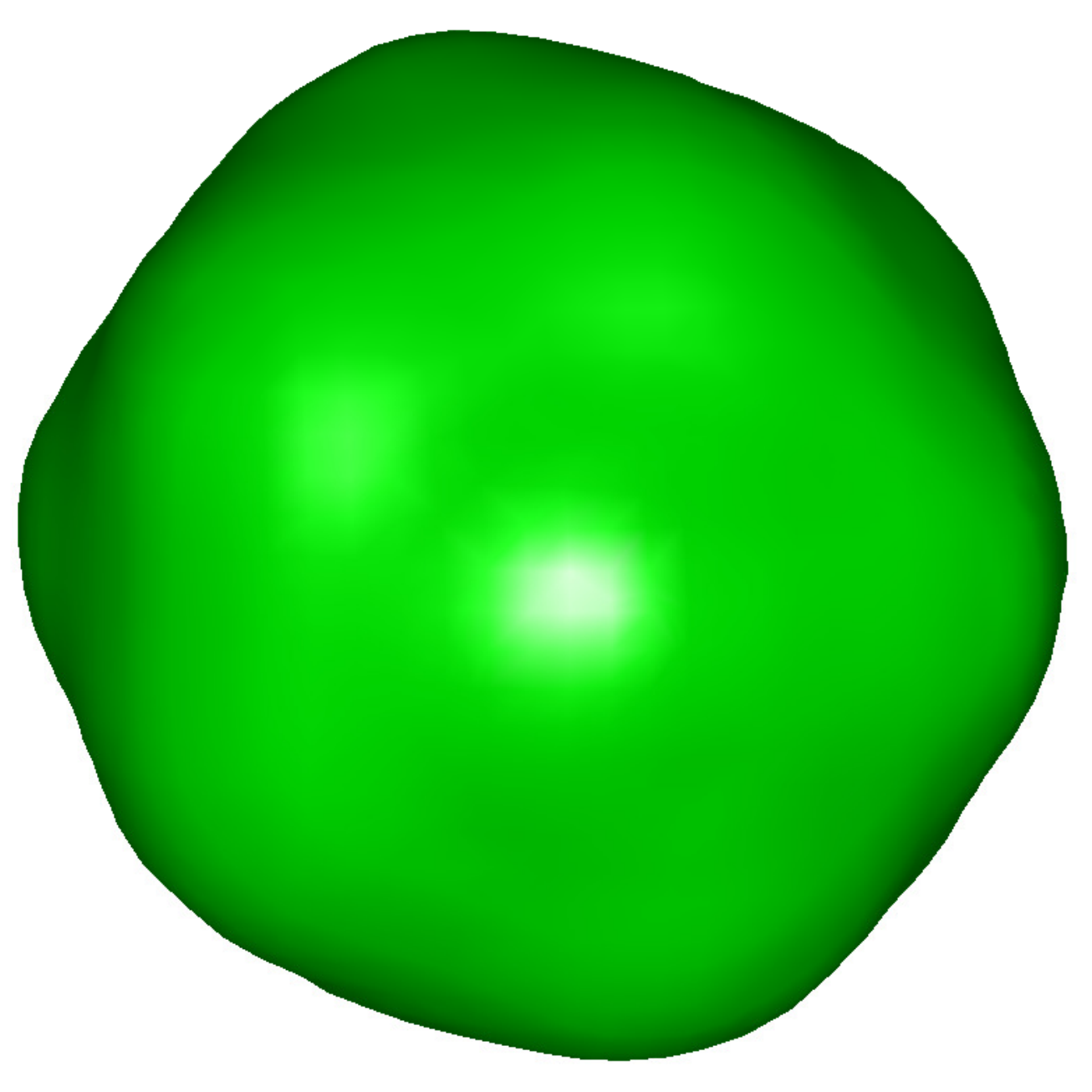} \label{fig:3D_3k_hann_iso}}
  \subfloat[]{\includegraphics[width=.2\textwidth]{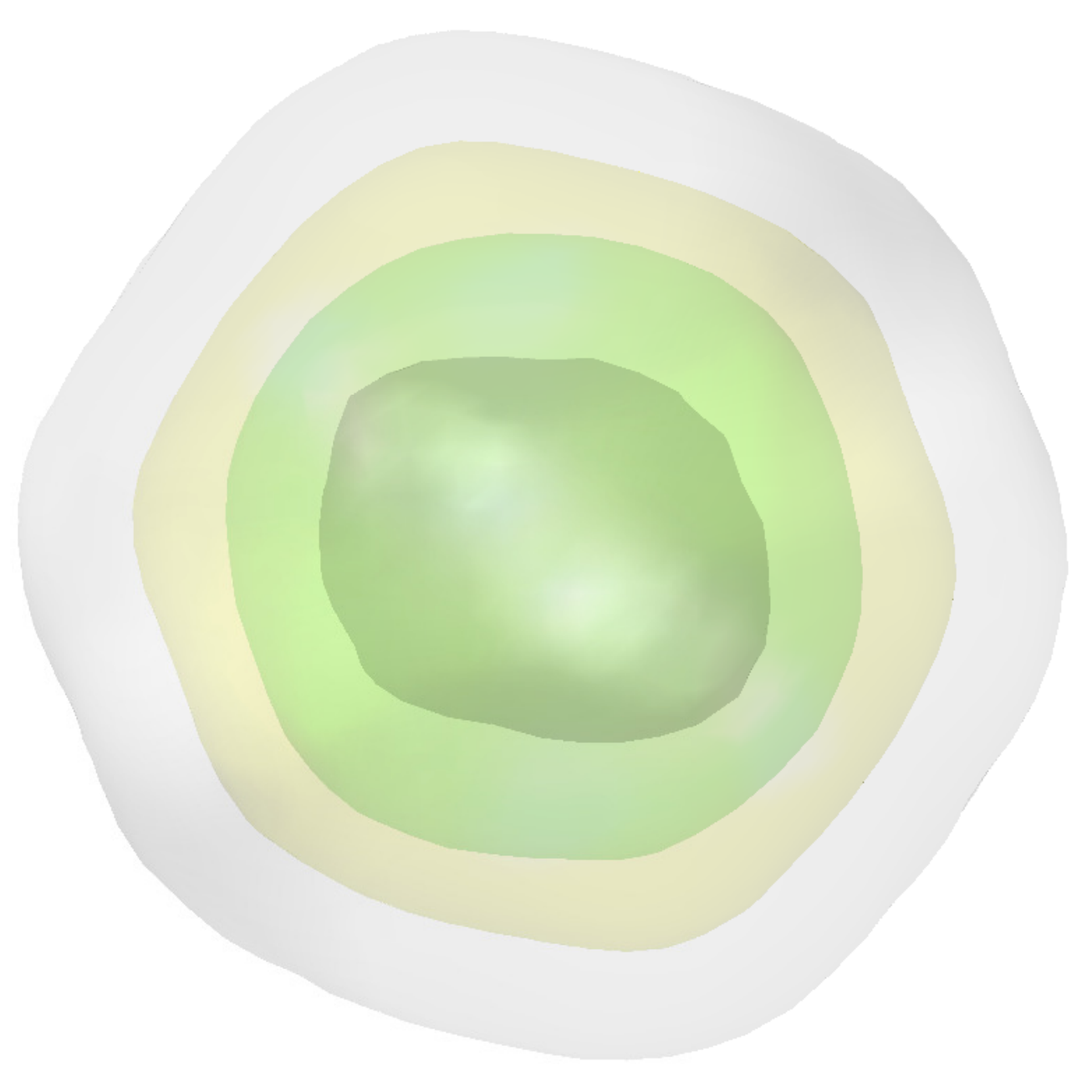} \label{fig:3D_3k_hann_real}} \hfill\null  \\
  \subfloat[]{\includegraphics[width=.2\textwidth]{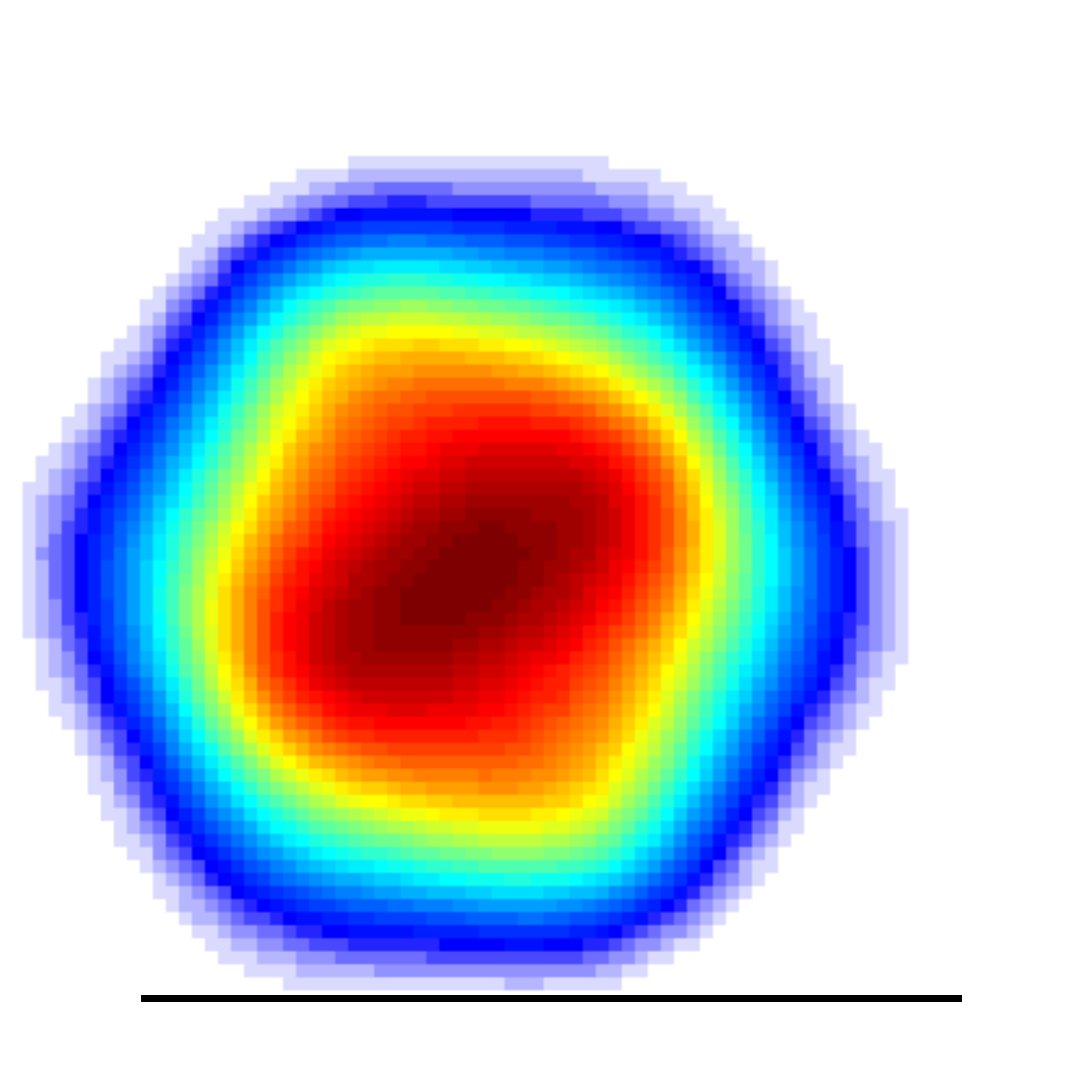} \label{fig:3D_1k_hann_proj}}
  \subfloat[]{\includegraphics[width=.2\textwidth]{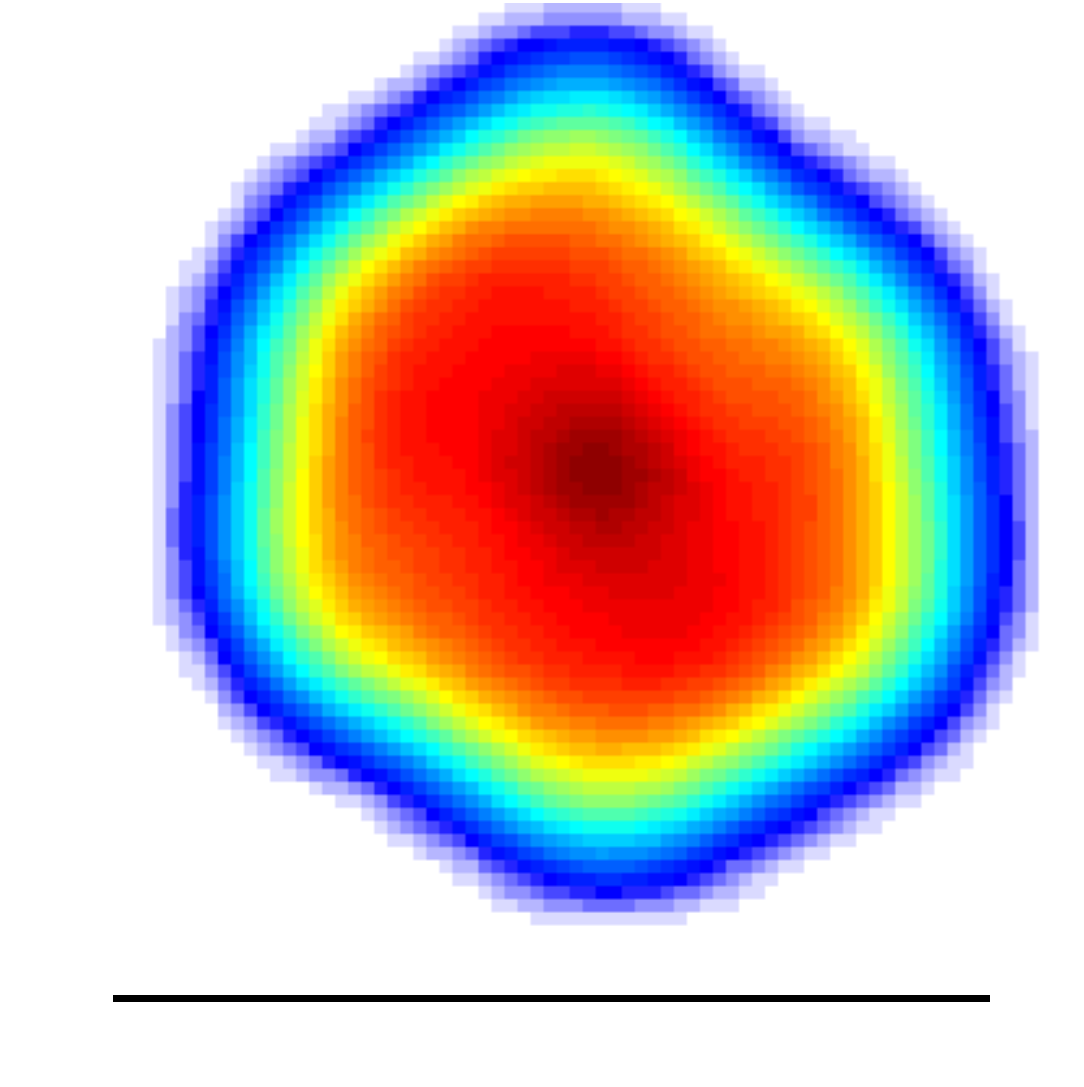} \label{fig:3D_1k_hann_slice}}
  \subfloat[]{\includegraphics[width=.2\textwidth]{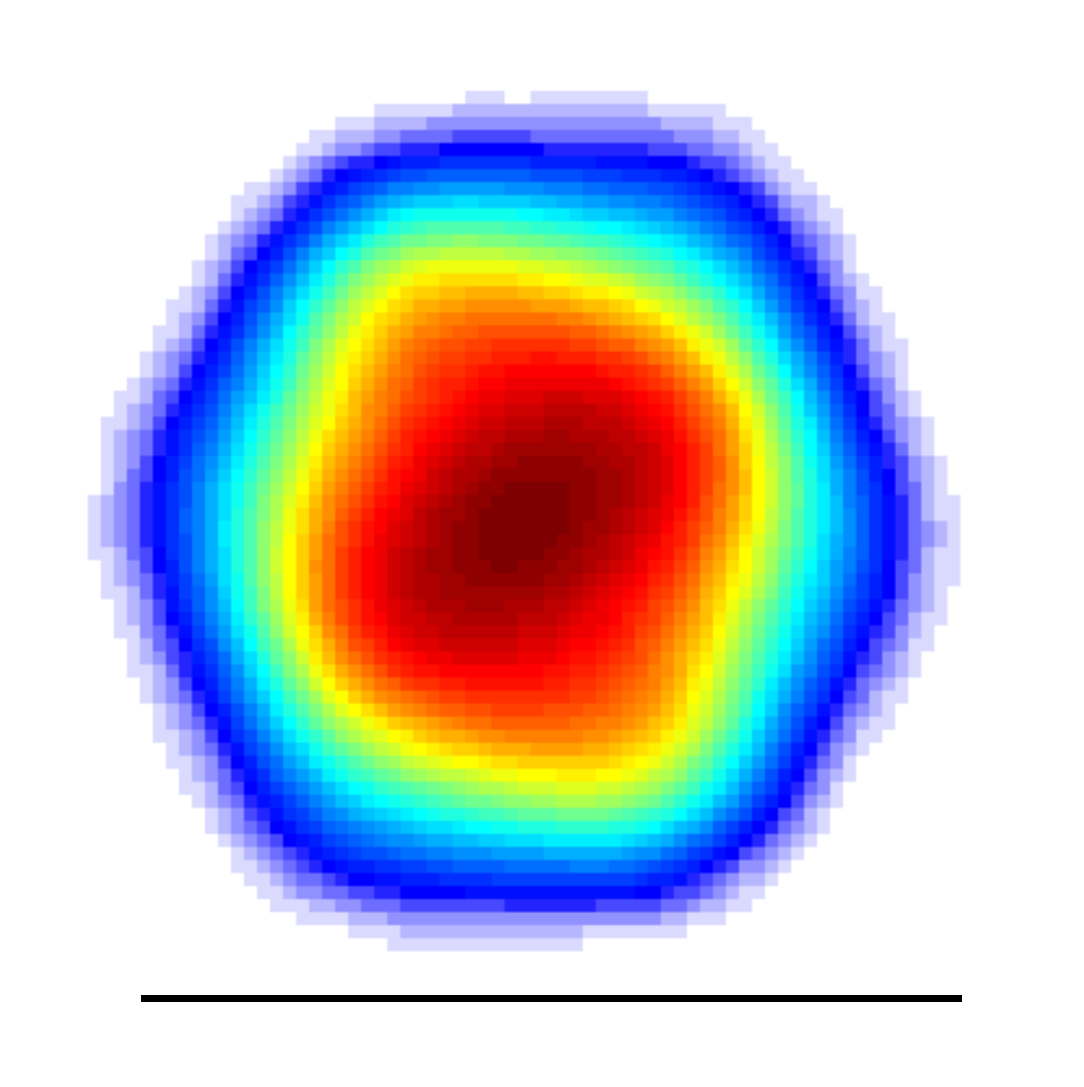} \label{fig:3D_3k_hann_proj}}
  \subfloat[]{\includegraphics[width=.2\textwidth]{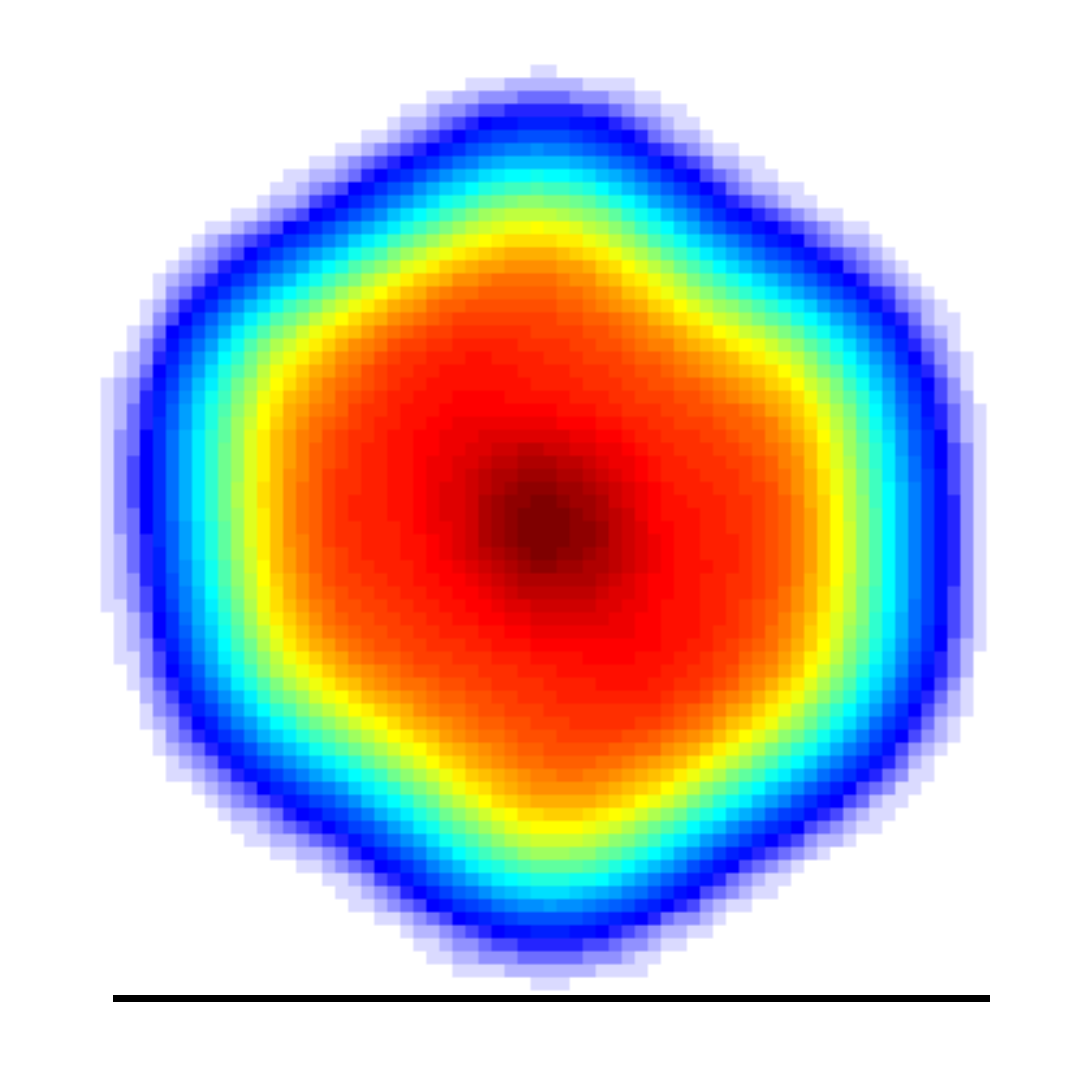} \label{fig:3D_3k_hann_slice}}
  \subfloat{\includegraphics[width=.2\textwidth]{fig/colorbar}}   \\
  \subfloat[]{\includegraphics[width=.2\textwidth]{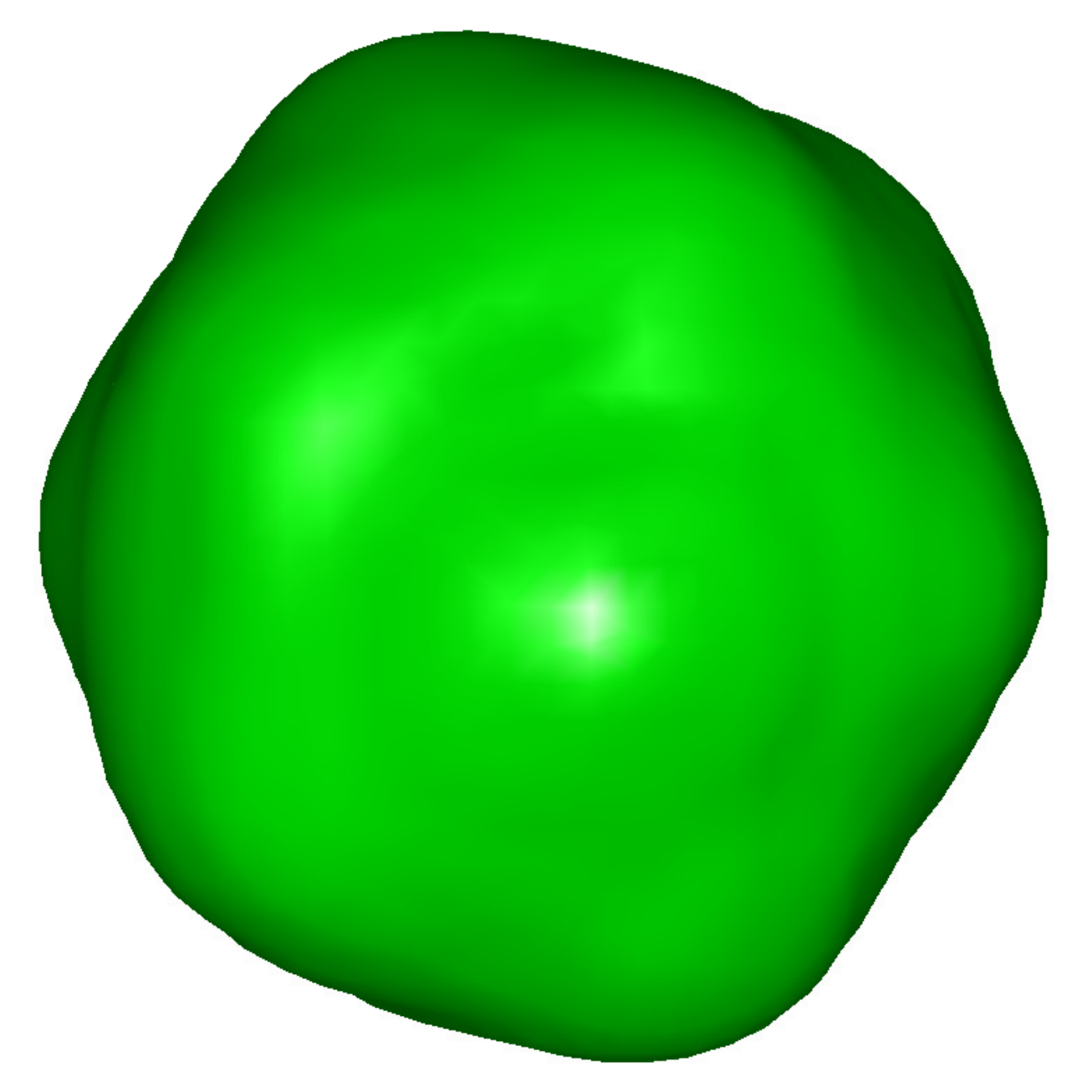} \label{fig:3D_1k_bgrmhann_iso}}
  \subfloat[]{\includegraphics[width=.2\textwidth]{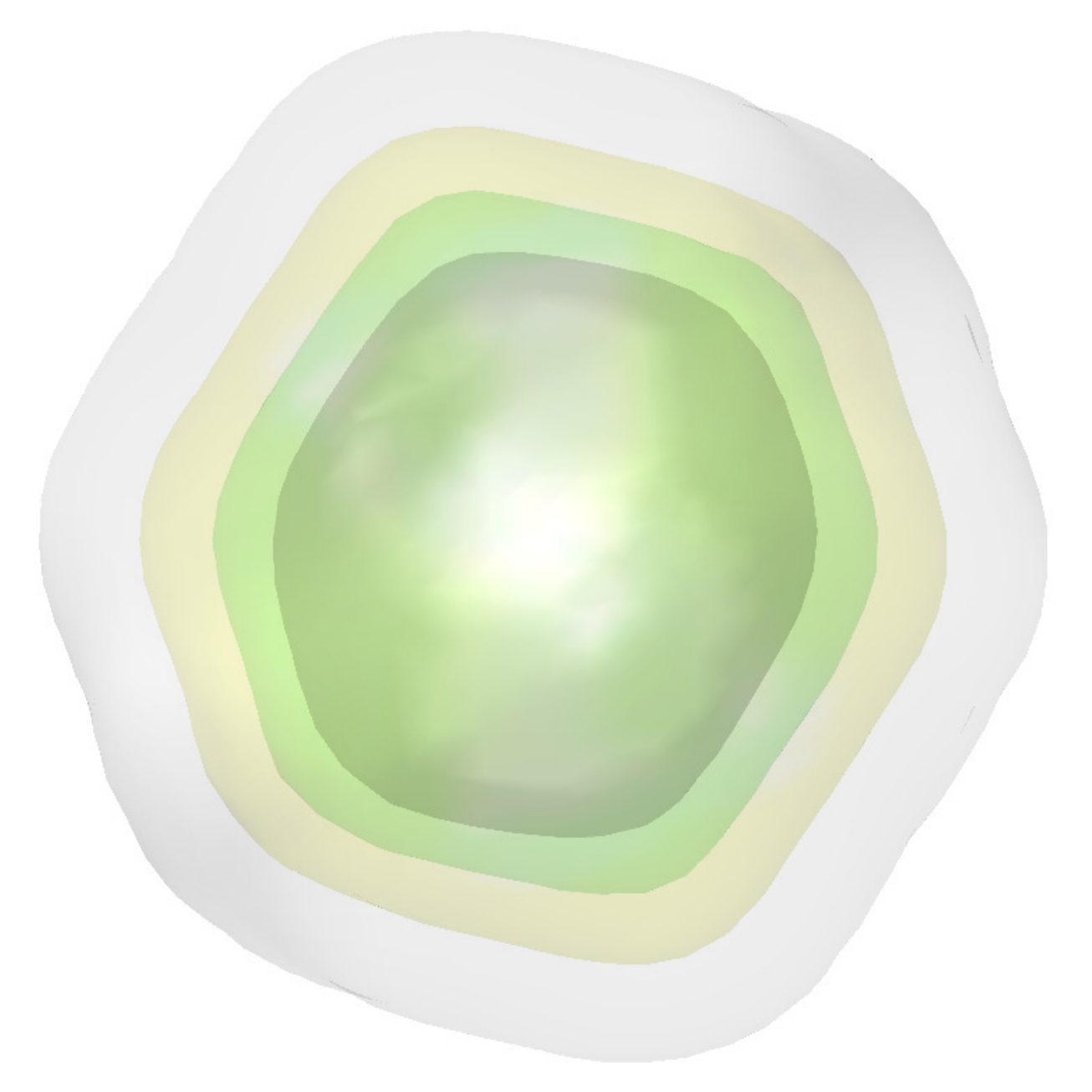} \label{fig:3D_1k_bgrmhann_real}}
  \subfloat[]{\includegraphics[width=.2\textwidth]{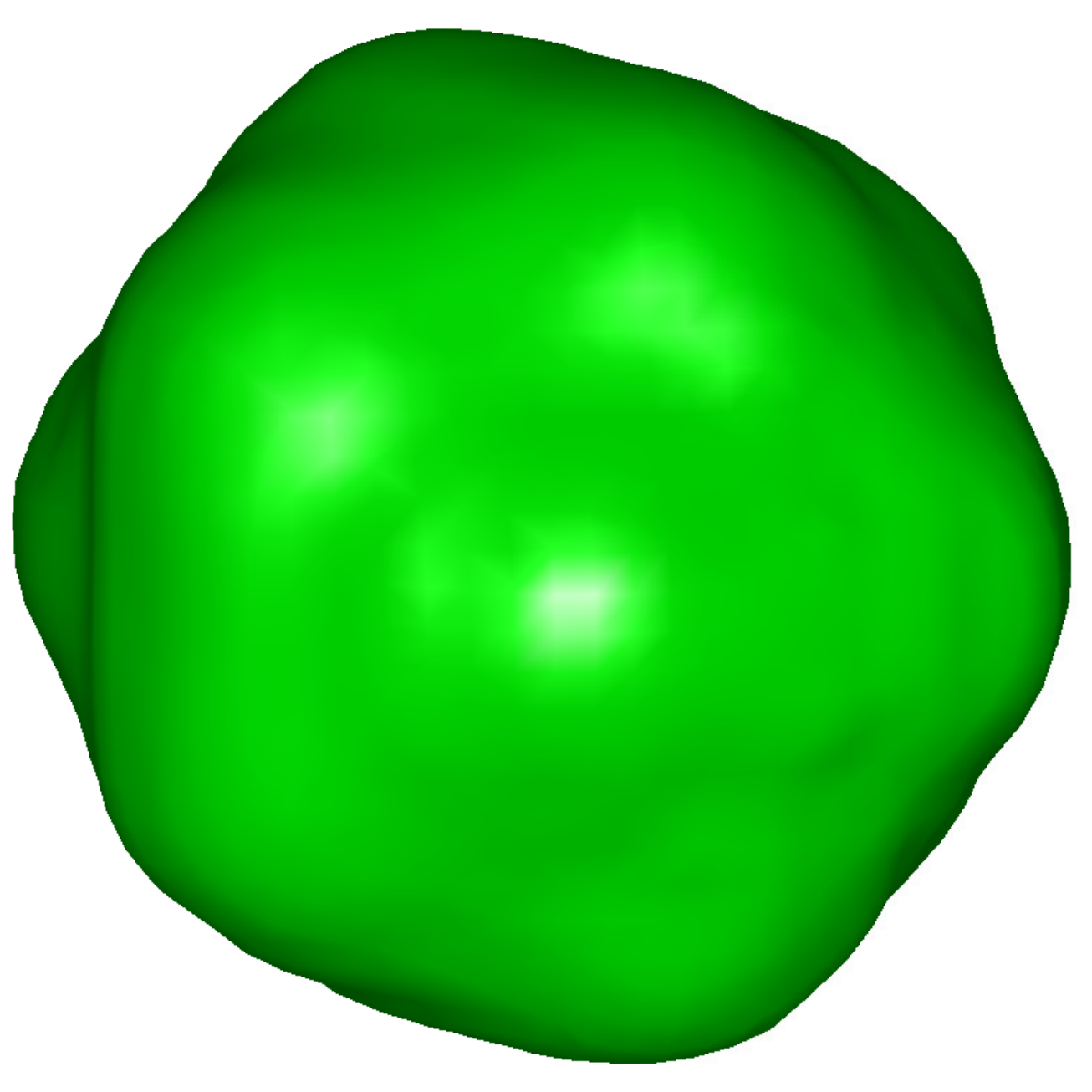} \label{fig:3D_3k_bgrmhann_iso}}
  \subfloat[]{\includegraphics[width=.2\textwidth]{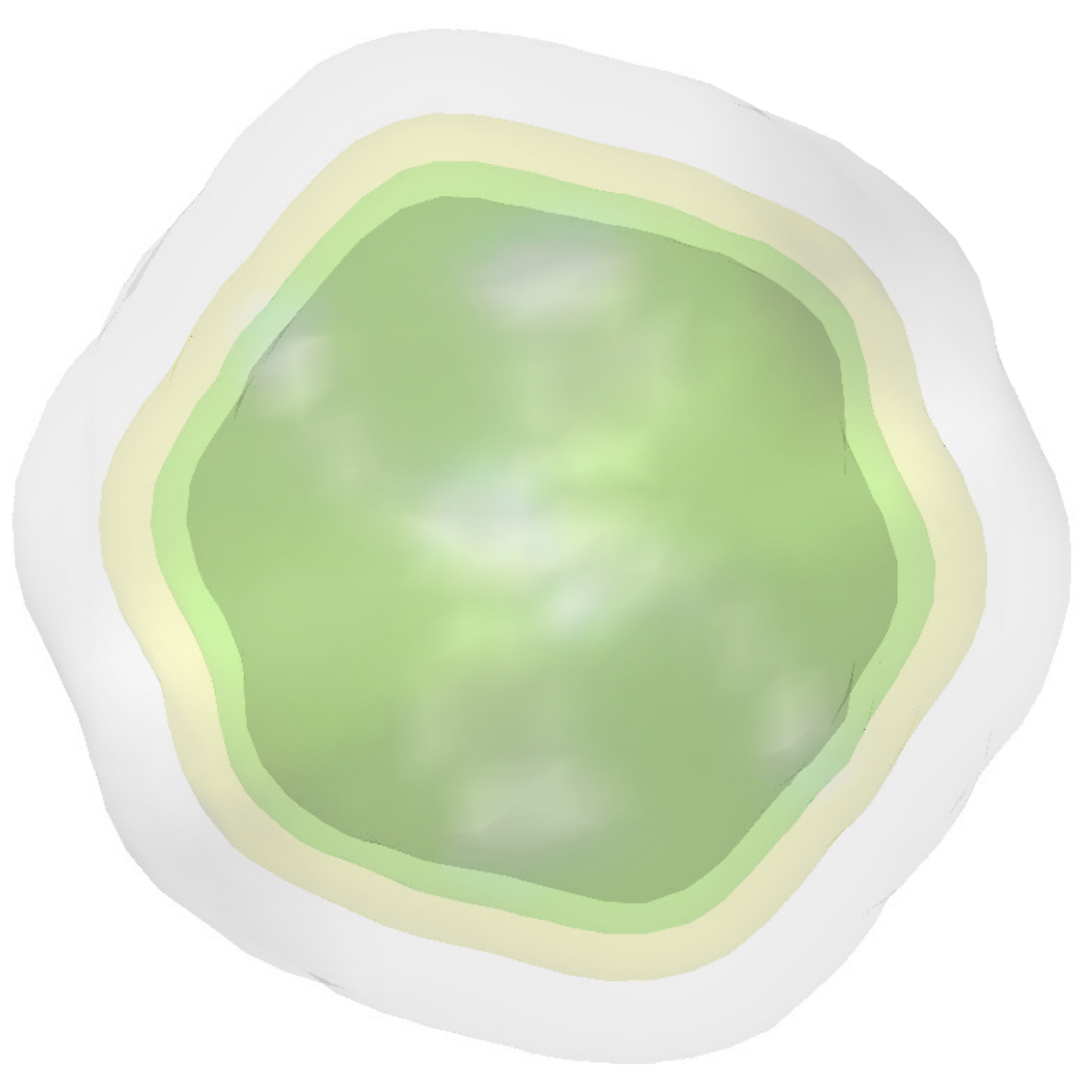} \label{fig:3D_3k_bgrmhann_real}} \hfill\null  \\
  \subfloat[]{\includegraphics[width=.2\textwidth]{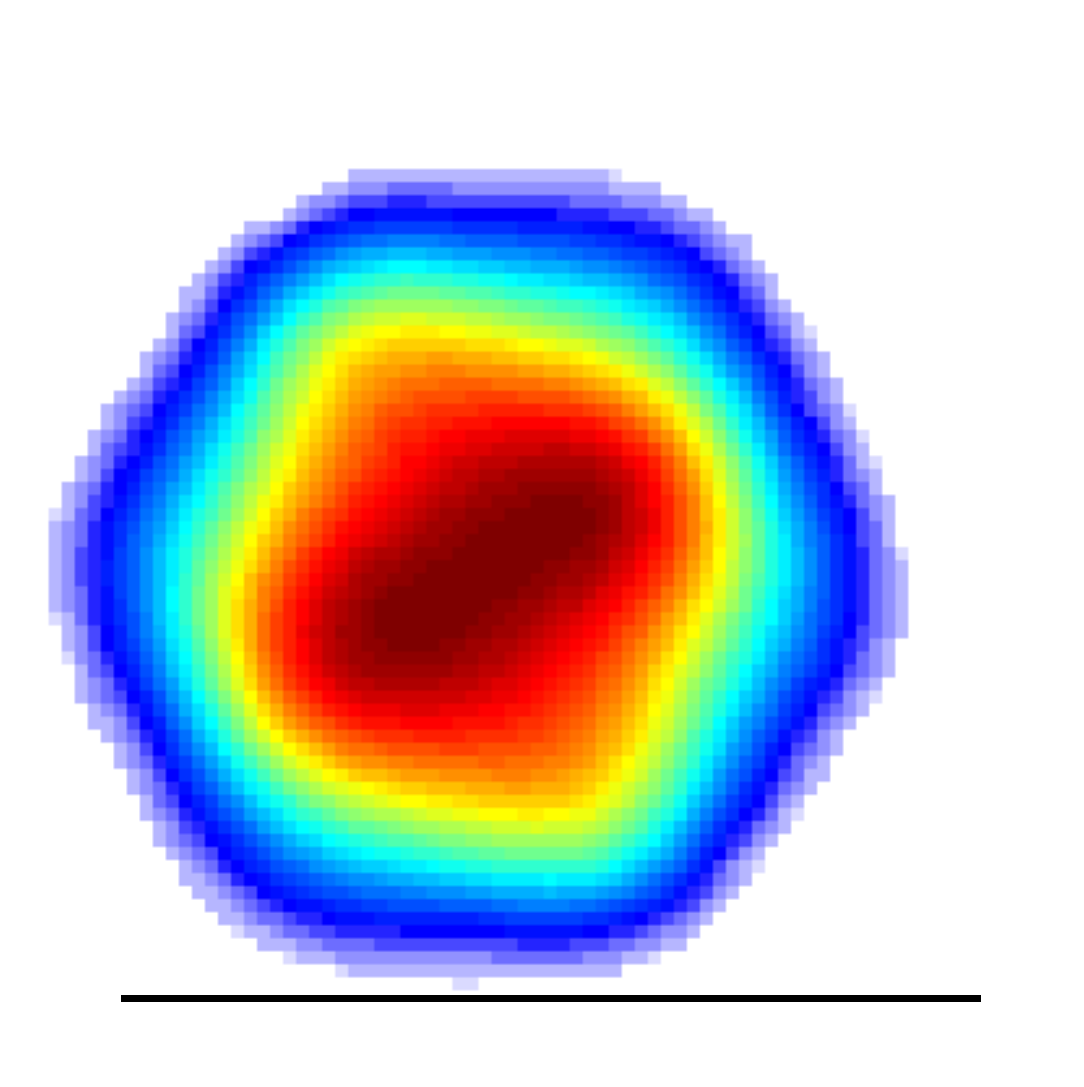} \label{fig:3D_1k_bgrmhann_proj}}
  \subfloat[]{\includegraphics[width=.2\textwidth]{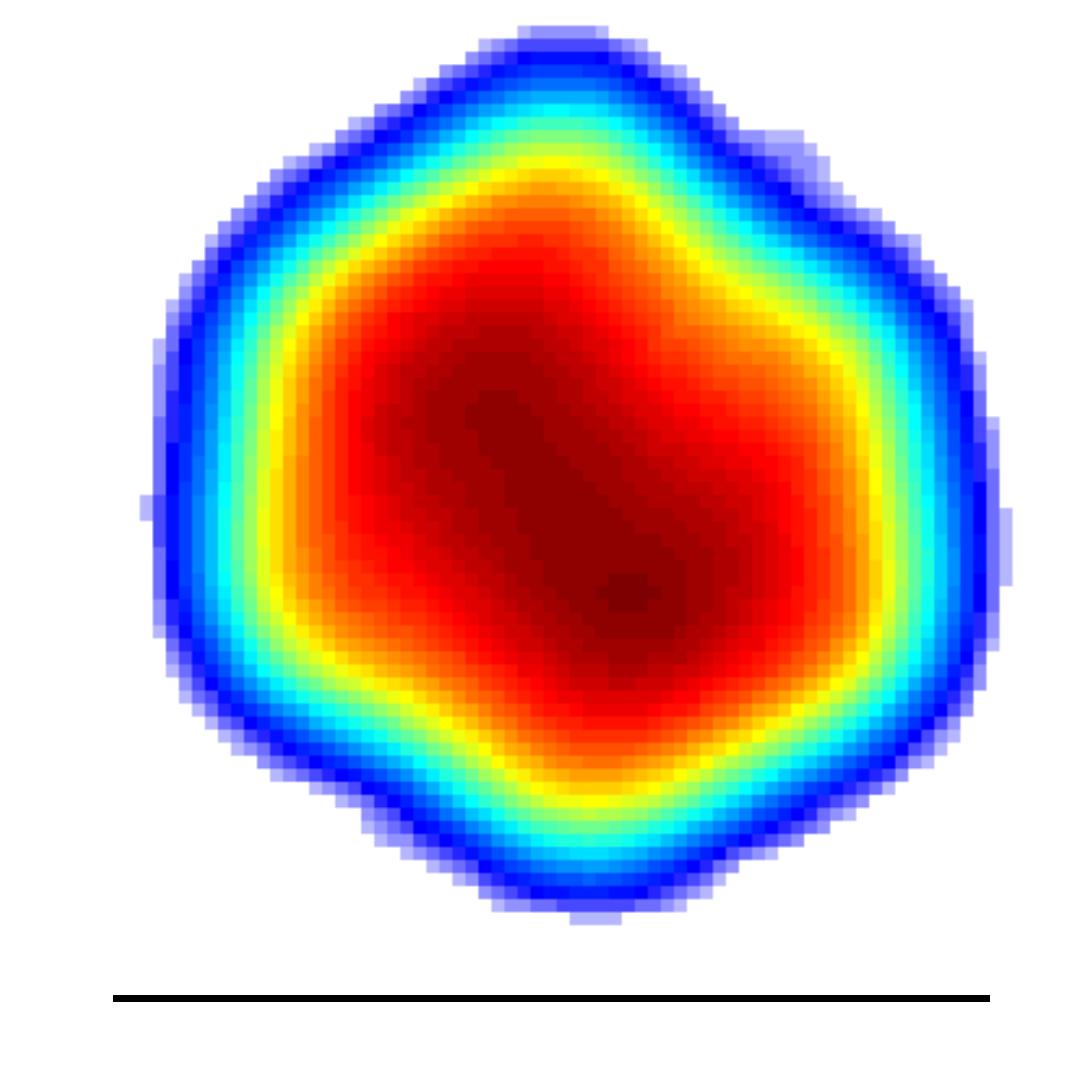} \label{fig:3D_1k_bgrmhann_slice}}
  \subfloat[]{\includegraphics[width=.2\textwidth]{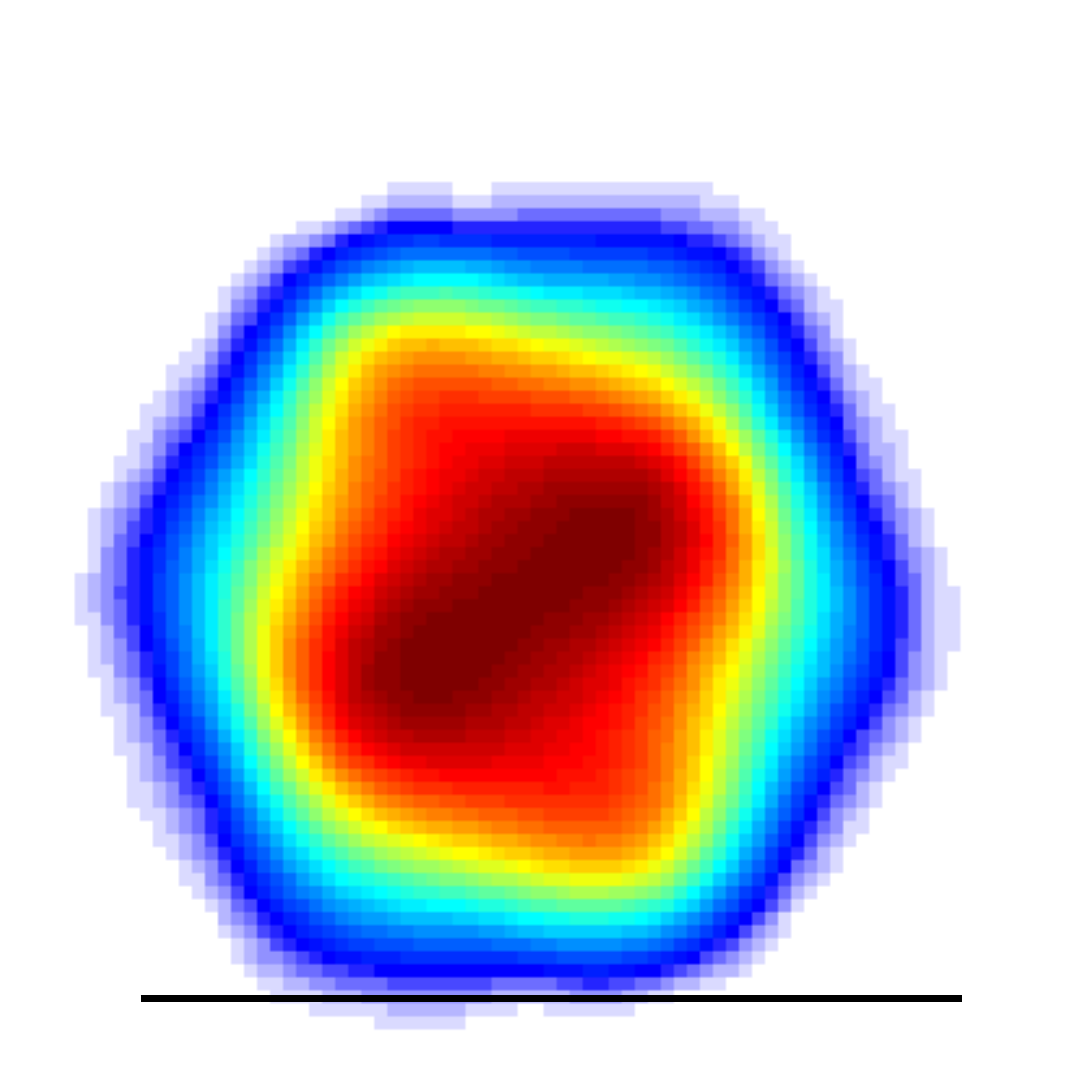} \label{fig:3D_3k_bgrmhann_proj}}
  \subfloat[]{\includegraphics[width=.2\textwidth]{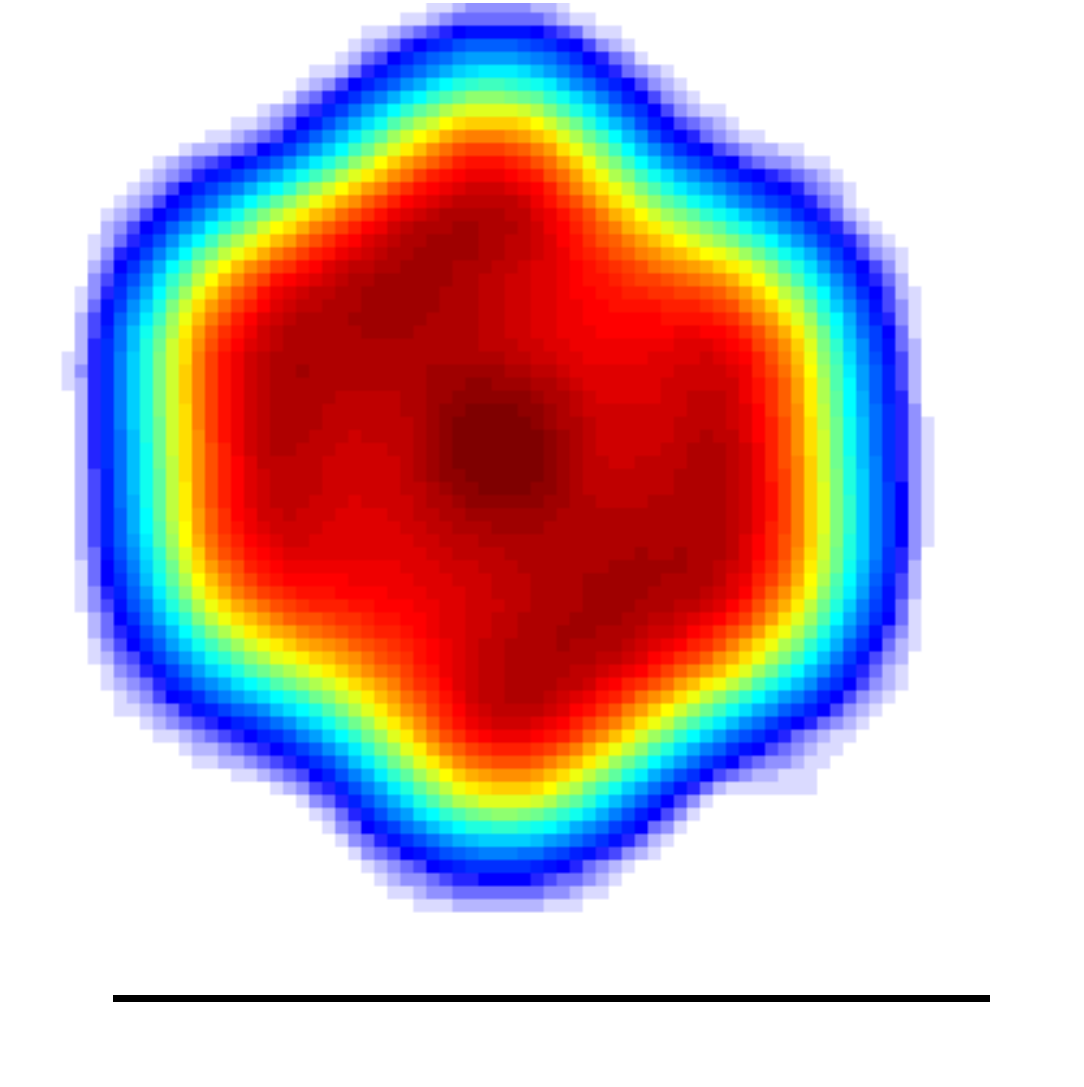} \label{fig:3D_3k_bgrmhann_slice}}
  \subfloat{\includegraphics[width=.2\textwidth]{fig/colorbar} }   \\
  \caption{The windowed 3D electron density distributions for $N_{1k}$
    (\textit{left two columns}) and $N_{3k}$ (\textit{right two
      columns}), respectively. The \textit{top two rows} are from
    reconstructions with background noises and the \textit{bottom two}
    are windowed electron density from reconstructions with subtracted
    background noise.  The Hann windows made the retrieved intensities
    much smother and effectively removed the three concentric layers,
    comparing with Figure~\ref{fig:phased_ori} and
    Figure~\ref{fig:bgrm_ori}.  Further, the intensities heavily
    concentrated at the particle centers with background noises, and
    by removing the background noises, the intensities spread out in
    the retrieved particles. The resolution was 10~nm from a PRTF
    analysis with threshold
    $e^{-1}$. [\protect\subref{fig:3D_1k_bgrmhann_iso} --
    \protect\subref{fig:3D_3k_bgrmhann_slice}]: the corresponding
    results with background removed. The resolution from $N_{1k}$ was
    here 11.2~nm and 8.7~nm for $N_{3k}$.  The combination of Hann
    window and background subtraction made the retrieved particle
    smooth and of comparably high-resolution.  }
  \label{fig:hann}
\end{figure*}

The finite capturing time for FXI experiments and the discrete digital
detector may lead to spectral leakage in diffraction patterns, which
will consequentially lead to artefacts in the phased object in real
space, such as very low intensities in the particle center and
aliasing effects. To compensate the potential energy leakage, we
applied a square 3D Hann window to the Fourier intensities before
phasing them. The phased objects were much smother in both the outer
capsid and the interior structure in Figure~\ref{fig:hann}, comparing
with the ones in Figure~\ref{fig:phased_ori} and
Figure~\ref{fig:bgrm_ori}. Moreover, the three concentric layers
vanished after applying the Hann window.  As expected, after applying
Hann windows, the intensities of the retrieved particles with
background noises were still more concentrated at their centers, while
the intensities of the ones with background subtraction spread
out. With background noises, the obtained resolutions were around
10~nm for both $N_{1k}$ and $N_{3k}$ datasets.  With background
subtraction, the resolutions were 11.2~nm and 8.7~nm for the
$N_{1k}$/$N_{3k}$ datasets, respectively. Note that the Hann window
efficiently removes aliasing effects, but it may also reduce the
resolution of the object, due to its effect on the higher
frequencies. Further, the window function enlarged the size of the
phased object by 0.5 pixels. The estimated sizes of the obtained
objects were around 69 nm after we deducted 0.5 pixels from our size
calculation. We suggest that subtracting background noise and applying
a square 3D Hann window before phasing improves the resolution without
introducing aliasing effects.

\subsection{Constrained Support}

The alternating phasing algorithms, i.e., RAAR and ER, solve a concave
optimization problem using convex optimization iteratively, and hence
may produce local optima. By applying the Convex Optimization of
Autocorrelation with Constrained Support (COACS) \cite{COACS} to the
3D Fourier intensity we may instead get global optimas and hopefully
achieve a higher resolution. Since COACS also uses windows, we may
also get less aliasing effects and avoid the low-intensity center.

\begin{figure*}[!htbp]
  \centering
  \subfloat[]{\includegraphics[width=.2\textwidth]{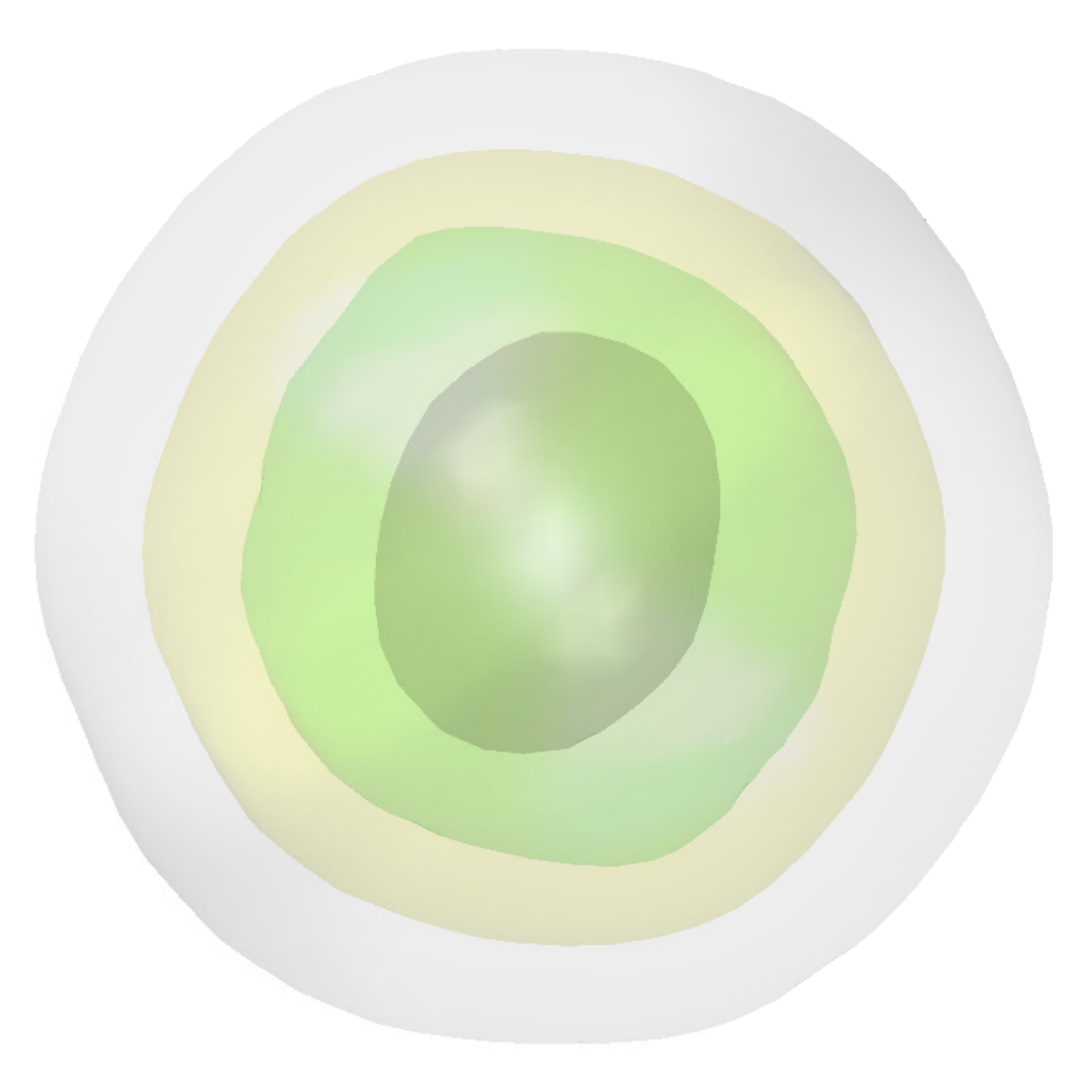} \label{fig:3D_Carl1908_noconstraints_3Dreal}}
  \subfloat[]{\includegraphics[width=.2\textwidth]{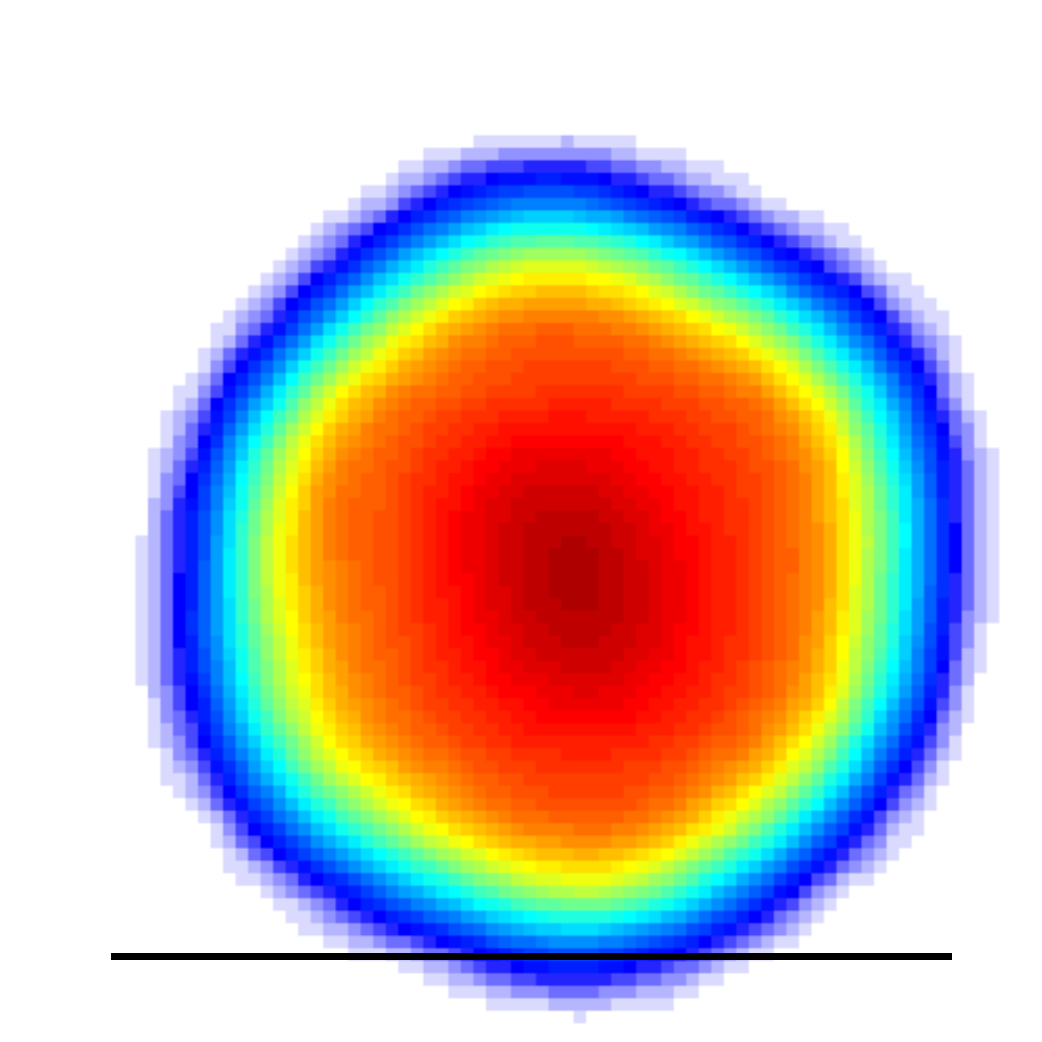} \label{fig:3D_Carl1908_noconstraints_slices1}}
    \subfloat[]{\includegraphics[width=.2\textwidth]{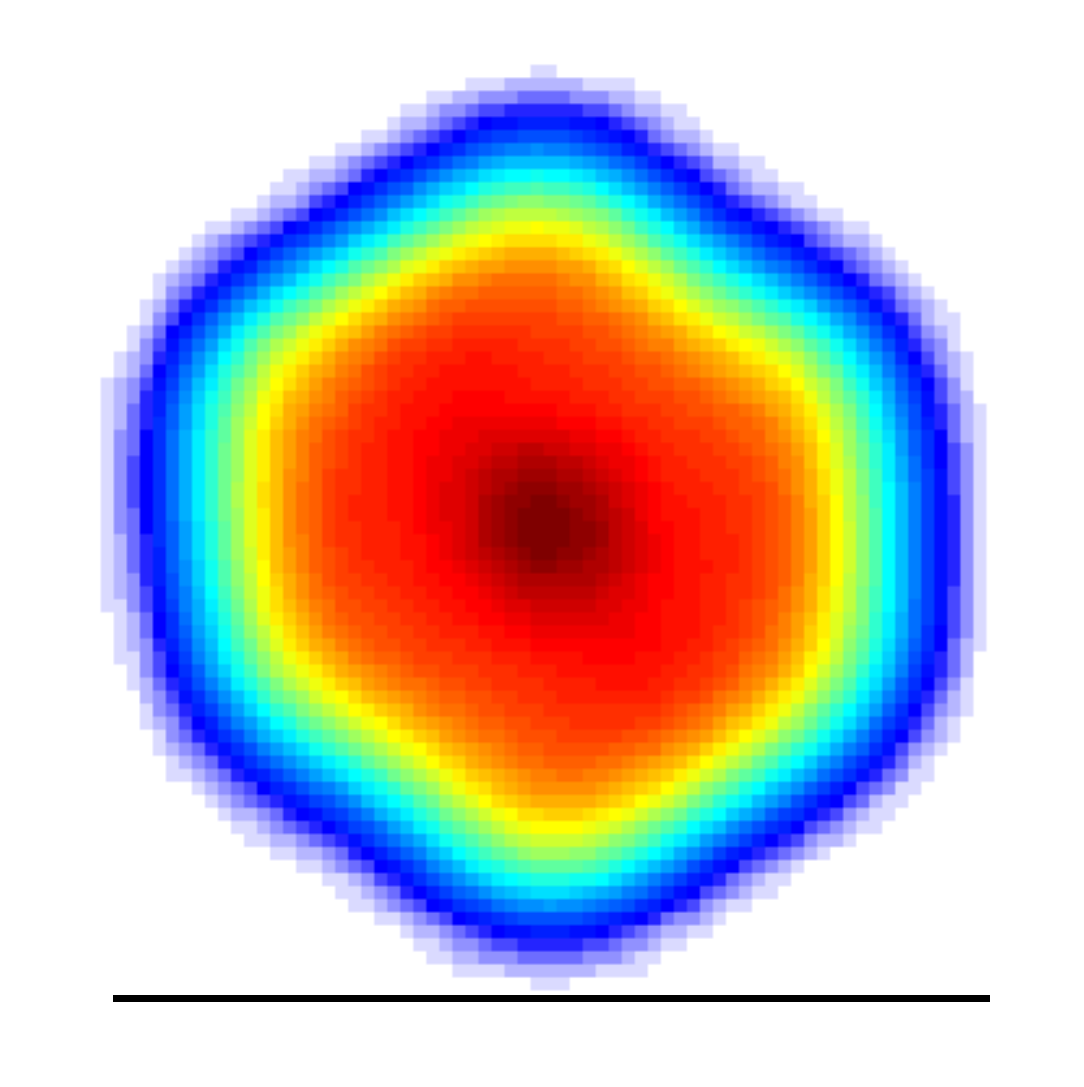} \label{fig:3D_Carl1908_noconstraints_slices2}}
      \subfloat[]{\includegraphics[width=.2\textwidth]{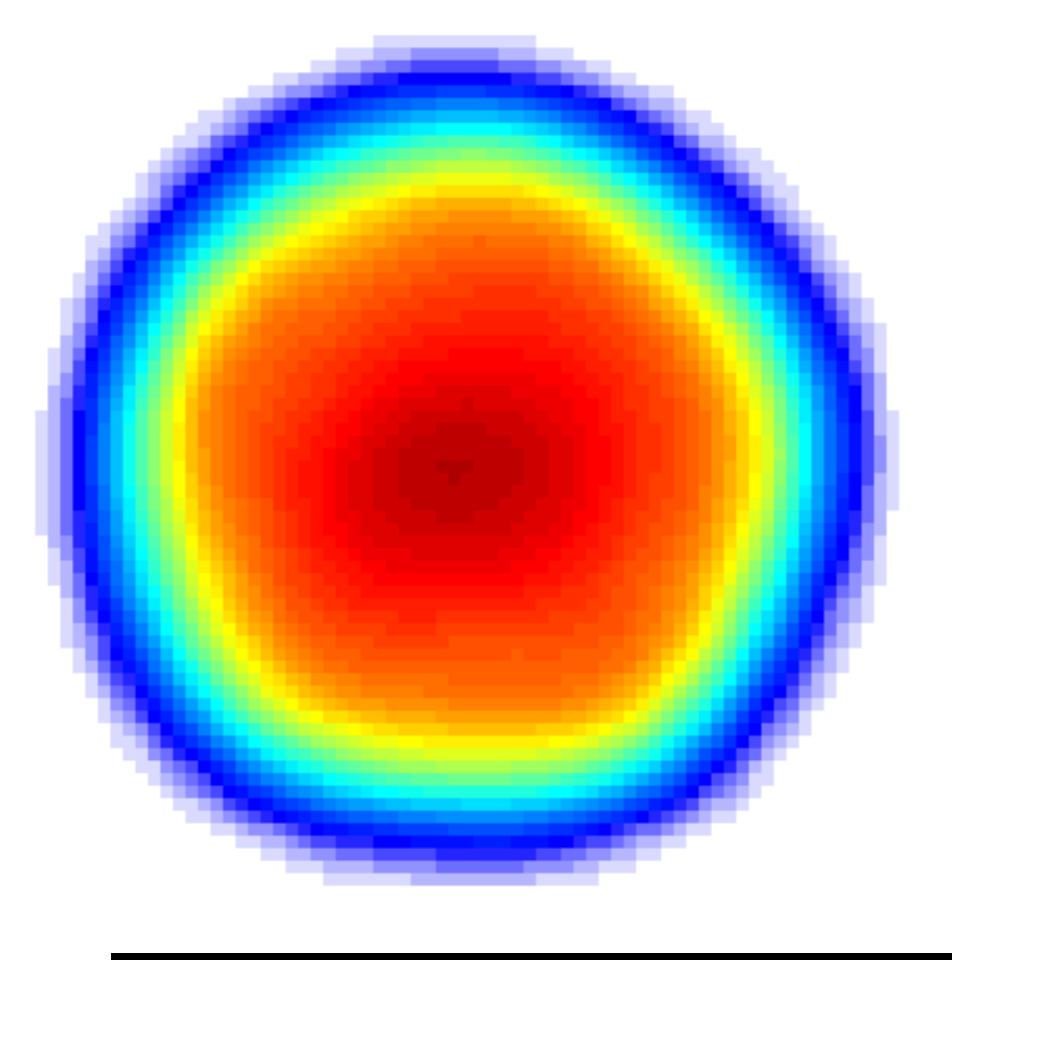} \label{fig:3D_Carl1908_noconstraints_slices3}}
  \subfloat  {\includegraphics[width=.2\textwidth]{fig/colorbar}} \\
  \caption{Phasing results of COACS healed
    $N_{3k}$. \protect\subref{fig:3D_Carl1908_noconstraints_3Dreal}
    shows the interior structures as isosurface plots at 10\%, 50\%,
    79\% and 89\% values of the maximum electron density, and here the
    electron density was phased without applying any
    constraint. [\protect\subref{fig:3D_Carl1908_noconstraints_slices1}
    -- \protect\subref{fig:3D_Carl1908_noconstraints_slices3} are
    three orthogonal slices through the particle center of
    \protect\subref{fig:3D_Carl1908_noconstraints_3Dreal}.}
  \label{fig:phased_COACS}
    \end{figure*}

Figure~\ref{fig:phased_COACS} shows the phasing results of the COACS
healed Fourier intensity for $N_{1k}$.  Compared with the original
phasing results for $N_{1k}$ in Figure~\ref{fig:phased_ori}, the COACS
healed results were much smoother and are slightly larger, due to the
Hann window used in COACS. Similar to the Hann window results in
Figure~\ref{fig:hann}, the healed results also removed the 3-layers
structure, and the central low intensity hole.  It also gave flatter
faces and smoothed vertices. We also slightly improved the resolution
to 9.1~nm compared to applying only a Hann window.

\subsection{Uncertainty Analysis}

\begin{figure*}[!htbp]
   \centering
   \subfloat[]{\includegraphics[width=.2\textwidth]{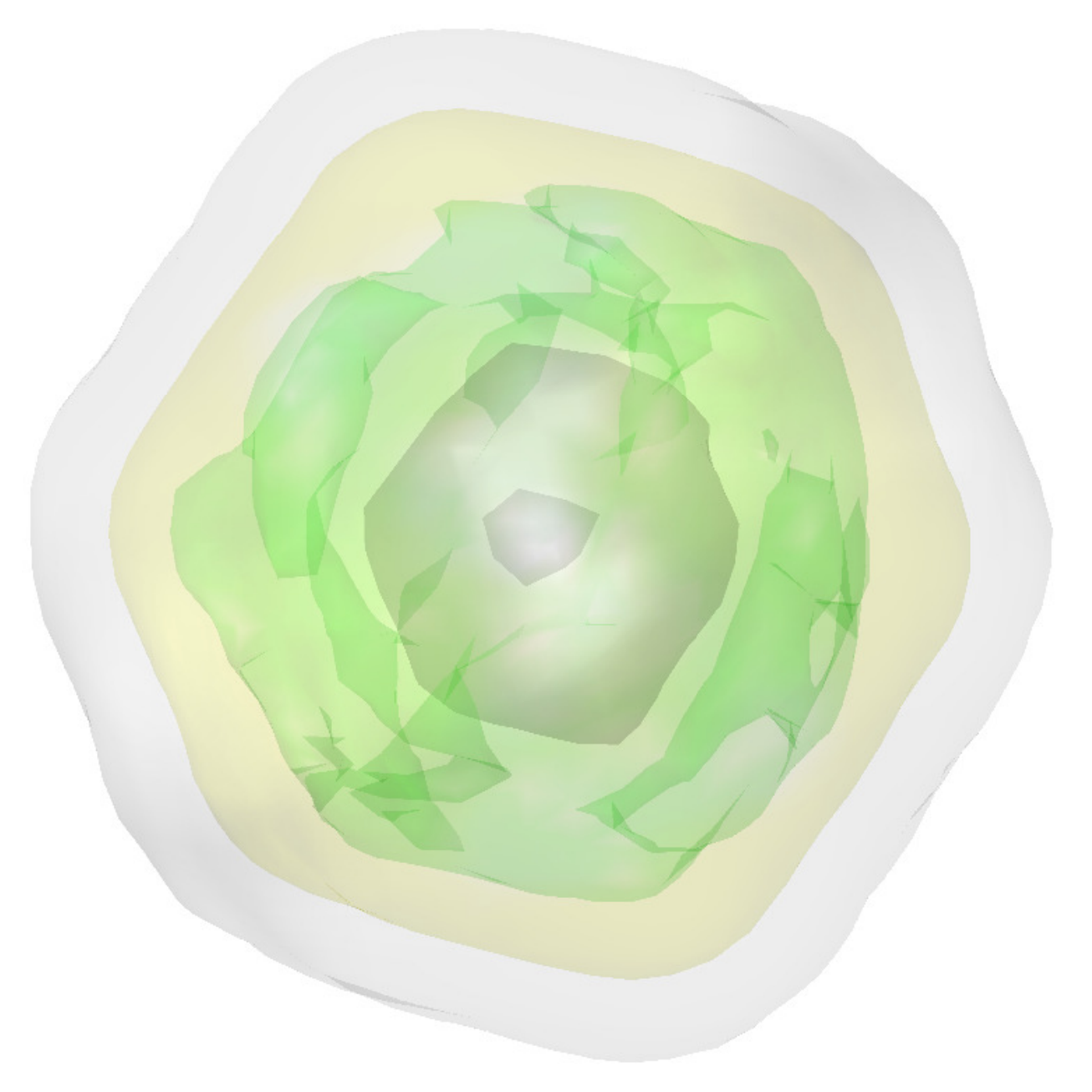} \label{fig:1k_boot}}
   \subfloat[]{\includegraphics[width=.2\textwidth]{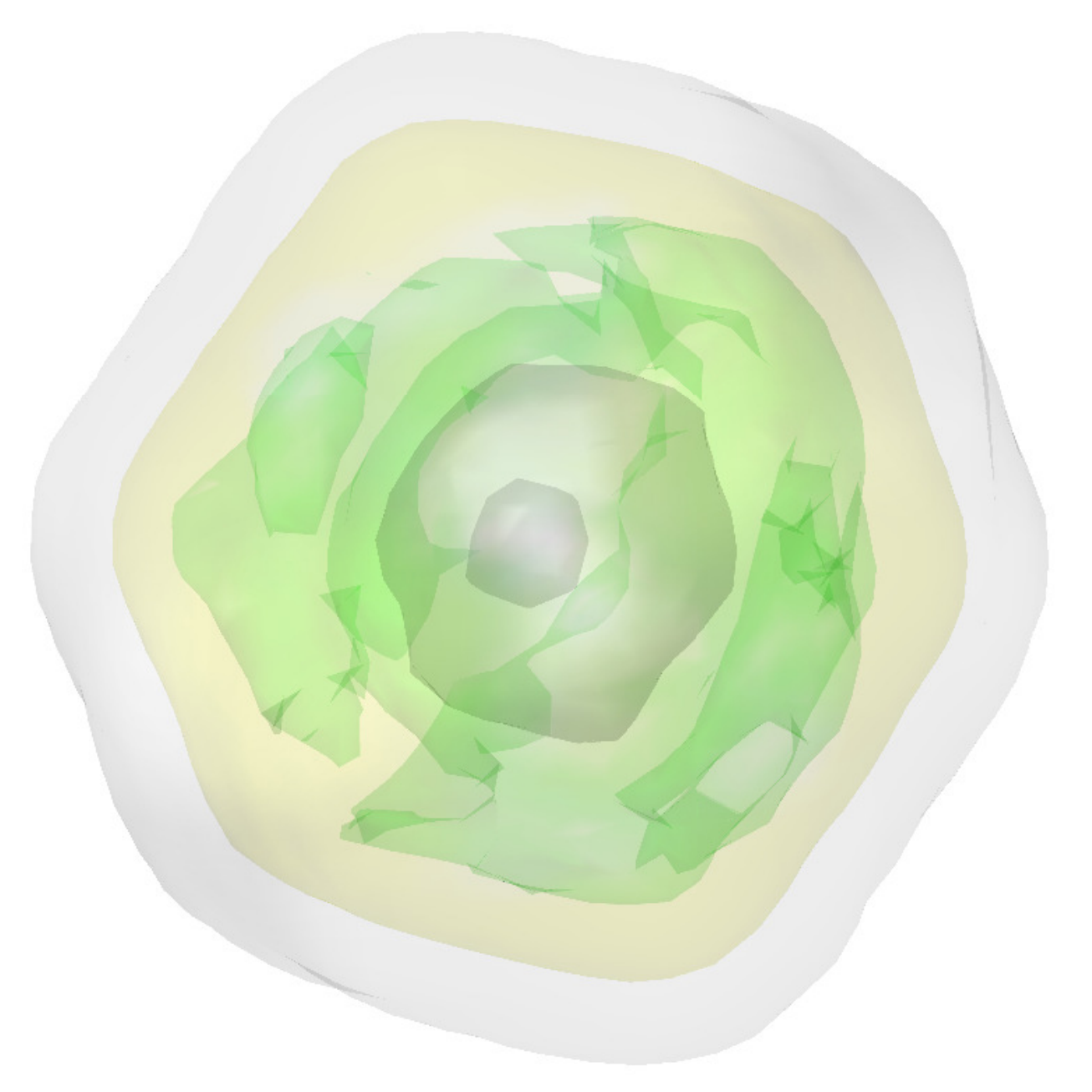} \label{fig:1k_boot_mean}}
   \subfloat[]{\includegraphics[width=.2\textwidth]{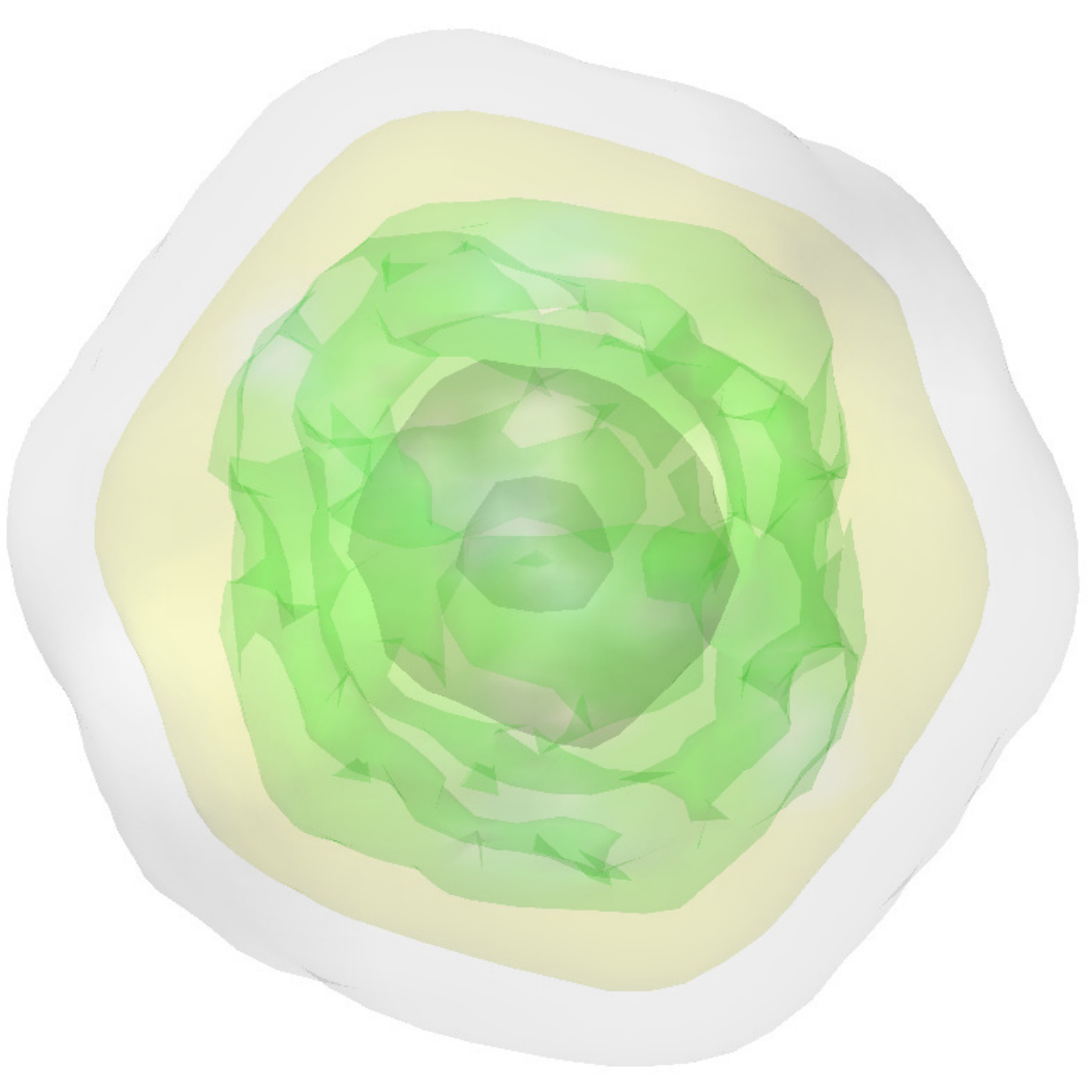} \label{fig:3k_boot}}
   \subfloat[]{\includegraphics[width=.2\textwidth]{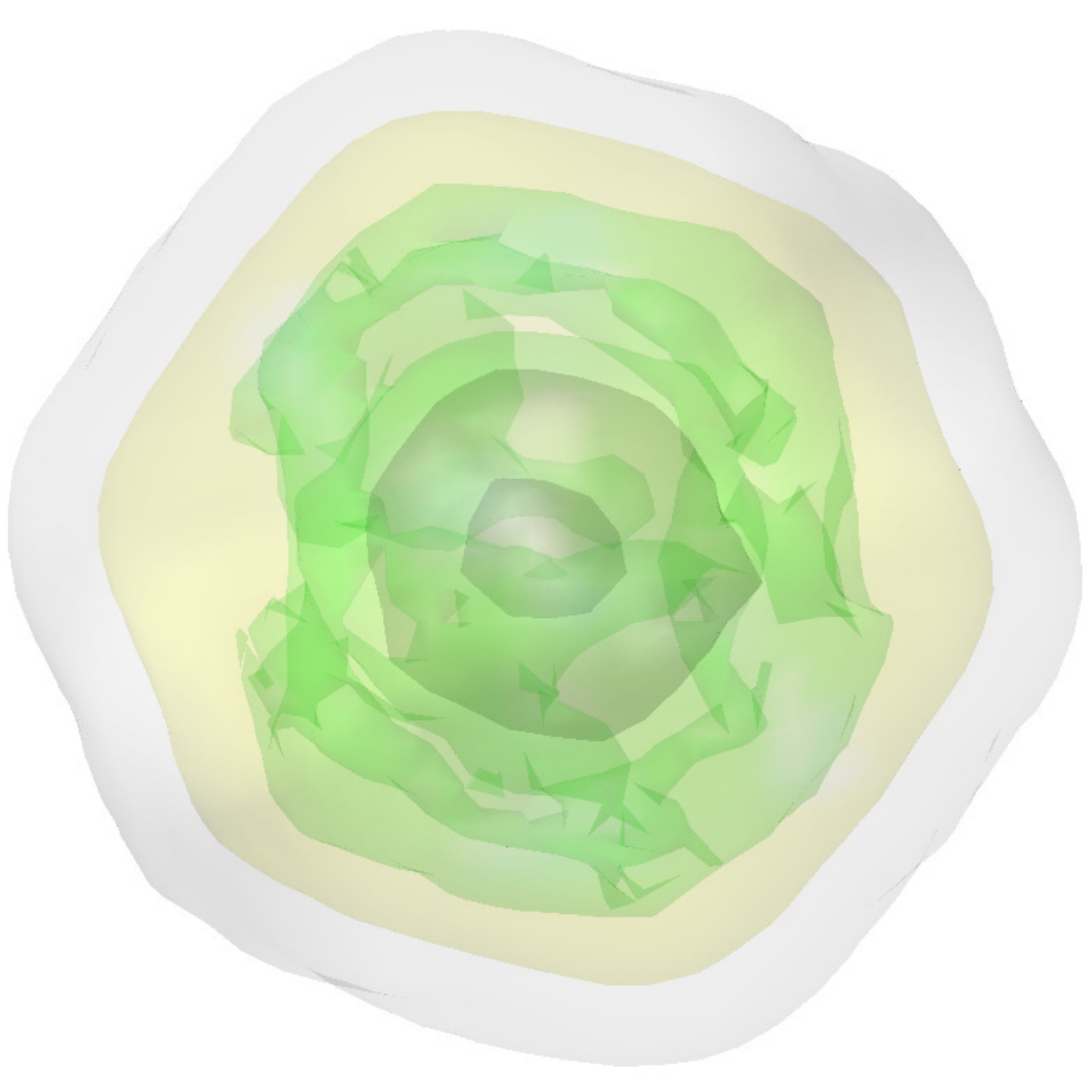} \label{fig:3k_boot_mean}} \hfill\null \\
   \subfloat[]{\includegraphics[width=.2\textwidth]{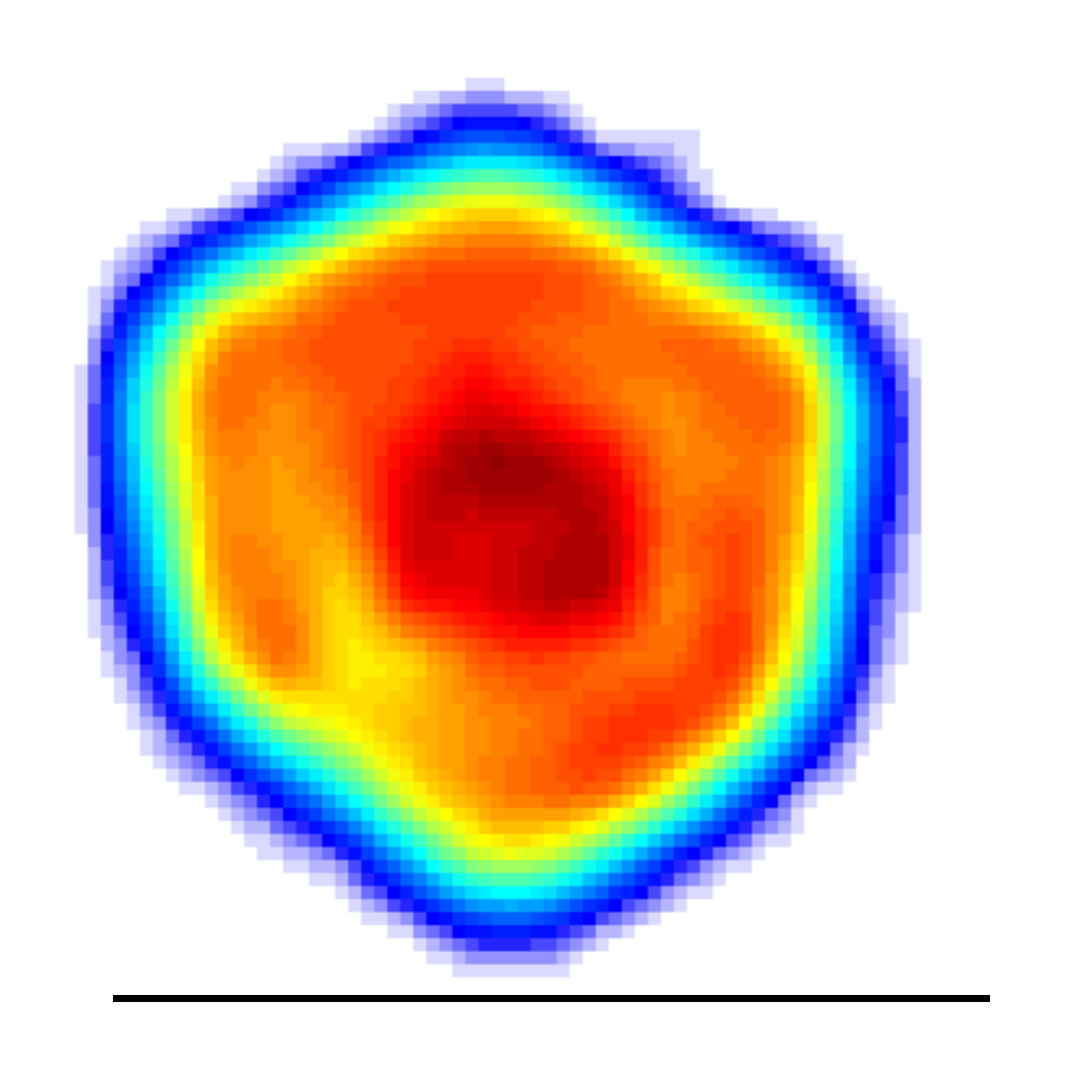} \label{fig:1k_boot_slice}}
   \subfloat[]{\includegraphics[width=.2\textwidth]{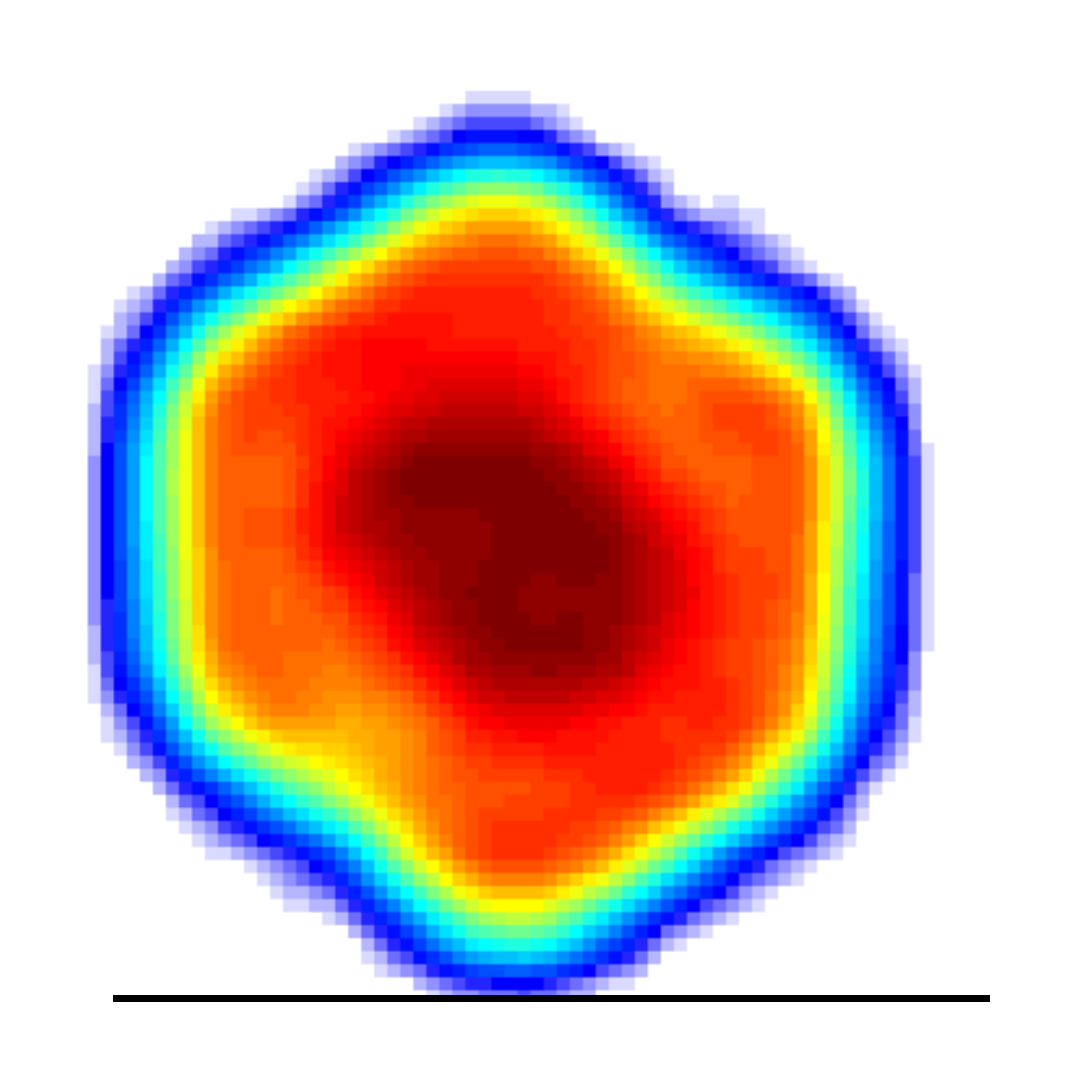} \label{fig:1k_bootmean_slice}}
   \subfloat[]{\includegraphics[width=.2\textwidth]{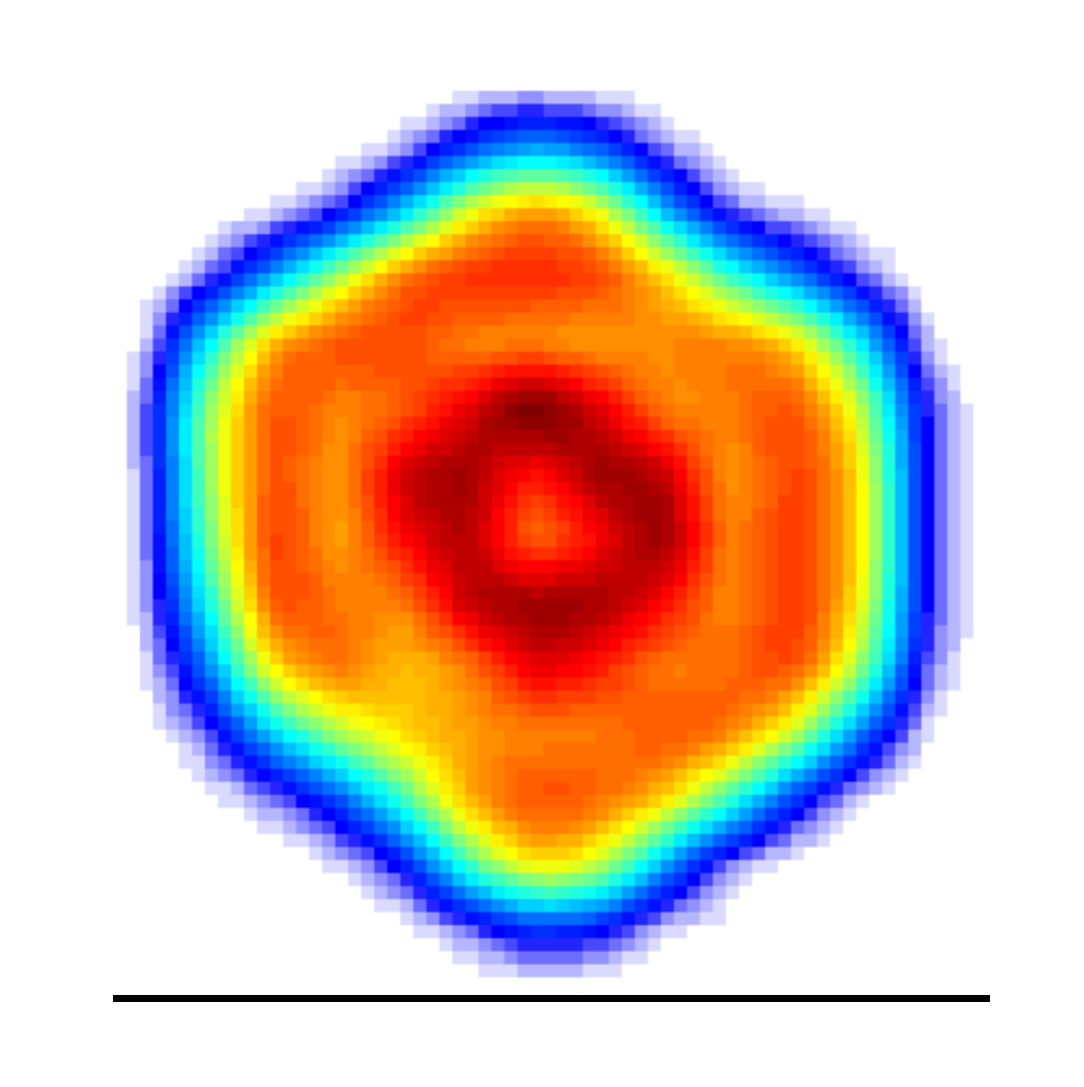} \label{fig:3k_boot_slice}}
   \subfloat[]{\includegraphics[width=.2\textwidth]{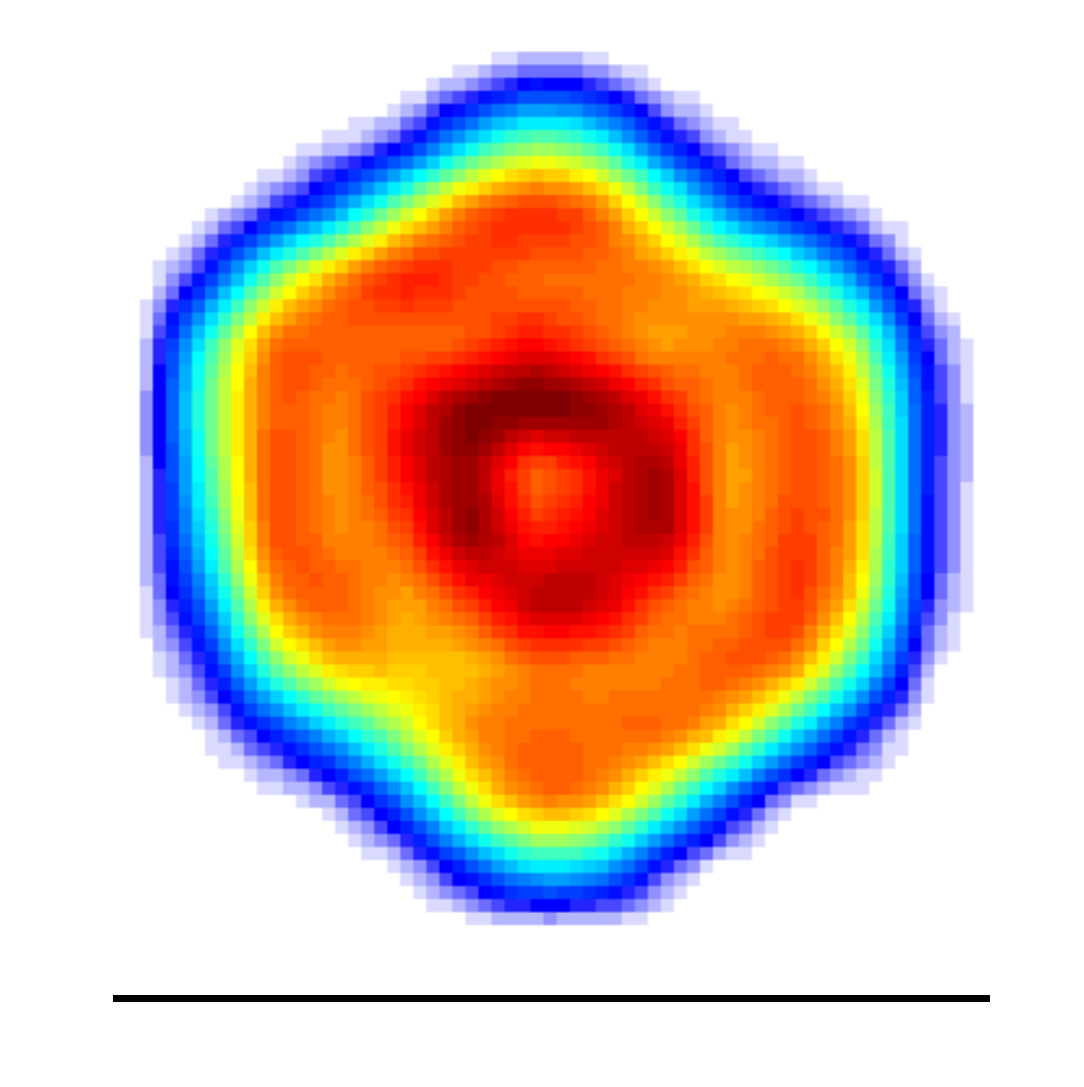} \label{fig:3k_bootmean_slice}}
   \subfloat  {\includegraphics[width=.2\textwidth]{fig/colorbar}} \\

  \subfloat[]{\includegraphics[width=.5\textwidth]{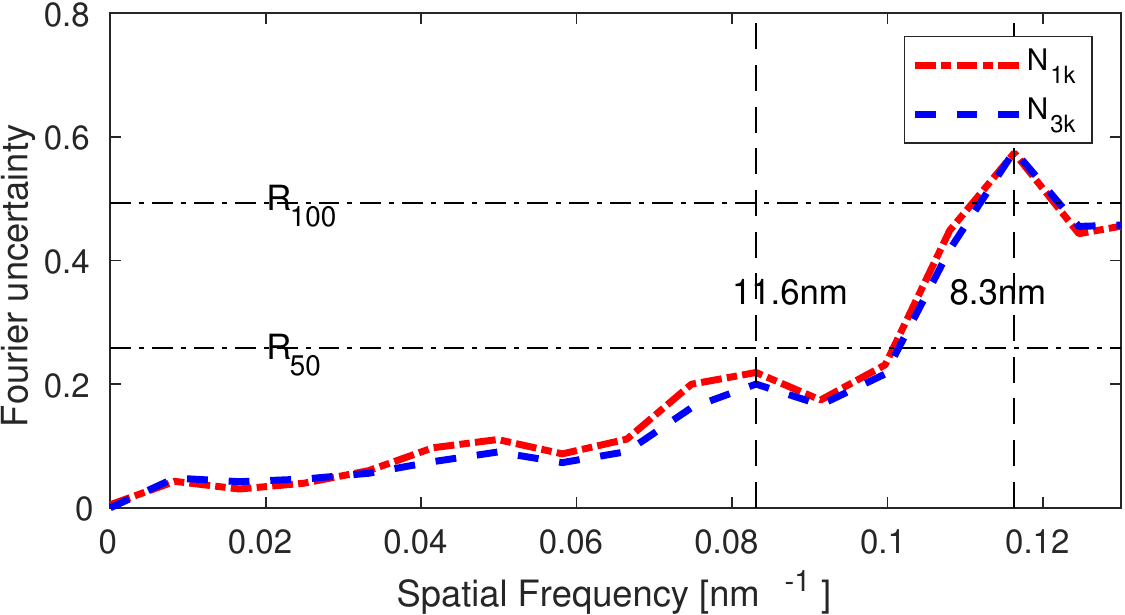} \label{fig:bootFourier}}
   \subfloat[]{\includegraphics[width=.5\textwidth]{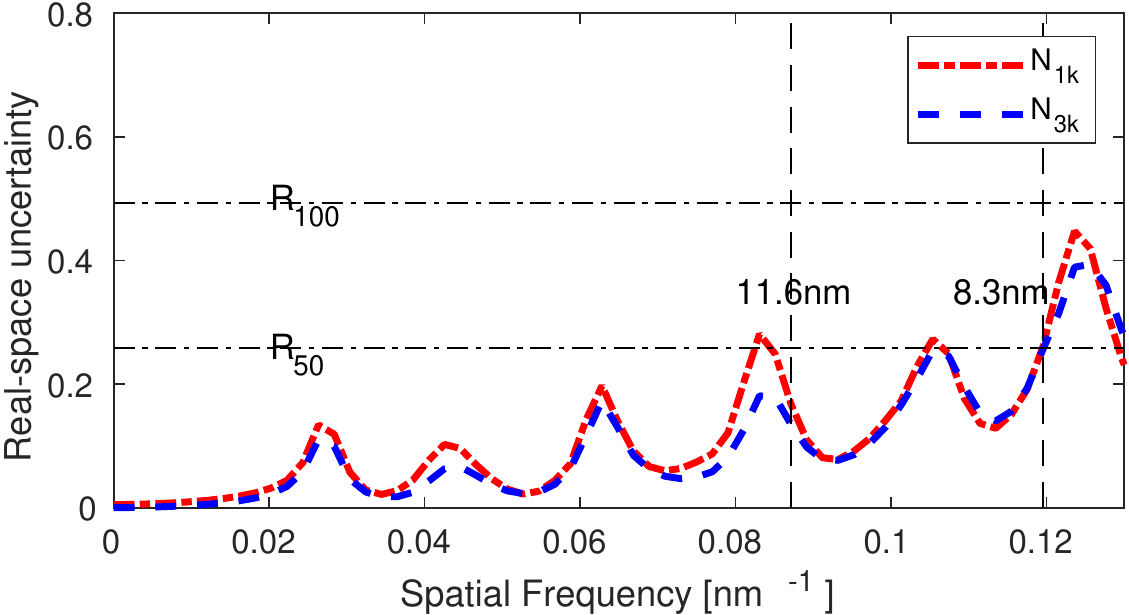}  \label{fig:bootReal}} \\
   \caption{Bootstrap analysis. [\protect\subref{fig:1k_boot} and
     \protect\subref{fig:1k_boot_mean}]: Isosurface at 10\%, 50\%,
     77\% and 87\% values of the maximum electron density of the
     average phased object of $\Wgrid_r$ and the phased object of
     $\Wgrid_m$ for $N_{1k}$. [\protect\subref{fig:3k_boot} and
     \protect\subref{fig:3k_boot_mean}]: Isosuface from the phased
     objects of $\Wgrid_r$ and $\Wgrid_m$ for $N_{3k}$.
     [\protect\subref{fig:1k_boot_slice} and
     \protect\subref{fig:3k_bootmean_slice}]: a slice through the
     particle center of [\protect\subref{fig:1k_boot} and
     \protect\subref{fig:3k_boot_mean}].
     [\protect\subref{fig:bootFourier} and
     \protect\subref{fig:bootReal}]: the estimated uncertainties in
     Fourier and real space, respectively.  }
  \label{fig:bootRf}
\end{figure*}

We have previously proposed a bootstrap procedure to estimate the
uncertainties of the EMC Fourier intensities in \cite{algEMC}, and in
this section, we extend the uncertainty-estimation procedure into
real space, see Figure~\ref{fig:bootRf}. By doing $B=100$ bootstrap
runs for the dataset $N_{1k}$ and $N_{3k}$ using the standard
bootstrap procedure from \cite{algEMC}, we obtained $B$ EMC Fourier
intensities $\Wgrid_r$, $r=1,\ldots,B$, from $B$ random initial
guesses, and the bootstrap mean of $\Wgrid_r$ was $\Wgrid_m$. By
analyzing $\Wgrid_r$ by the method in \cite{algEMC}, we obtained the
Fourier uncertainties, see Figure~\subref*{fig:bootFourier}. Our
uncertainty analysis evolve around the error limits $R_{50}=0.248$ and
$R_{100}=0.485$, respectively, determined as follows. From synthetic
data, $R_{50}$ was the average radial difference between a certain 3D
ground truth and a 3D Fourier intensity, obtained by inserting 1,500
frames (50\%) in their correct rotations and the remaining $1,500$
frames in random rotations. The latter value $R_{100}$ was the
difference between the 3D truth and a Fourier intensity obtained from
3,000 patterns inserted in random rotations.
 
We also measured the uncertainties in the real domain, see
Figure~\subref*{fig:bootReal}.  Let $O_a$ be the average phased
structure in the real domain for the dataset $N_{1k}$ (or
$N_{3k}$). Let $A_a$ be the 3D Fourier intensity computed from $O_a$.
Let $O_r$ ($r= 1,\ldots,B$) be $B$ phased structures from $B$ EMC
Fourier intensities, and $A_r$ be the corresponding Fourier intensity
computed from $O_r$. We also denote the average of $A_r$ by $A_m$, and
hence, the real-space uncertainty can be defined as follows:
\begin{align}
  R_{total}^2 = R_{bias}^2 + R_{std}^2 = |A_a - A_m|^2 + Var(A_r),
\end{align}
where $Var$ is the variance. Note that all data are aligned into the
same orientation during the analysis.

\begin{figure*}[!htbp]
  \centering
  \subfloat[]{\includegraphics[width=.5\textwidth]{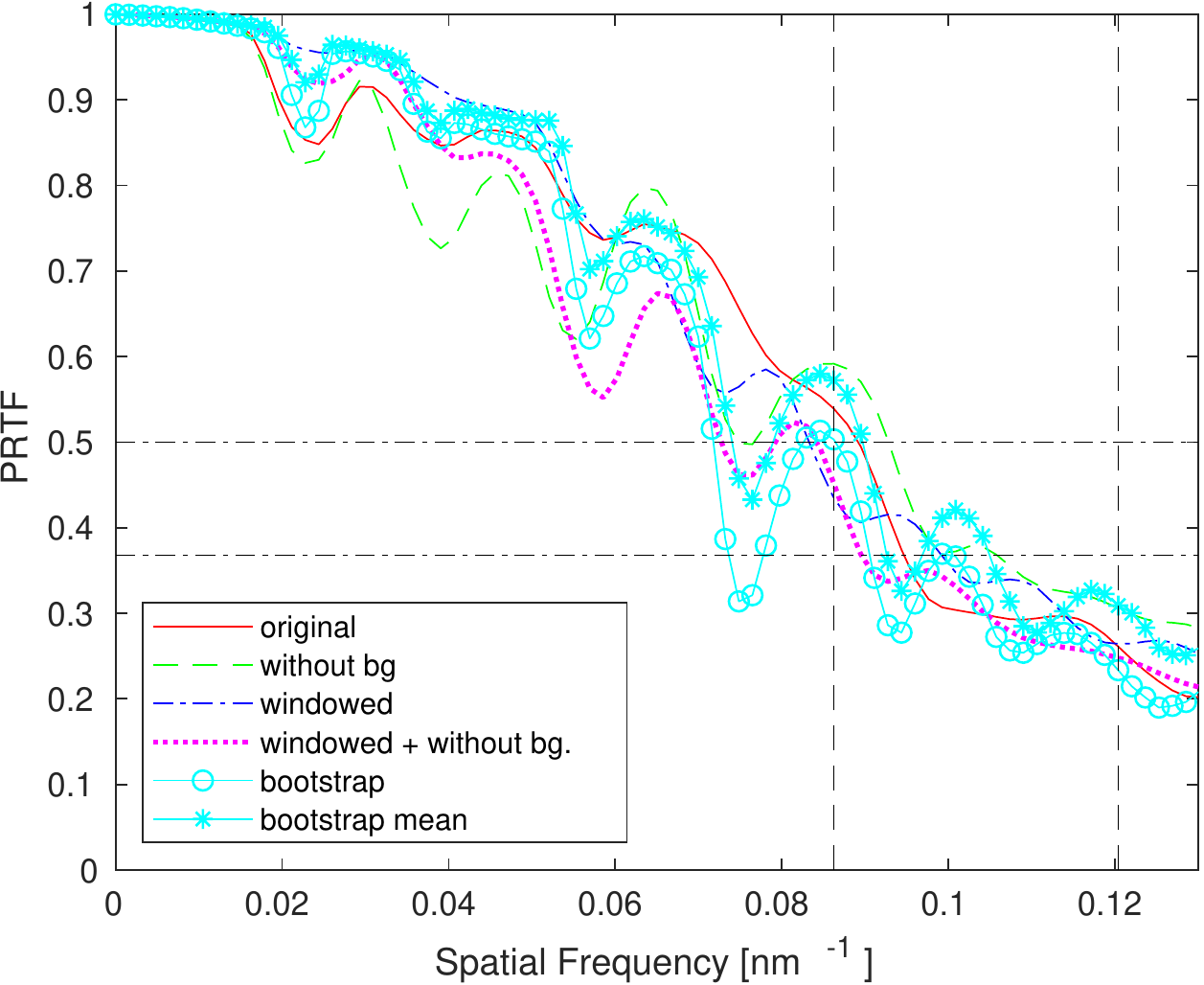} \label{fig:1k_prtf}}
  \subfloat[]{\includegraphics[width=.5\textwidth]{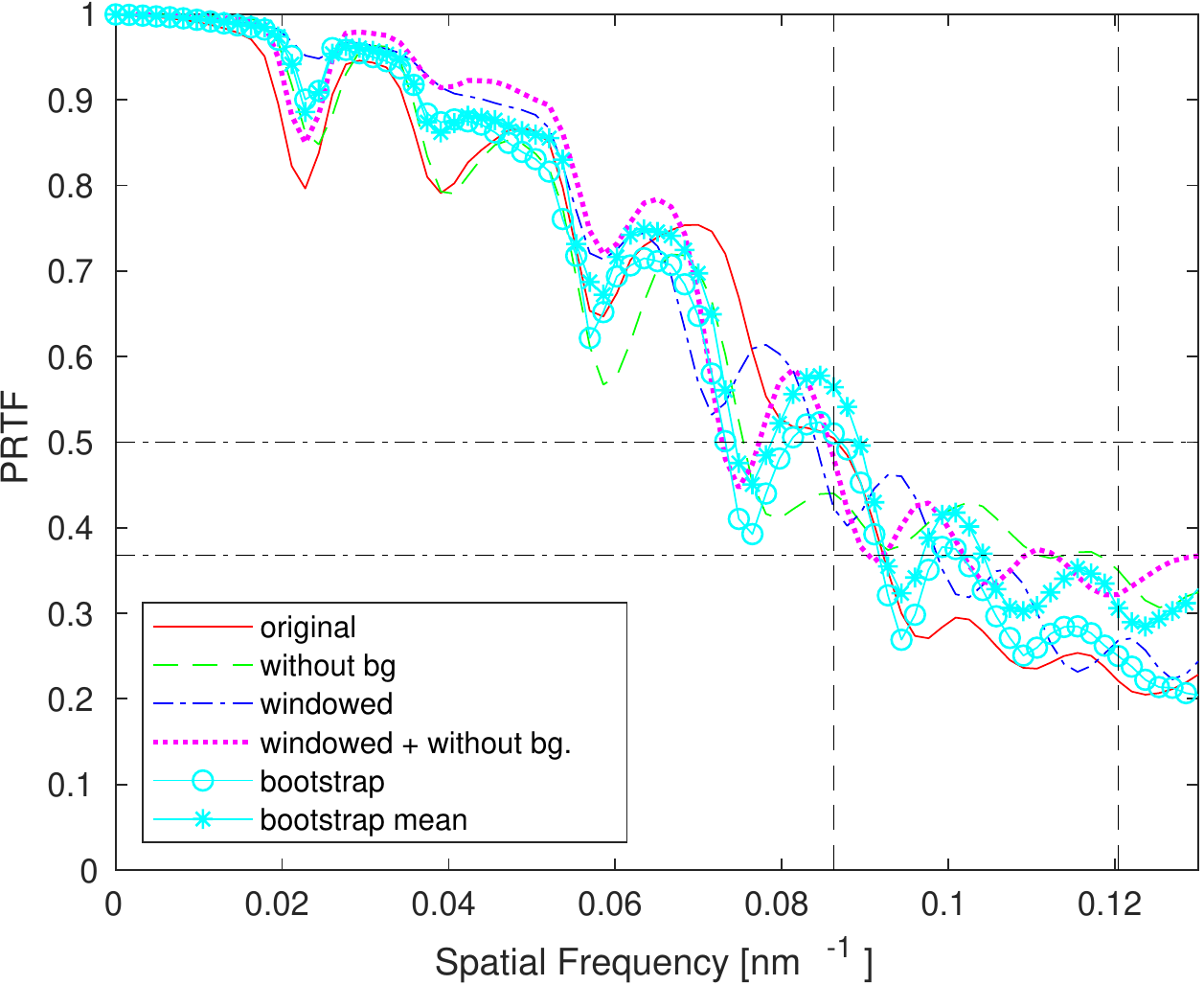} \label{fig:3k_prtf}}\\
  \caption{PRTF analysis for different reconstruction of dataset
    $N_{1k}$ and $N_{3k}$.}
  \label{fig:prtfs}
\end{figure*}

Figure~\ref{fig:bootRf} illustrates the results from the bootstrap
analysis in both real and Fourier space. Both the phased
bootstrap mean $O_m$ and the mean of $O_r$ gave similar results as
$O_a$, see Figure~\subref*{fig:1k_boot} and
Figure~\subref*{fig:1k_boot_mean} (or Figure~\subref*{fig:3k_boot} and
Figure~\subref*{fig:3k_boot_mean}). As expected, the Fourier bootstrap
analysis showed a sharp increase outwards the detector edge and
crossed $R_{50}$ at a resolution of 10~nm. The real-space bootstrap
analysis gave a similar results. However, the uncertainty from the
dataset $N_{1k}$ crossed $R_{50}$ at 12.5~nm, which may be due to the
smaller number of frames in $N_{1k}$. We obtained similar resolution
from the PRTF analysis, see Figure~\ref{fig:prtfs}. Again, PRTF showed
that the resolution from $\Wgrid_r$ of $N_{3k}$ was 10.7 nm and
12.8~nm for $N_{1k}$. The phased objects from the bootstrap analysis
maintained the concentric capsid layer structure, and the low
intensity area close to the facets. However, the central ring got a
higher intensity and the phased particle was much smoother compared
with the original particle (see Figure~\ref{fig:phased_ori}).

\subsection{Shape Analysis and Error Validation}

Phase retrieval algorithms such as RAAR \cite{rra,Li2017} and ER
\cite{Fienup:82}, are iterative algorithms with concave Fourier
constraints, and hence we validate our phasing procedure with Fourier
error $E_f$ and real error $E_r$,
\begin{align}
  E_f = \sqrt{\dfrac{\sum_{i=1}^{\Mpix}(|\tilde{h}_i| - \sqrt{I_i}) ^2} {\sum_{i=1}^{\Mpix} I_i}},
  \label{eq:phase_error_F}
\end{align}
and
\begin{align}
  E_r = \sqrt{\dfrac{\sum_{i \in \bar{S}}(|h_i|) ^2} {\sum_{\in \bar{S} \cup S} |h_i|^2}},
  \label{eq:phase_error_R}
\end{align}
where $S$ is the object support, $\bar{S}$ is the area outside of
$S$. Further, $h$ is the recovered real-space intensity, while
$\tilde{h}$ is the recovered wave in Fourier space, $\Mpix$ is the
number of pixels of the detector, and $I_i$ the $i$th pixel value of a
diffraction pattern.  We also explored the particle shapes and sizes
of the phased objects as follows.

For calculating sizes, we chose an intensity threshold of 10\% of the
maximum value, and the object diameter $D_r$ is defined as
\begin{align}
  D_r = 2m\left(\dfrac{3V}{4\pi}\right)^{-3}  ,
  \label{eq:dr}
\end{align}
where $V$ is the number of pixels with intensities above the intensity
threshold, $m$ is the pixel size in the real space, and
$m \approx 5.2$~nm for our retrieved objects.  We also investigated
the maximum $D_{max}$, the minimum $D_{min}$ and the mean $D_{mean}$
distances of $300$ opposite points pairs of the virus capsid. With
these values, we may quantitatively analyse the particle shape and
phasing errors in Table~\ref{tab:er}.

\begin{table}[!htbp]
  \centering
  \caption{Shape and error analysis for the phased objects from the
    dataset $N_{1k}$ and $N_{3k}$. The measured sizes ( $D_r$,
    $D_{mean}$, $D_{max}$, $D_{min}$) are in nanometers and so are the
    resolutions (from a PRTF analysis with a threshold of $e^{-1}$,
    see Figure~\ref{fig:1k_prtf} and Figure~\ref{fig:3k_prtf}).  The
    object diameter was defined in \eqref{eq:dr}.  The phasing errors
    $E_r$ and $E_f$ were calculated in the last phase-retrieval
    iteration by \eqref{eq:phase_error_R} and
    \eqref{eq:phase_error_F}, respectively. The abbreviations Win.,
    w/o, bg., and Res.~mean windowed signal, without, background and
    resolution, respectively.}
\begin{tabular}{l|l|c|c|c|c|c|c|c}
\hline
                          &                & $D_r$ & $D_{mean}$ & $D_{max}$ & $D_{min}$ & $E_r$   & $E_f$  & Res  \\ \hline
\multirow{6}{*}{$N_{1k}$} & Original       & 64.8  & 63.2       & 68.9      & 58.5      & 6.25e-7 & 3.4e-3 & 10.6 \\
                          & Win.            & 64.7  & 64.6       & 69.1      & 58.5      & 9.9e-7  & 3.2E-3 & 10   \\
                          & w/o bg.         & 65.7 & 63.4 &  68.6 & 57.0 & 8.0E-6 & 5.5E-3 & 9.5    \\
                          & Win. w/o bg.   & 65.8& 63.6& 68.7 & 58.0 & 6.1E-6 & 4.0E-3 & 11.2     \\
                          & Bootstrap      & 65.3& 63.0 & 69.1 & 58.1 & 5.4E-6& 3.1E-3& 12.8     \\
                          & Bootstrap mean & 65.2 & 63.0& 68.5&57.8& 1.1E-6& 1.5E-3&  10.7      \\
                          &COACS & 65.1 & 64.3& 68.4&59.3& 4.3E-6& 1.7E-4&  9.1 \\  \hline
\multirow{7}{*}{$N_{3k}$} & Original      & 64.9 &  63.2 &69.4 & 58.6 & 6.3E-6 &3.2E-3 & 10.7      \\
                          & Win.            & 64.4 & 64.4 & 69.2 & 58.5 & 9.9e-5 & 3.4E-3 &  10    \\
                           &w/o bg. & 64.7 & 64.4 &  68.9 & 56.6 & 4.6e-5 & 5.1E-3 & 8.4 \\  
   & Wind. w/o bg. & 65.2& 64.0& 69.0 & 58 & 4.6e-5 & 5.0E-3 & 8.7\\ 
    &Bootstrap & 65.3& 66.5& 69.2&57.1& 5.4E-6& 2.7E-3&10.7  \\ 
    &Bootstrap mean & 65.3 & 63.3& 69.4&57.8& 5E-6& 1.5E-3&  10.7 \\
   \hline
\end{tabular}
\label{tab:er}
\end{table}

\begin{figure*}[!htbp]
  \centering
  \subfloat[]{\includegraphics[width=.25\textwidth]{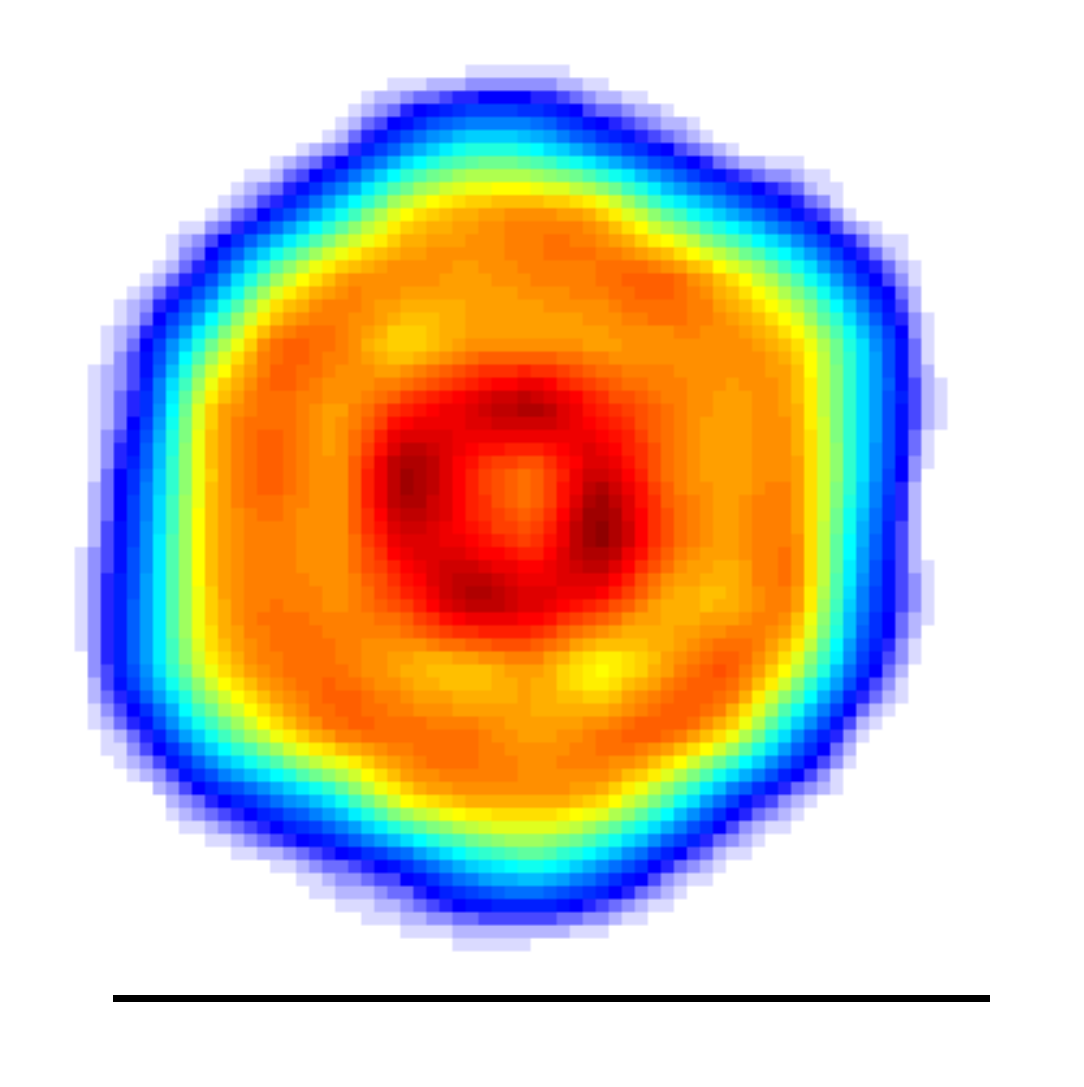} \label{fig:3k_slice1}}
  \subfloat[]{\includegraphics[width=.25\textwidth]{fig/3k_ave_slices2} \label{fig:3k_slice2_again}}
  \subfloat[]{\includegraphics[width=.25\textwidth]{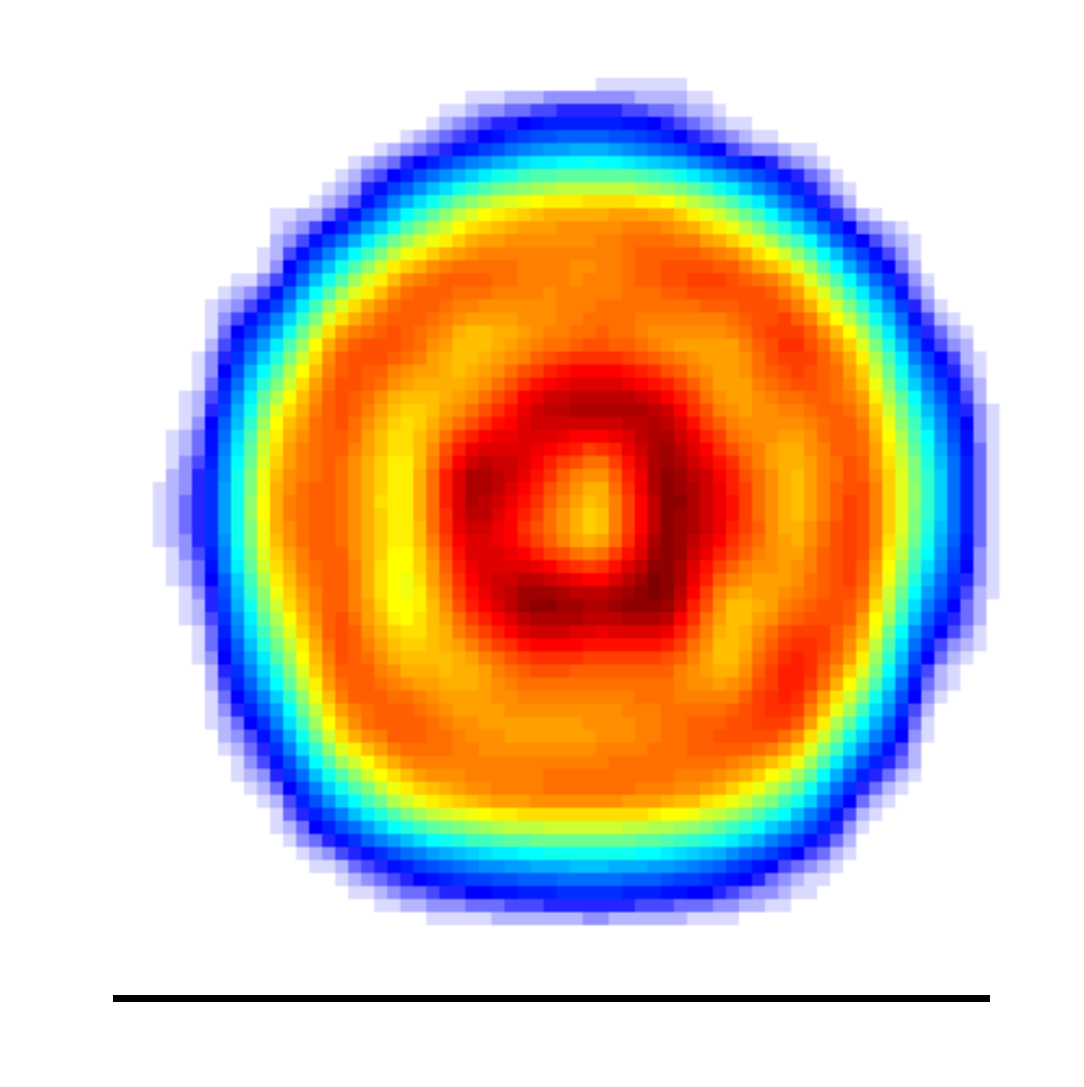} \label{fig:3k_slice3}}    
  \subfloat{\includegraphics[width=.25\textwidth]{fig/colorbar}}\\
  \subfloat[]{\includegraphics[width=.25\textwidth]{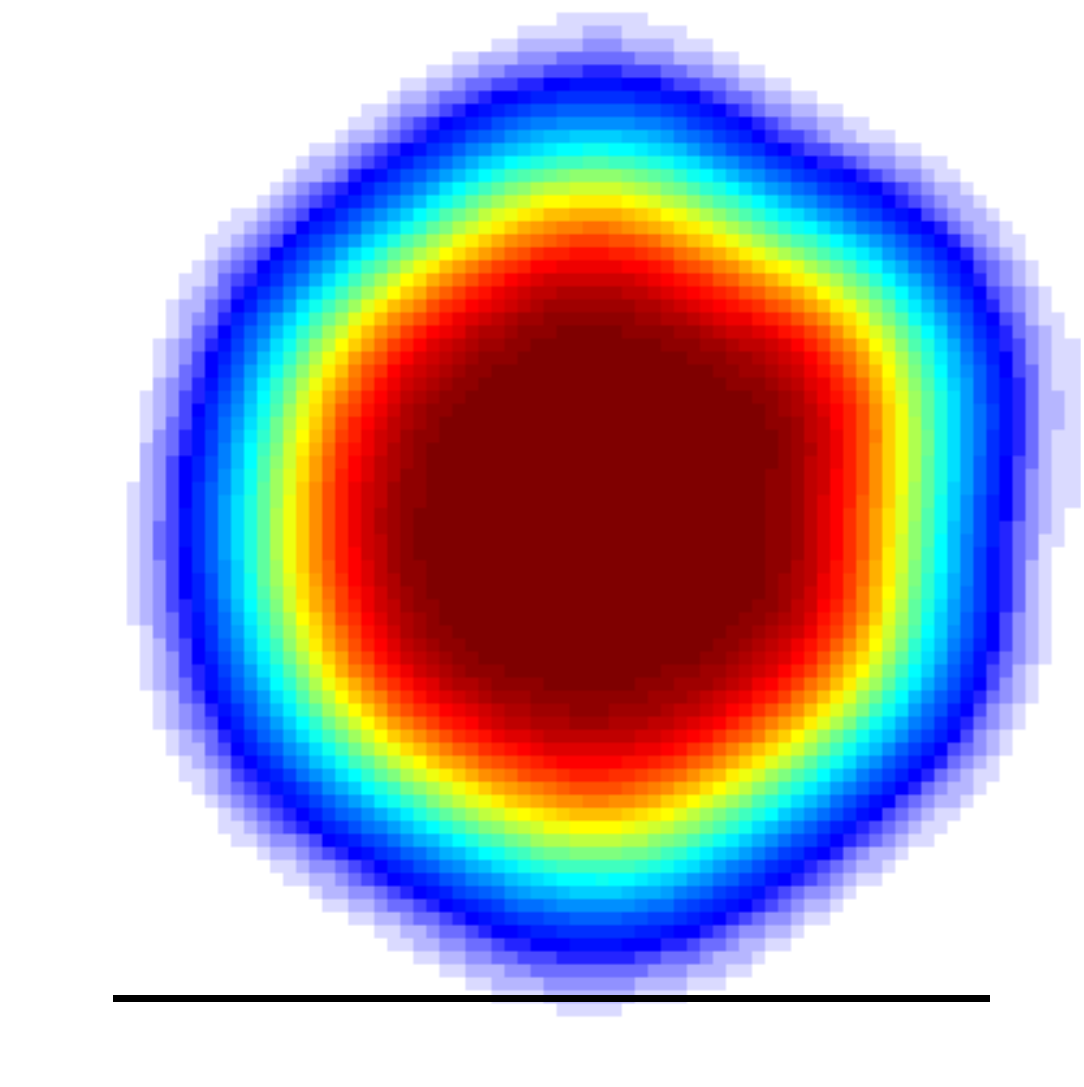} \label{fig:ico_slice1}}
  \subfloat[]{\includegraphics[width=.25\textwidth]{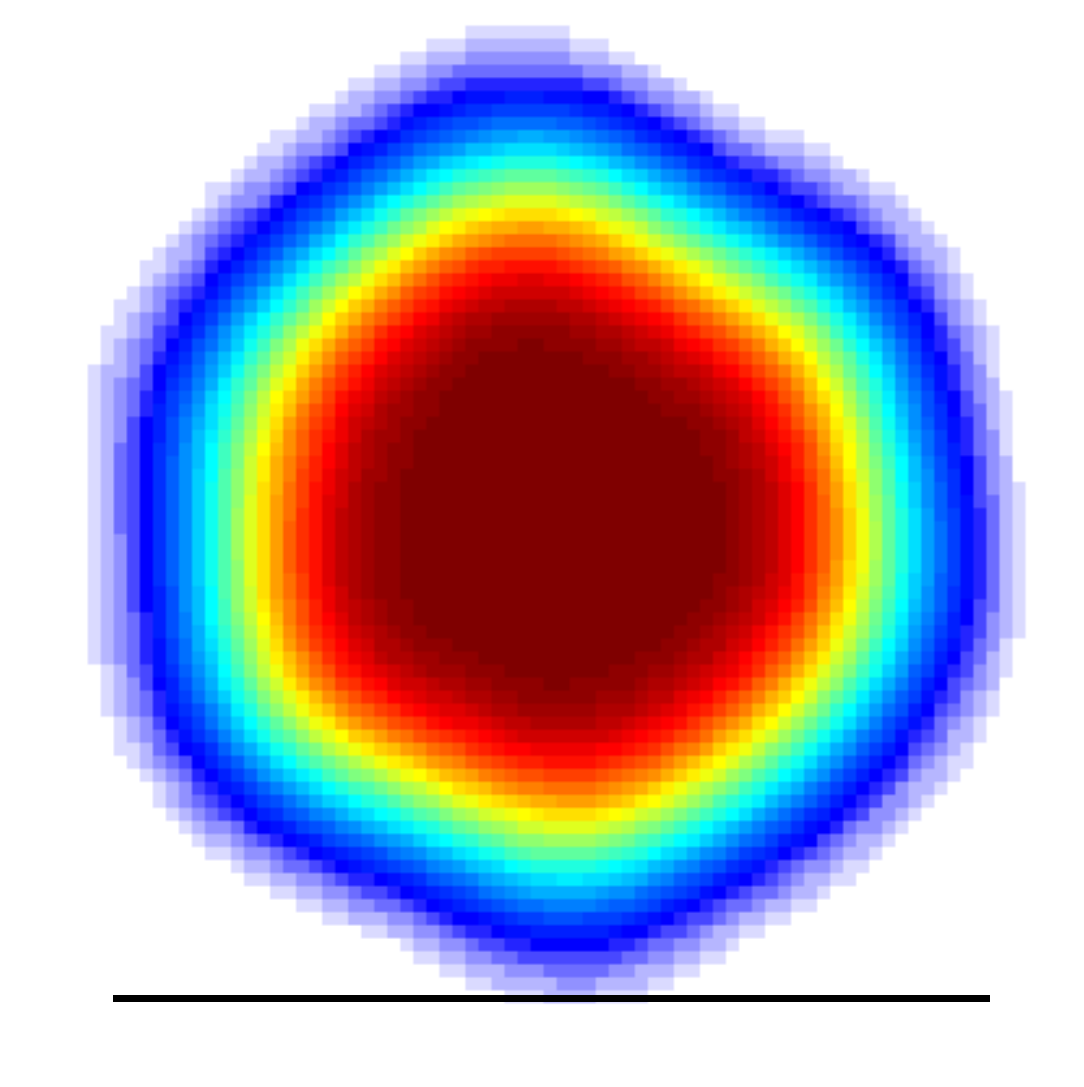} \label{fig:ico_slice2}}
  \subfloat[]{\includegraphics[width=.25\textwidth]{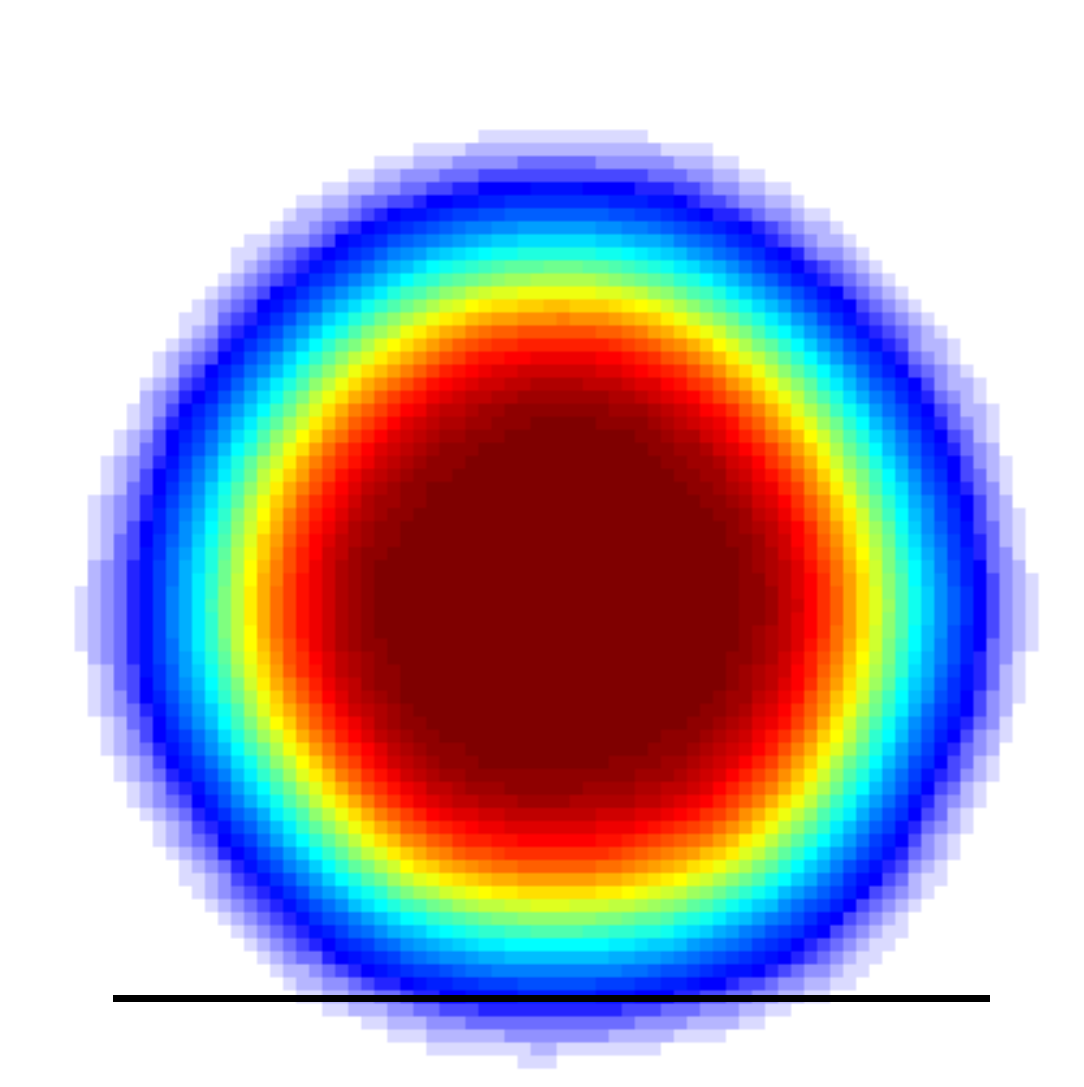} \label{fig:ico_slice3}}   
  \subfloat{\includegraphics[width=.25\textwidth]{fig/colorbar}} \hfill
  \subfloat[]{\includegraphics[width=.25\textwidth]{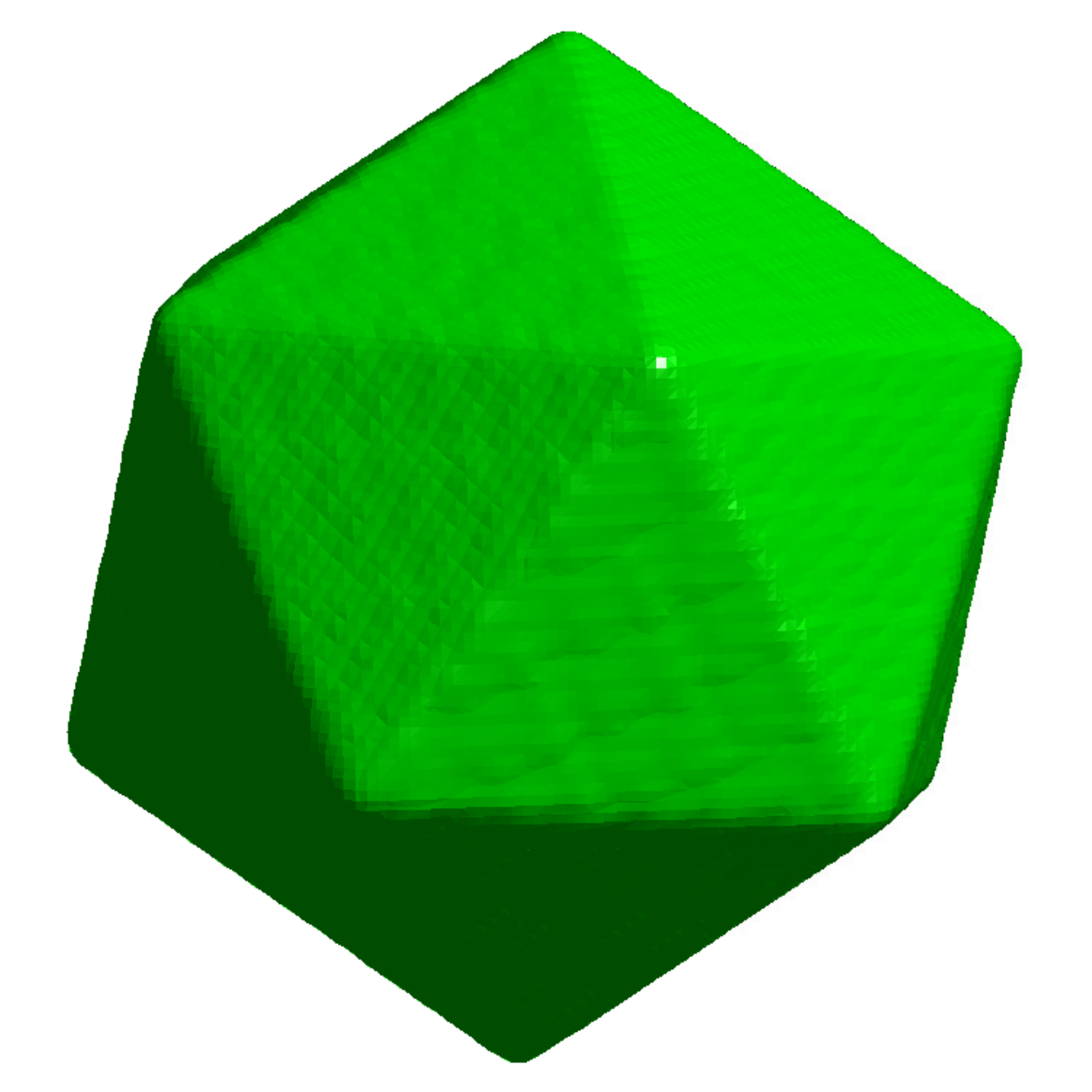} \label{fig:ico_surf1}}
  \subfloat[]{\includegraphics[width=.25\textwidth]{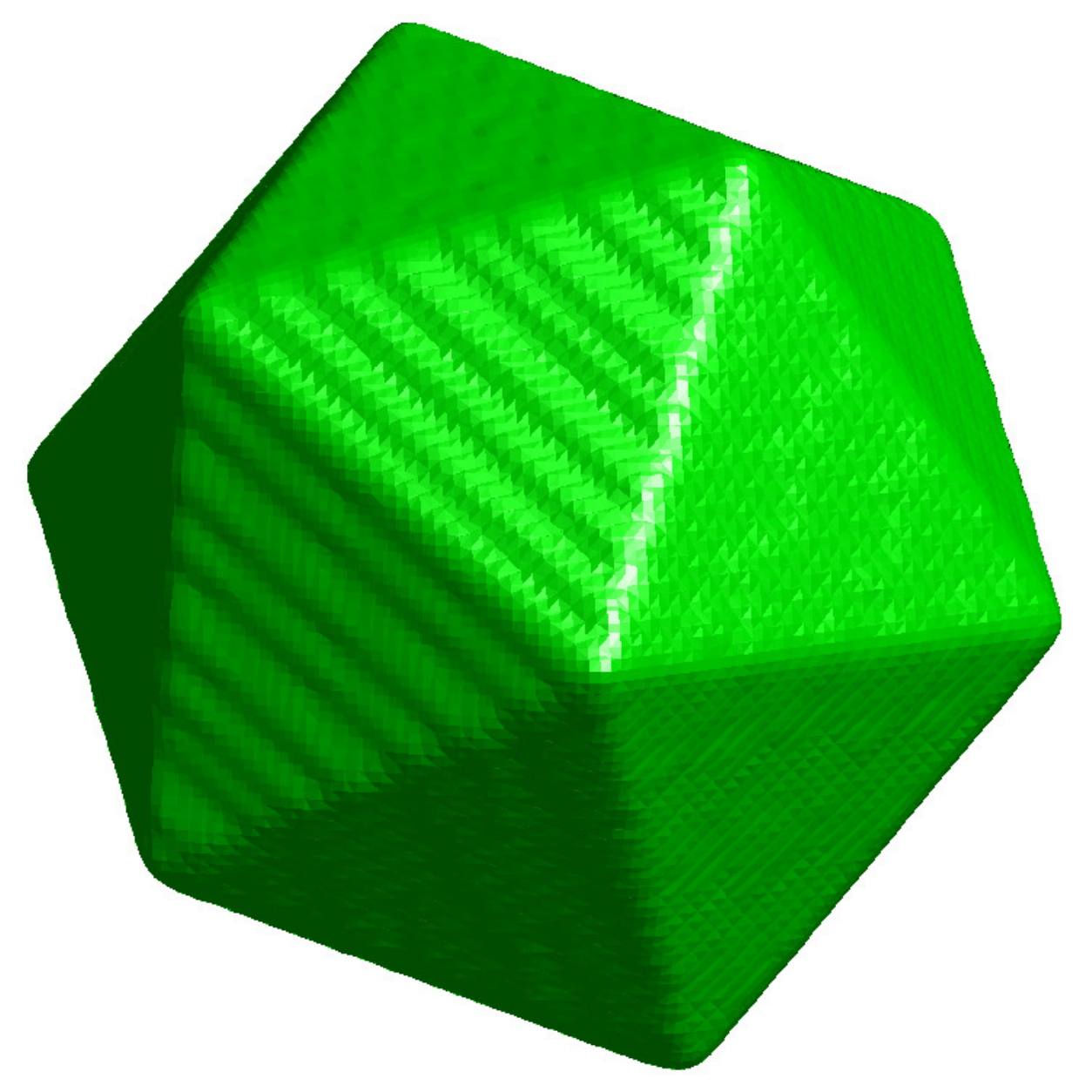} \label{fig:ico_surf2}}
  \subfloat[]{\includegraphics[width=.25\textwidth]{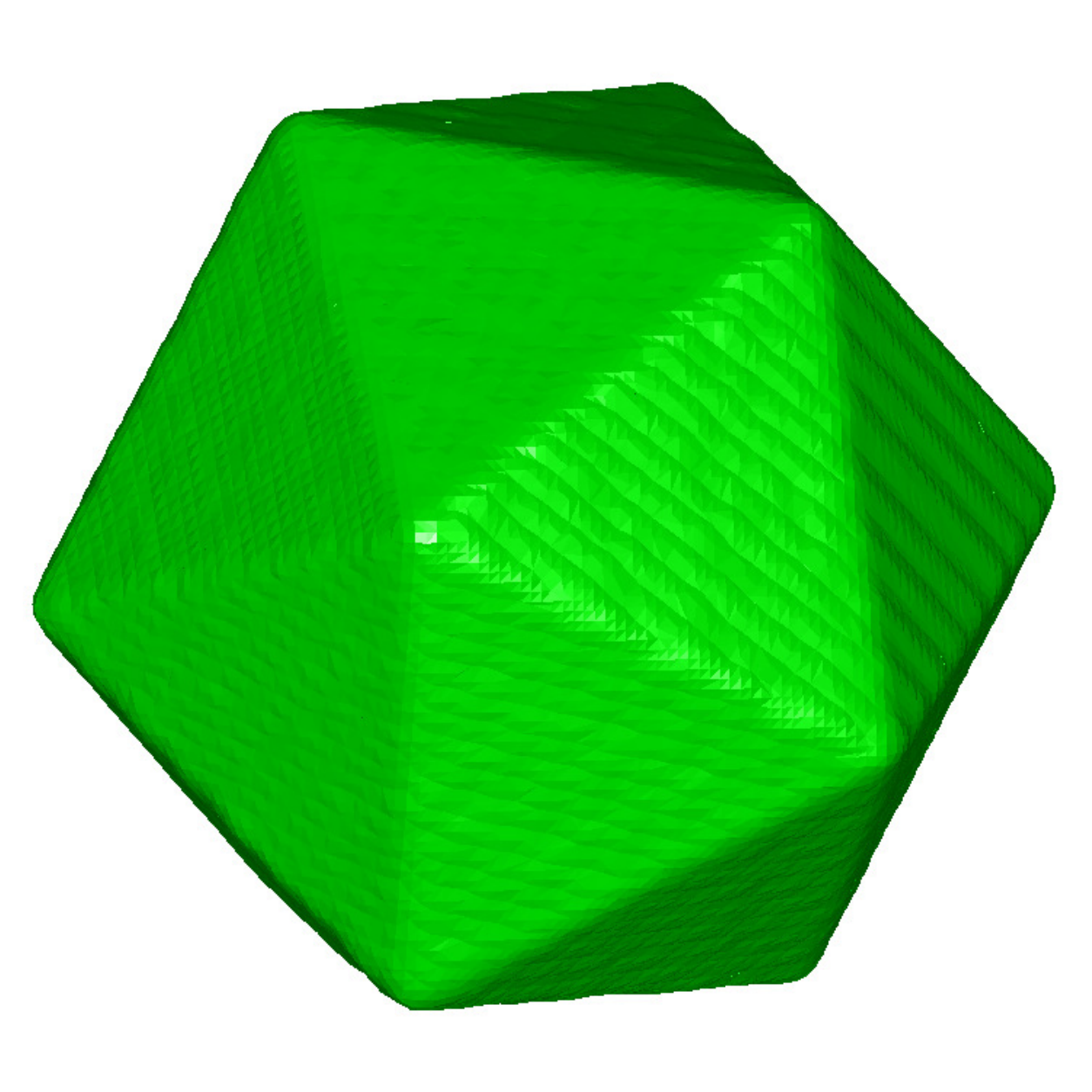} \label{fig:ico_surf3}}         \hfill
  \null 
  \caption{\textit{Top row}: Cross-section images of the retrieved
    particles from the dataset $N_{3k}$ in
    Figure~\ref{fig:phased_ori}.  \textit{Middle row}: Cross-section
    images from a smooth icosahedron object at the same rotations.  To
    create the smooth icosahedron, we first generated an ideal
    icosahedron, whose outer radius was 12 pixels, and then smoothed
    it using two Gaussian convolution kernels.
    [\protect\subref{fig:ico_surf1}--\protect\subref{fig:ico_surf3}]:
    The surfaces of a ideal icosahedron object at the same
    rotations. The retrieved particle (the \textit{top} row) matched
    well to the smoothed icosahedron (the \textit{middle} row), but
    the vertices were much smoother than vertices of an ideal
    icosahedron (the \textit{bottom} row). }
  \label{fig:all3slices}
\end{figure*}

The volume of a regular icosahedron, whose distance between two
vertices is 69~nm is about $1.04\times10^5$ voxels, which is
equivalent to the volume of a sphere of diameter 58.36~nm. However,
the estimated diameter $D_r$ of the retrieved particle was around
64.9~nm, suggesting that the retrieved particle was smoother than an
ideal icosahedron, and it is hard to observe vertices in some
directions, see the cross-section images in
Figure~\ref{fig:all3slices}. As references, we compared our retrieved
virus with (smoothed/non-smoothed) regular icosahedra. To generate
smoothed icosahedron we convoluted two Gaussian kernels (window size
was $3\times3\times3$ pixels and the standard deviation was 1 pixel)
over an 69~nm regular icosahedron. We argue that, at the current
resolution, we can not observer clear features in the retrieved object
as the object was blurred, and therefore further investigation of
higher resolution diffraction patterns is necessary to determine any
interior features.

\section{Conclusion}
\label{sec:conclusion}

We have proposed a working data-analysis pipeline for FLASH X-ray
single particle imaging experiments. This pipeline works with the raw
FXI diffraction patterns, and retrieves high-resolution 3D electron
density of the sample particles together with the associated
uncertainty analysis. The workflow consists of several steps,
including a) selection of single hit diffraction data, b)
classification, c) 3D reconstruction in Fourier space, and d)
post-analysis with reconstruction of the particle electron density by
phase retrieval methods, `healing' the 3D Fourier intensity, handling
background noise, bootstrapping analysis and shape/size analysis.

As an applied demonstration we used the data from the AMO beamline at
LCLS \cite{pr772_sd}. We used the results of the diffusion map
embedding in \cite{pr772_sd}, as the collection of single hit
diffraction data, and used a template-based classification method
\cite{psort}, the EigenImage classifier, to select high-quality
patterns. As a result of this classification, we pushed slightly more
than 1000 and, respectively, 3000 diffraction patterns into the
reconstruction procedure, based on a scaled Possonian model
\cite{algEMC}. With 3$\times$Nvidia GeForce GTX 680, we were able to
align our datasets into 3D Fourier intensities within the hour.

At the last step, we performed the 3D phase retrieval to reveal the
electron density of the virus.  As expected, the size of the retrieved
particle was around 69~nm with asymmetric shape and the obtained
resolution was in between the detector-edge resolution and the
detector-corner resolution. We conclude that the proposed
data-analysis pipeline is able to handle raw FXI data properly and
obtain 3D electron densities at the design resolution. We also argue
that increasing the scattering angle will further improve the
resolution. In addition to performing phase retrieval directly on the
Fourier intensities, we also suggested a 3D background removal
procedure, the use of 3D Hann window, and the use of the Convex
Optimization of Autocorrelation with Constrained Support (COACS)
method \cite{COACS} for achieving better resolution and lower aliasing
effects. Further, we also employed a previously suggested bootstrap
procedure to reconstruct and measure Fourier/real intensities in a
more robust way with a consistent uncertainty analysis.

The newer XFEL facilities, such as the European XFEL and the LCLS II,
can produce a sufficient amount of single hits at larger diffraction
angles with higher photon fluences. We may thus obtain several
billions of diffraction patterns for one FXI experiment in the near
future, and hence a proper data analysis pipeline is absolutely
necessary to determine the 3D electron density of sample particles
reliably, rapidly, and robustly. With our proposed pipeline the 3D
electron density may be practically obtained \emph{during} the FXI
experiment along with the appropriate uncertainty analysis.

	
\section*{Funding}

This work was financially supported by the Swedish Research Council
within the UPMARC Linnaeus center of Excellence (S.~Engblom, J.~Liu)
and by the Swedish Research Council, the R\"ontgen-\AA ngstr\" om
Cluster, the Knut och Alice Wallenbergs Stiftelse, the European
Research Council (J.~Liu).


\section*{Disclosures}

The authors declare no conflicts of interest.
 

\newcommand{\available}[1]{Available at \url{#1}}
\bibliographystyle{abbrvnat} 
\bibliography{pr772}

\clearpage
\appendix


\section{Orientation determination}

\begin{figure*}[!htbp]
  \centering
  \subfloat[]{\includegraphics[width=.33\textwidth]{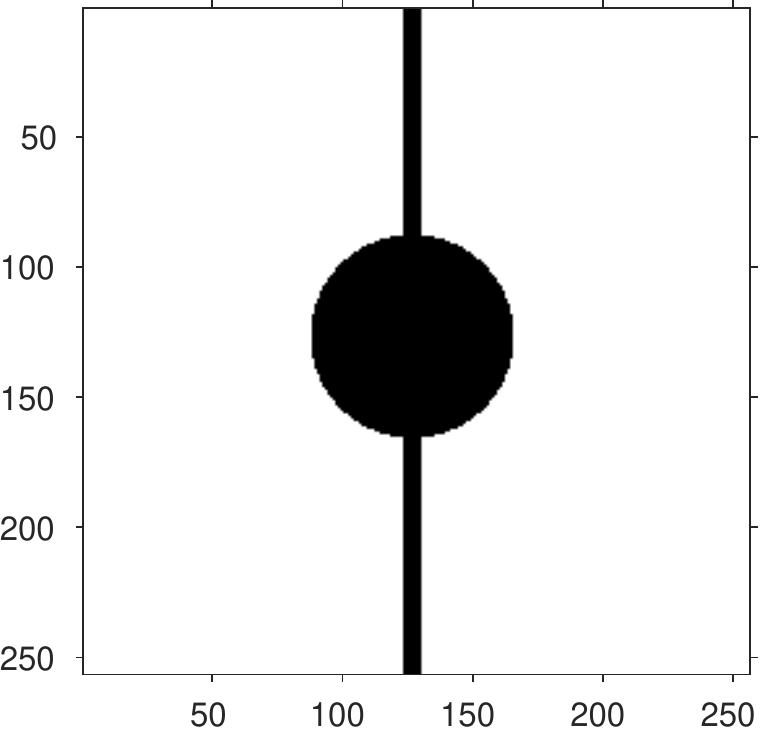} \label{fig:mskB}}
  \subfloat[]{\includegraphics[width=.33\textwidth]{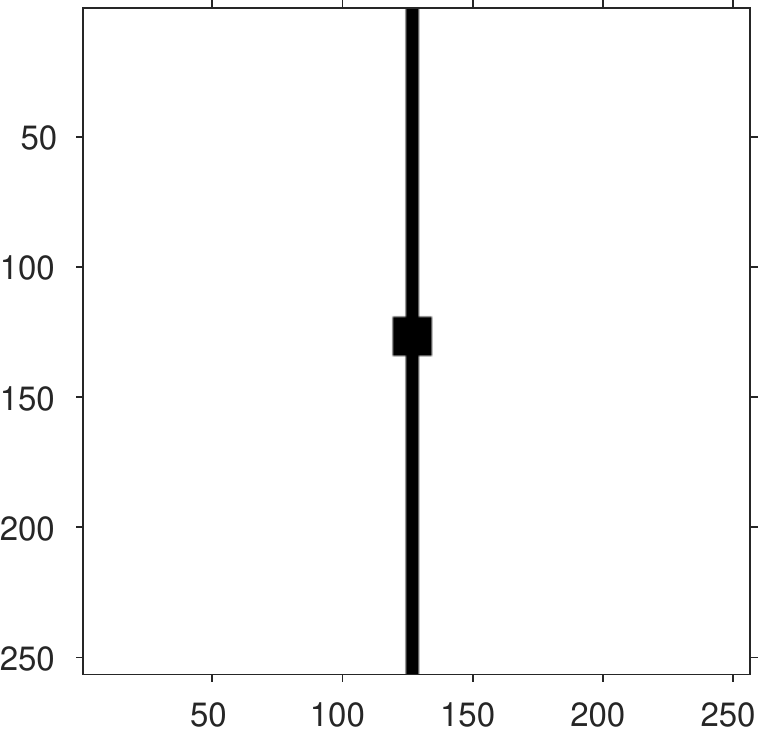}  \label{fig:mskS}} 
  \subfloat[]{\includegraphics[width=.33\textwidth]{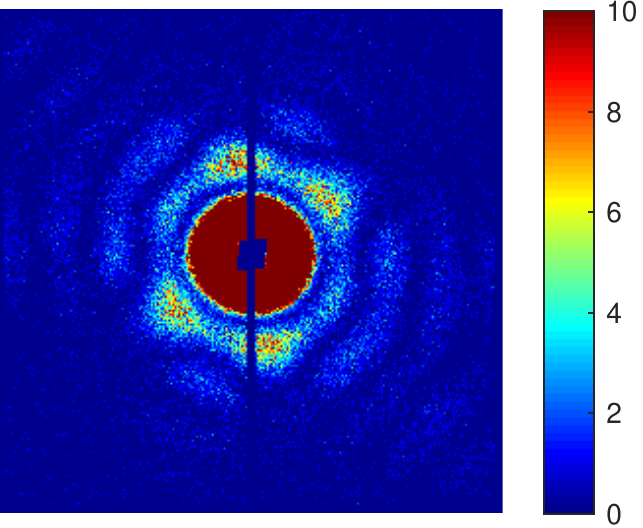} \label{fig:pattern}} \\
  \subfloat[]{\includegraphics[width=.33\textwidth]{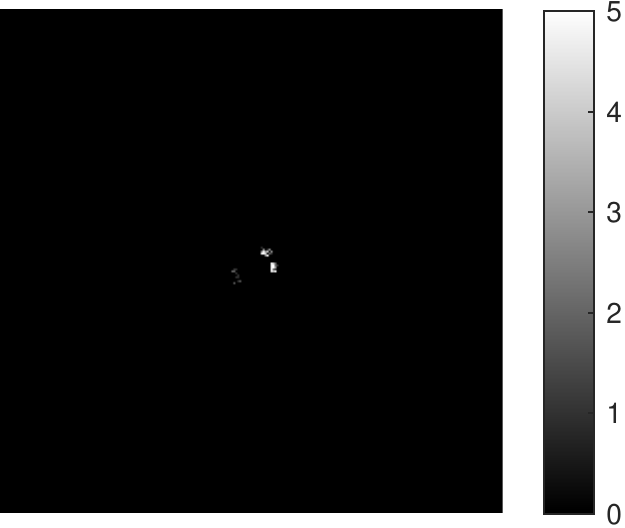} \label{fig:bg}}
  \subfloat[]{\includegraphics[width=.33\textwidth]{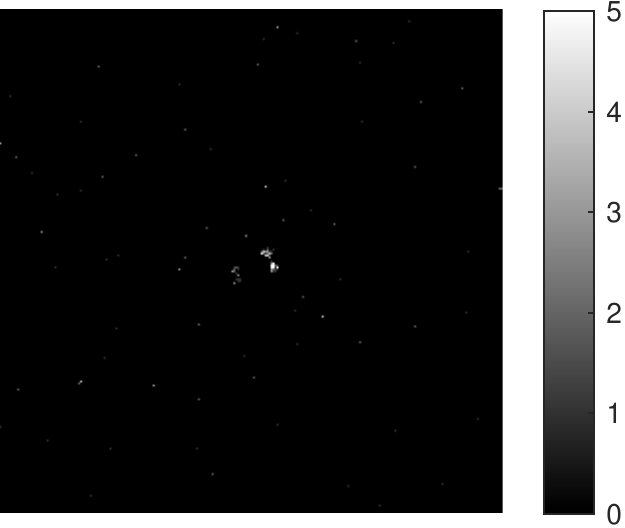} \label{fig:bge}}
  \subfloat[]{\includegraphics[width=.33\textwidth]{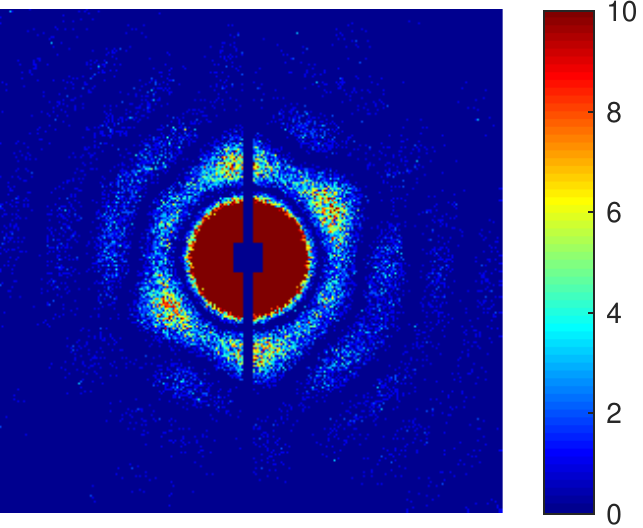} \label{fig:bgepattern}}
  \caption{Masks used in the EMC algorithm. \protect\subref{fig:mskB}:
    The masks used for computing the rotational probability in each
    EMC iteration.  \protect\subref{fig:mskS}: The mask used in the
    final EMC compression step. \protect\subref{fig:pattern} is a
    random pattern from our selected dataset
    $N_{1k}$. \protect\subref{fig:bg} is the background scattering
    estimated from dark runs. \protect\subref{fig:bge} is an estimated
    background with extra scattering. \protect\subref{fig:bgepattern}
    is a synthetic pattern with the small mask in
    \protect\subref{fig:mskS}. }
  \label{fig:masks}
\end{figure*}

The orientation determination was performed for both data sets
$N_{1k}$ and $N_{3k}$ using the scaled Poissonian model. In order to
make the orientation determination robust wih background noise and
with possible detector saturation, a mask of 39 pixels in diameter,
see Figure~\subref*{fig:mskB}, was used for computing the rotational
probability. In the final iteration, a smaller mask, which covered
only the gap between two detectors, was used in merging the 3D
diffraction volume, see Figure~\subref*{fig:mskS}.

\begin{figure}[!htbp]
  \includegraphics[width=.8\textwidth]{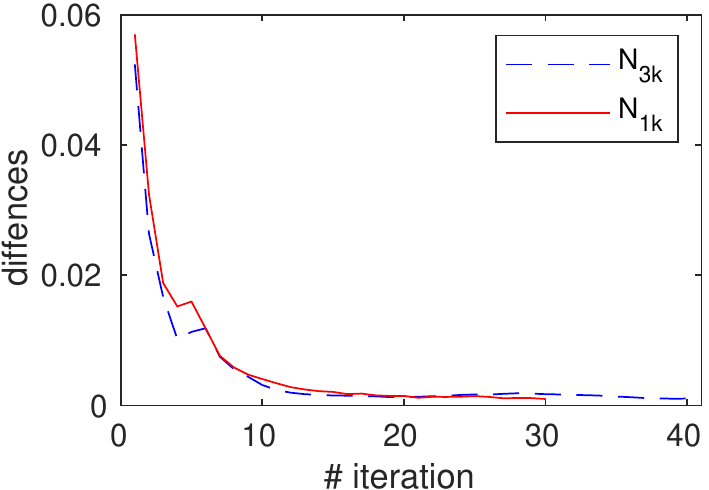}
  \caption{The sum absolute differences of two EMC models from two
    adjacent iterations.}
  \label{fig:emcconv}
\end{figure}

We stopped the EMC algorithm when the sum absolute differences of the
two adjacent models was less than 0.001, and hence 41 EMC iterations
were performed for $N_{3k}$ and 30 iterations for
$N_{1k}$. Figure~\ref{fig:emcconv} illustrated the sum absolute
differences of the two adjacent 3D Fourier intensities. As can be
seen, a peak appeared around the fifth EMC iteration for both
datasets, suggesting the EMC algorithm has formed an icosahedron, and
further iterations perfected the model. For synthetic dataset, EMC
stopped after around 15 iterations.

\section{Retrieved objects from Synthetic data}
 
\begin{figure*}[!htbp]
  \centering
  \subfloat[]{\includegraphics[width=.25\textwidth]{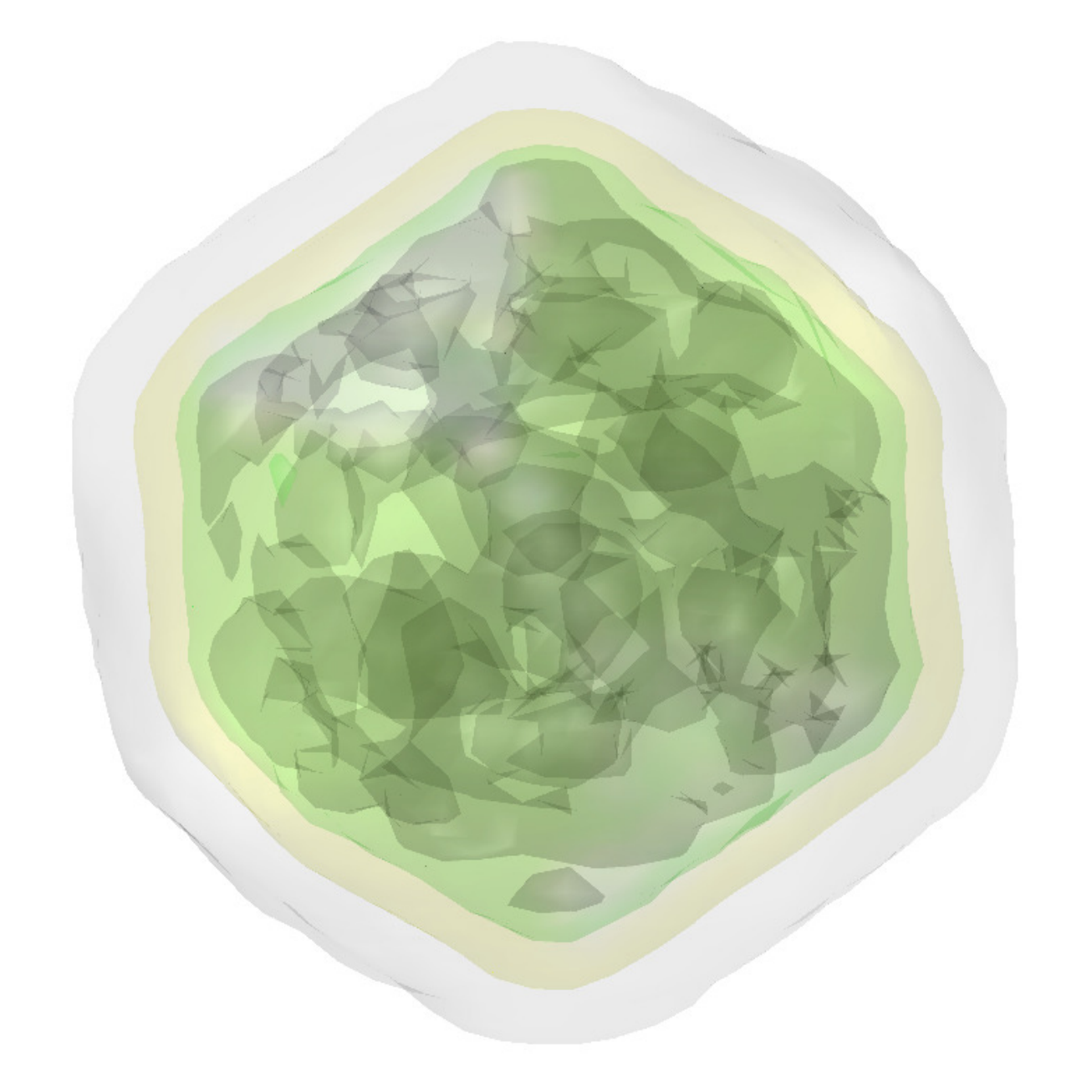}
    \label{fig:ori3d}}
  \subfloat[]{\includegraphics[width=.25\textwidth]{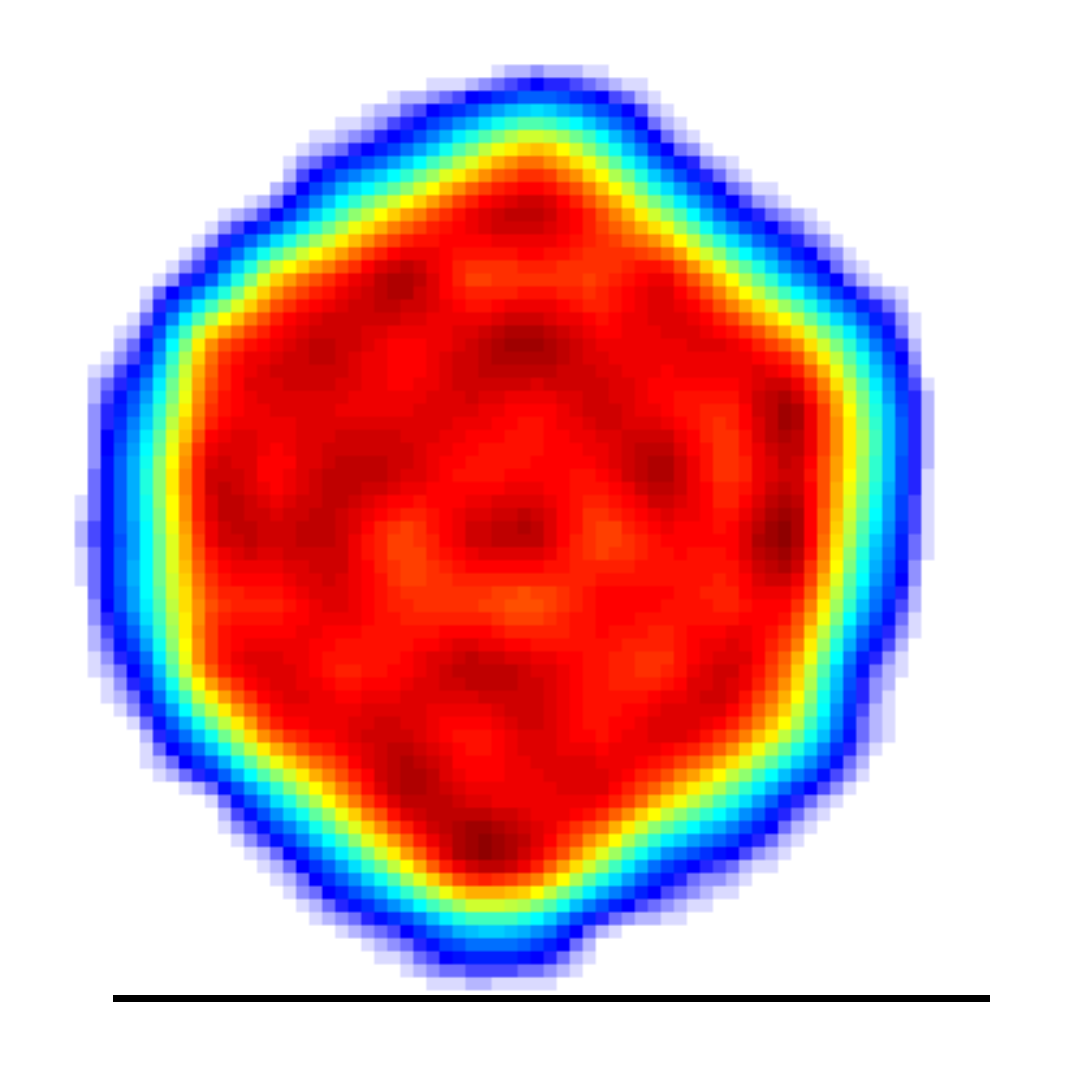}
    \label{fig:ori}}
  \subfloat[]{\includegraphics[width=.25\textwidth]{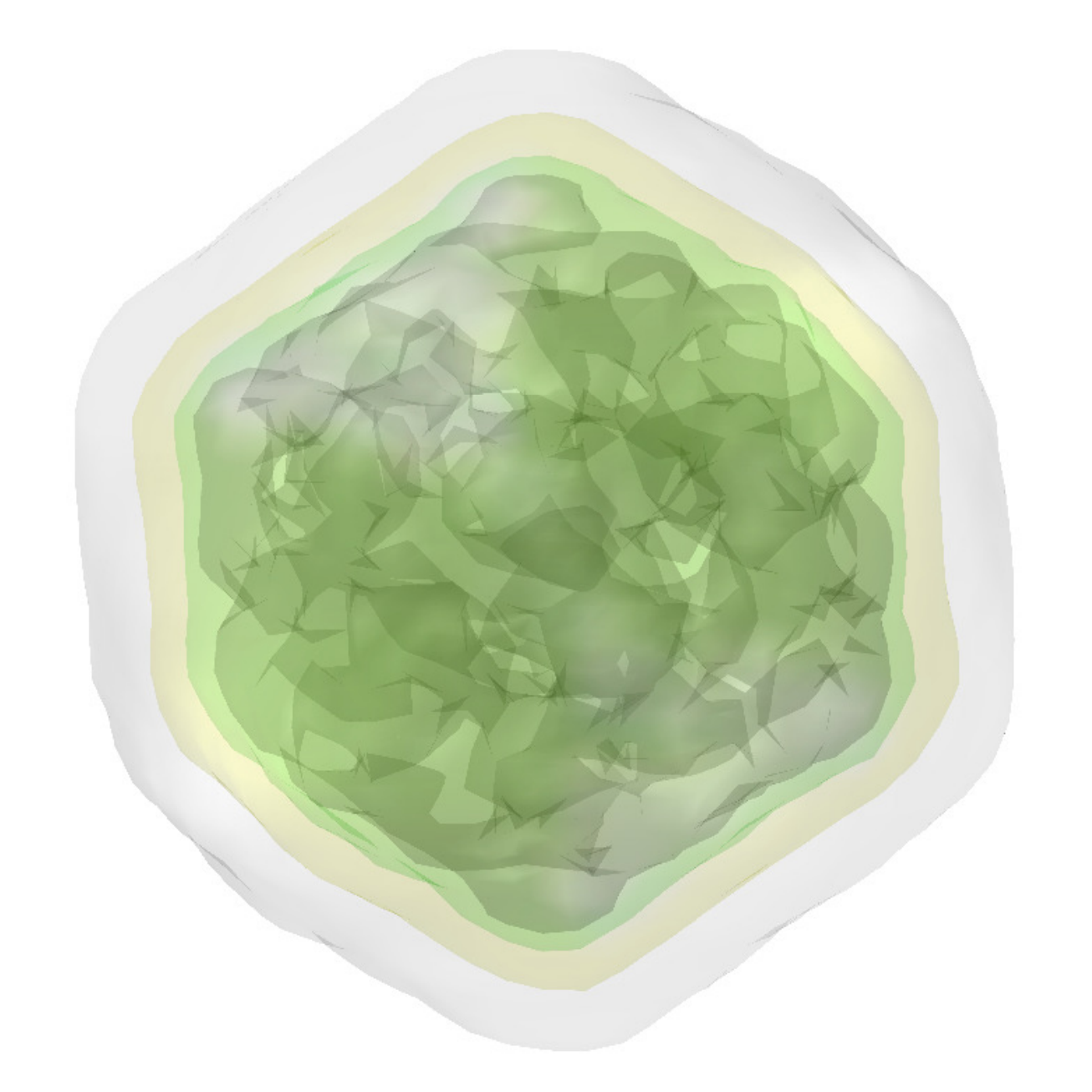}
    \label{fig:fl3d}}
  \subfloat[]{\includegraphics[width=.25\textwidth]{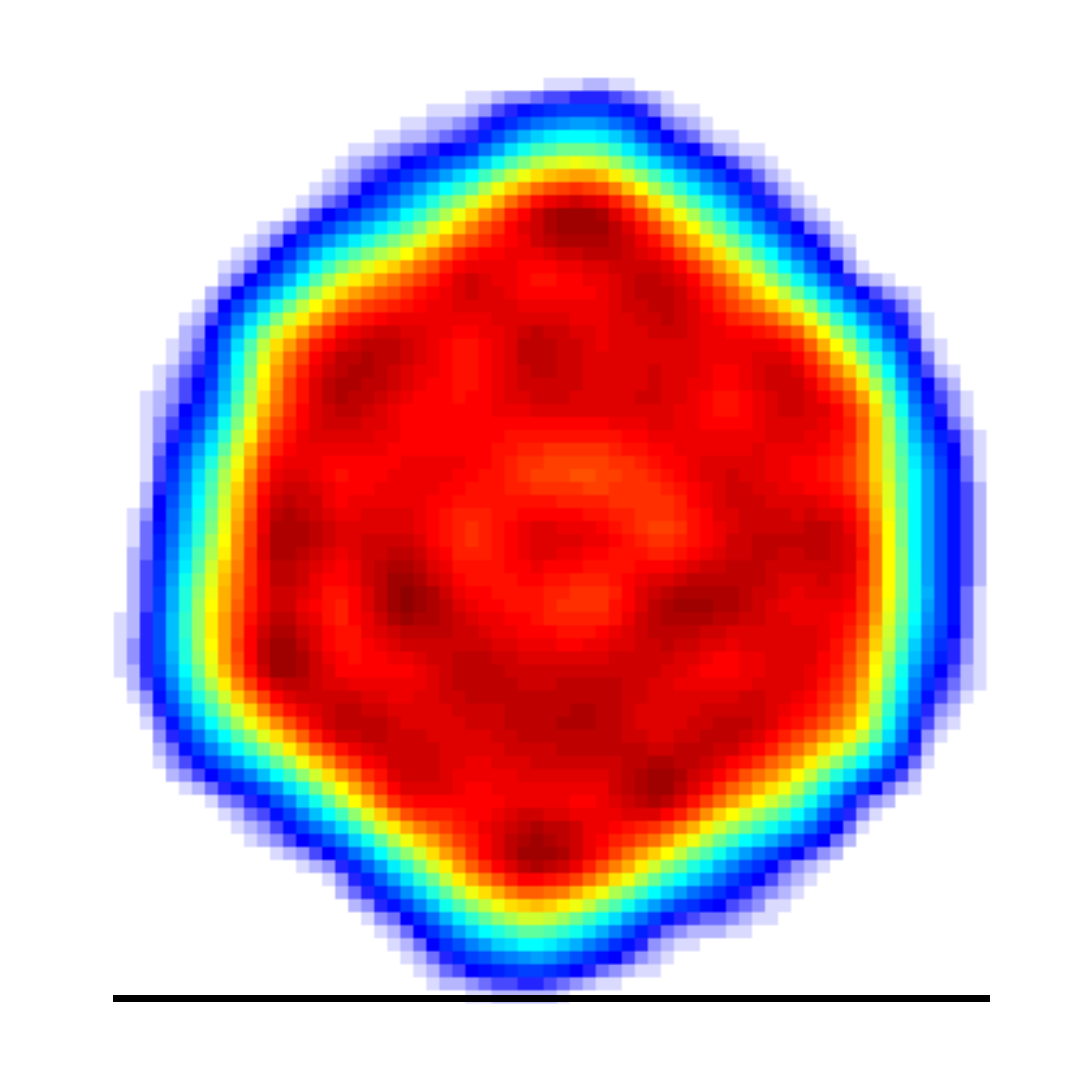}
    \label{fig:fl}} \\
  \subfloat[]{\includegraphics[width=.25\textwidth]{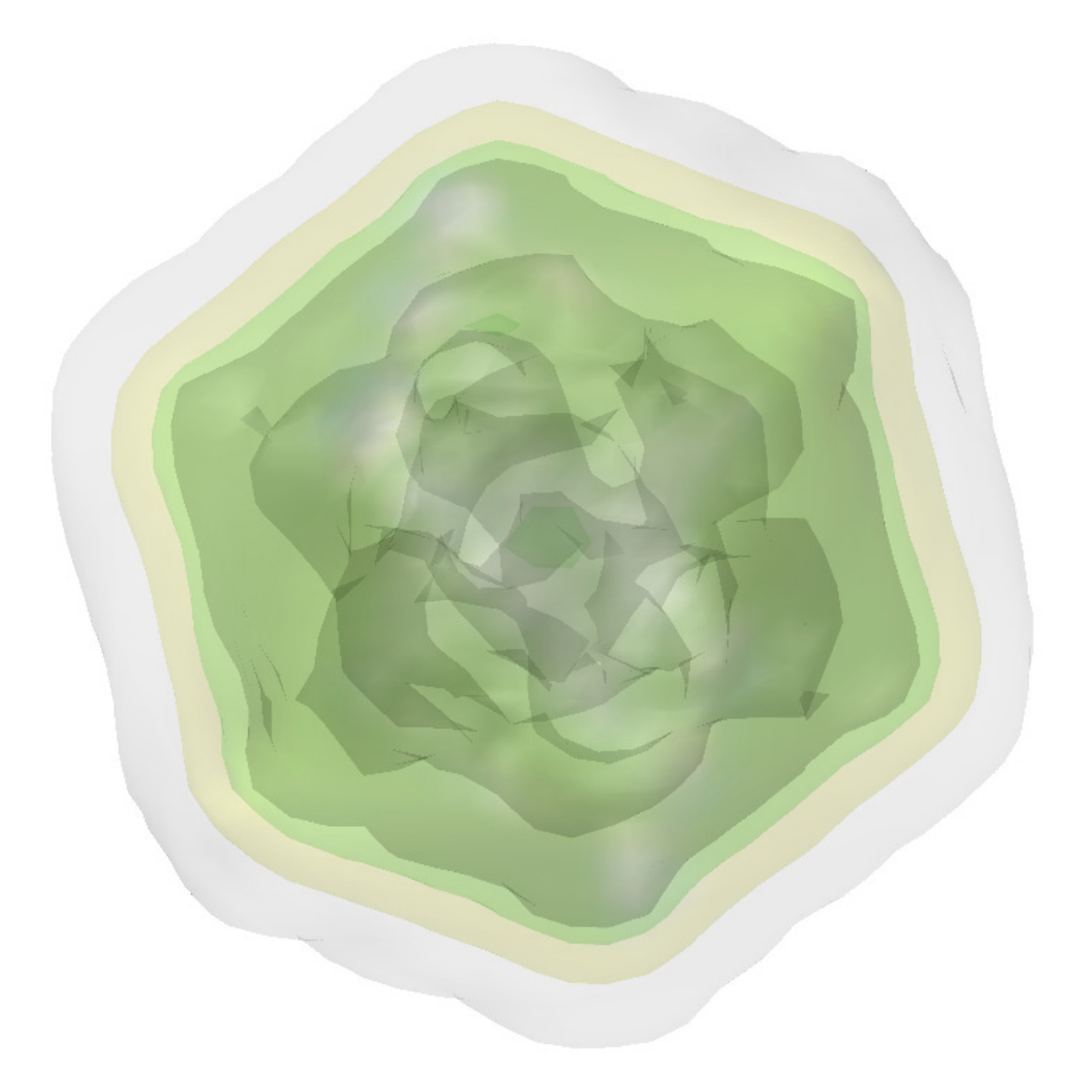}
    \label{fig:flbg3d}}  
  \subfloat[]{\includegraphics[width=.25\textwidth]{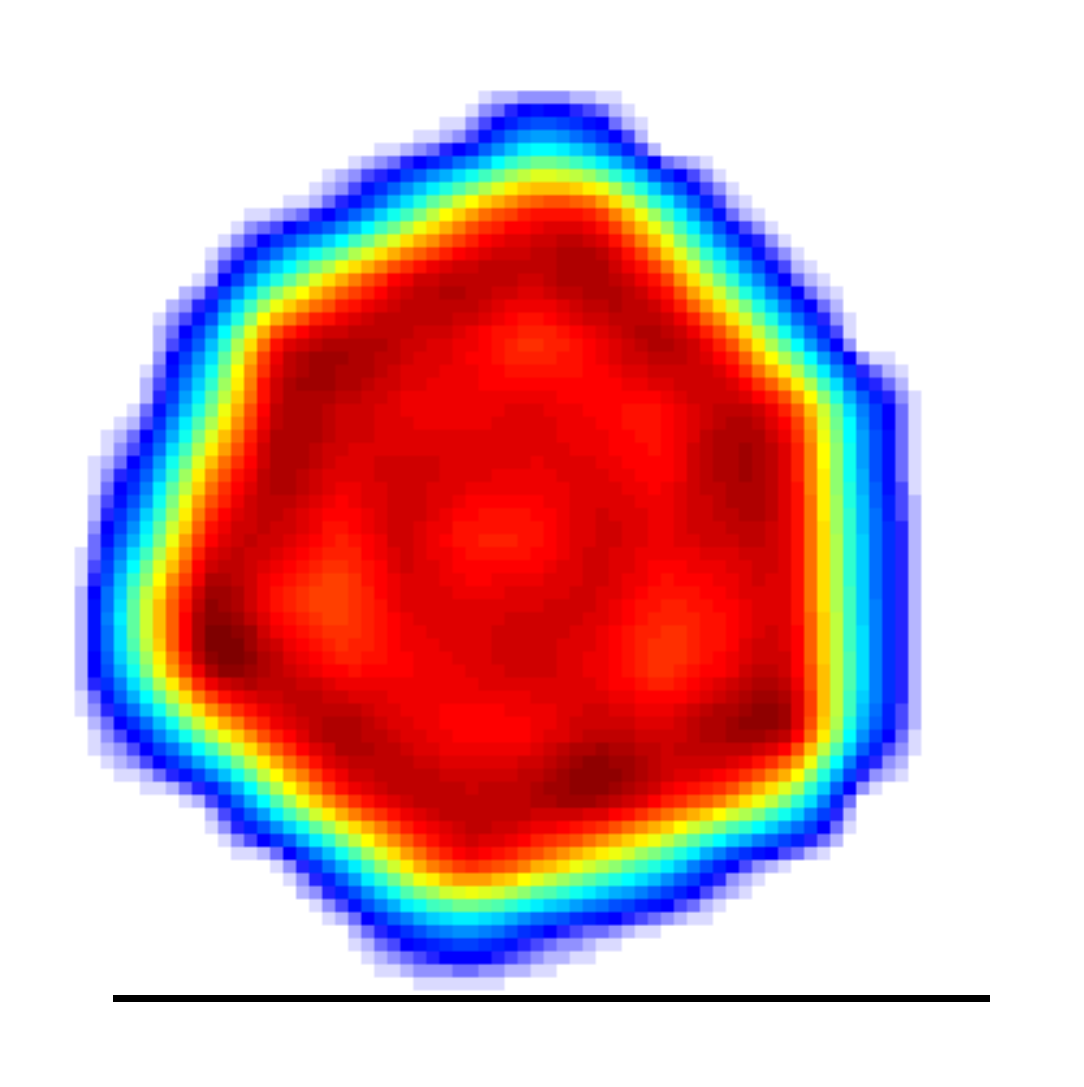}
    \label{fig:flbg}} 
  \subfloat[]{\includegraphics[width=.25\textwidth]{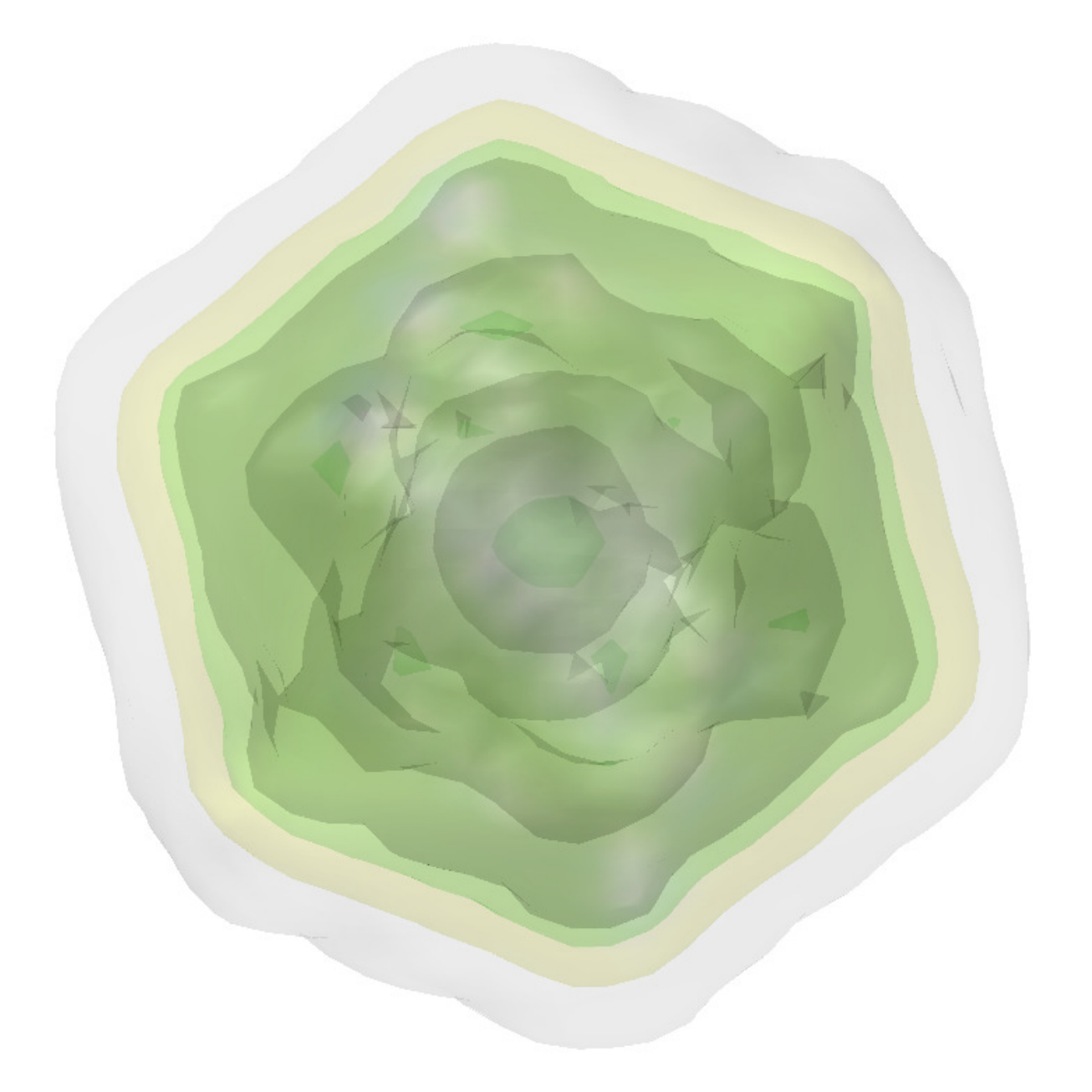}
    \label{fig:flbge3d}}
  \subfloat[]{\includegraphics[width=.25\textwidth]{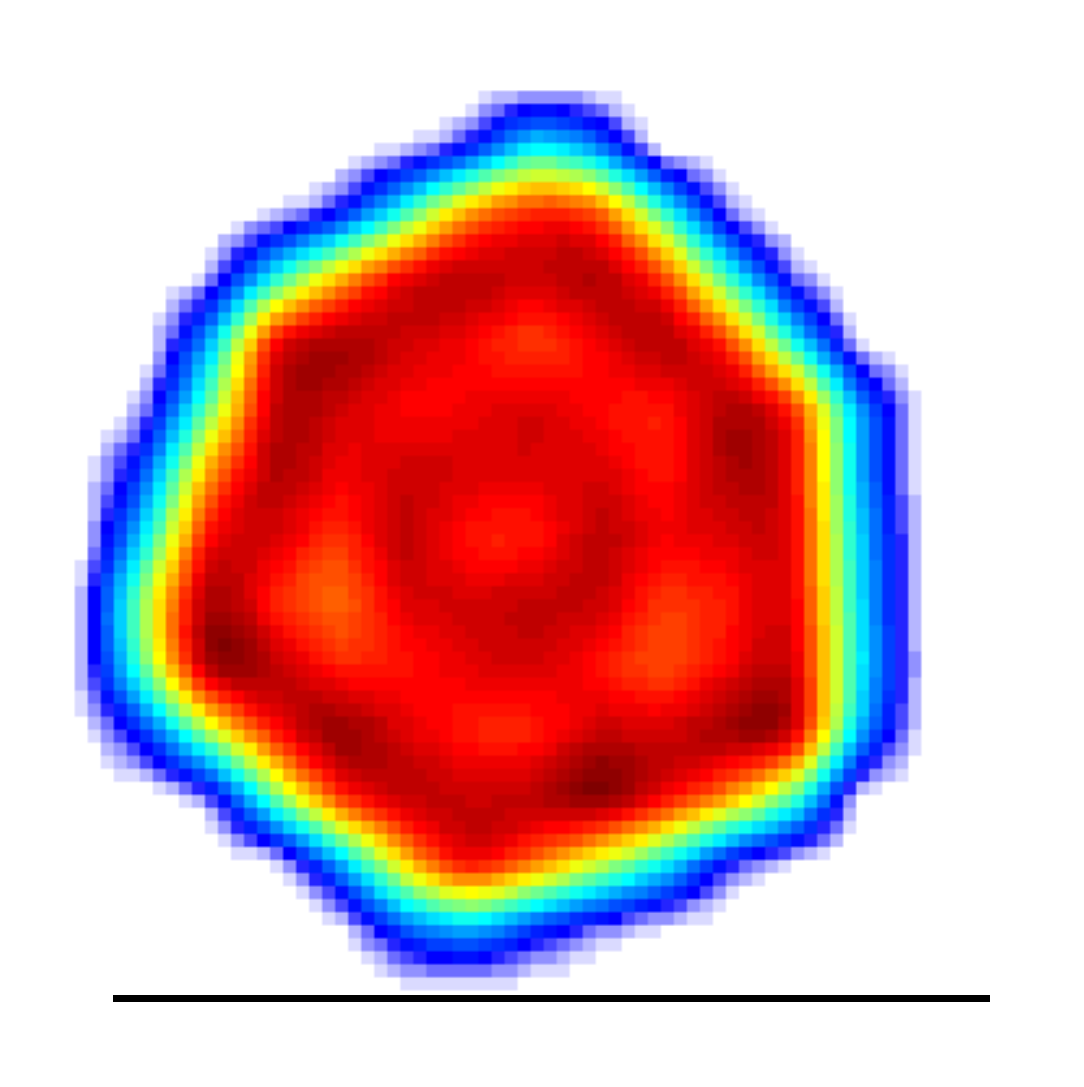}
    \label{fig:flbge}}\\
  \caption{ \protect\subref{fig:ori3d} illustrates interior structures
    as isosurface plots at 10\%, 50\%, 79\% and 89\% values of the
    maximum electron density for a synthetic Poissonian dataset $K$,
    and \protect\subref{fig:ori} is a random cross-section image from
    \protect\subref{fig:ori3d}.  [\protect\subref{fig:fl3d},
    \protect\subref{fig:fl}], [\protect\subref{fig:flbg3d},
    \protect\subref{fig:flbg}] and [\protect\subref{fig:flbge3d},
    \protect\subref{fig:flbge}] are the corresponding interior
    structures and cross-section images from the synthetic datasets
    $K^{f}$, $K^{b}$, and $K^{e}$, respectively. As can be seen,
    aliasing effects existed in all retrieved electron intensities,
    since a uniform-intensity icosahedron was used for generating the
    diffraction patterns. The background noise and the extra
    scattering contributed to the high-intensity central ring as shown
    in \protect\subref{fig:flbg} and \protect\subref{fig:flbge}.}
  \label{fig:syn}
\end{figure*}

Let $W_j$ be the ground truth of a 2D diffracted pattern in the $j$th
rotation, i.e.~$W_j$ can be simulated as a slice of a Fourier
transformed particle through the particle center at the $j$th
rotation. If photon counting is a Poisson process, we may write the 2D
diffraction patterns without background noise as
$(K_j^*)_{j=1}^{M} = Poisson(W_j)$, where $M$ is the number of sampled
rotations, where $M = 3000$ was used in our simulations. Considering a
varying photon fluence, we write $K^{f} = Poisson(W \phi)$, where
$\phi$ is constant for all pixels in one pattern, but differs from
shot to shot. For our synthetic data we took $\phi$ a uniform random
number in the range $[0.9,1.1]$. We also added a background signal,
related to the X-ray beam configuration,
i.e.~$(K_j^b)_{j=1}^{M} = Poisson(W_j \phi_j+K_{b})$. Note that
$K_{b}$ is a 2D pattern that is the same for all patterns under the
same beam configuration, and we illustrate $K_b$ in
Figure~\subref*{fig:bg}. Moreover, an extra source of scattering, was
considered as $(K_j^3)_{j=1}^{M} = Poisson(W_j \phi_j+K_{b} + B(p))$,
where $B$ takes the value 1 with probability $p = 1000^{-1}$, and zero
otherwise, and we illustrate an example of $K_b + B(p)$ in
Figure~\subref*{fig:bge}.  Different real-space objects reconstructed
from different datasets are shown in Figure~\ref{fig:syn}.

\section{Background Subtraction}

\begin{figure*}[!htbp]
  \centering
  \subfloat[]{\includegraphics[width=.5\textwidth]{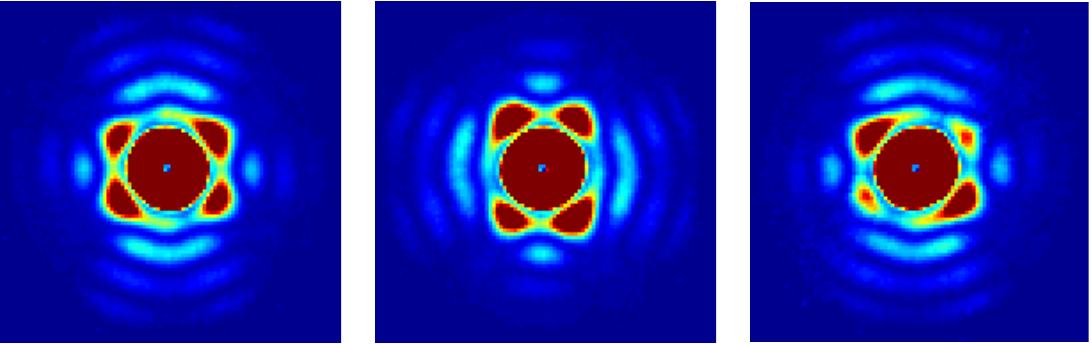}
    \label{fig:ori1k}}
  \subfloat[]{\includegraphics[width=.5\textwidth]{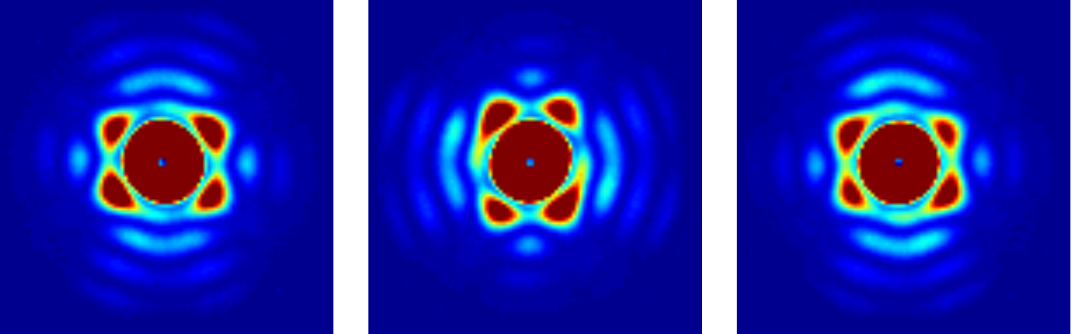}
    \label{fig:bgrm1k}}\\
  \subfloat[]{\includegraphics[width=.5\textwidth]{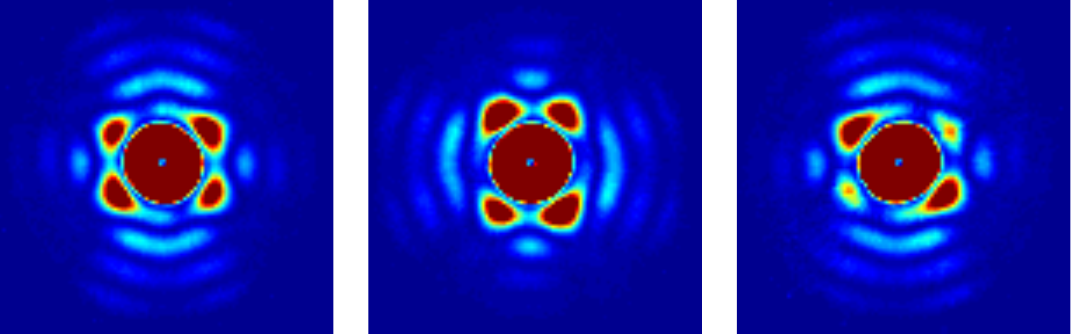}
    \label{fig:ori3k}}
  \subfloat[]{\includegraphics[width=.5\textwidth]{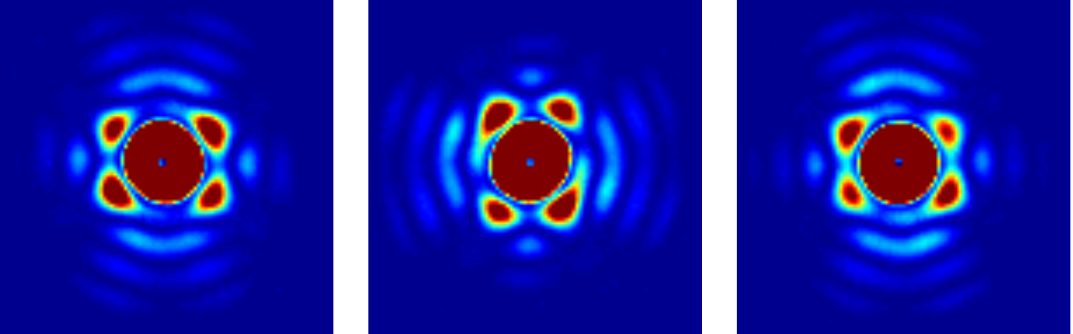}
    \label{fig:bgrm3k}}\\
  \subfloat[]{\includegraphics[width=.5\textwidth]{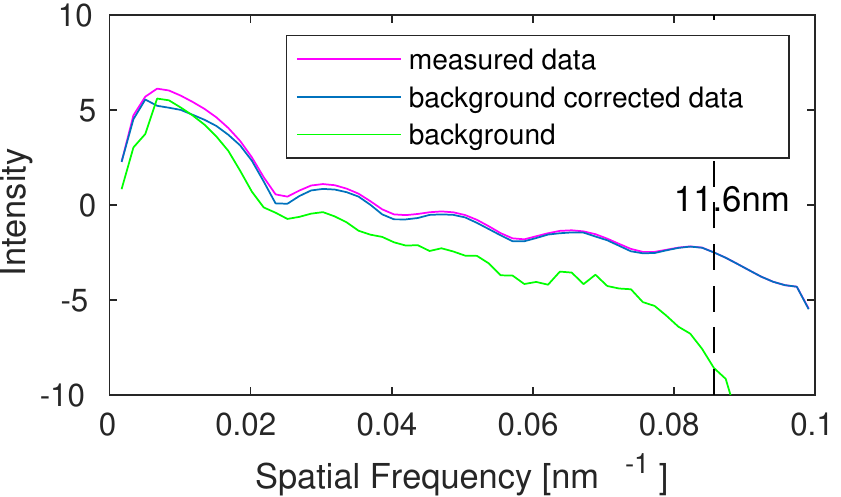}
    \label{fig:bg1k}}
  \subfloat[]{\includegraphics[width=.5\textwidth]{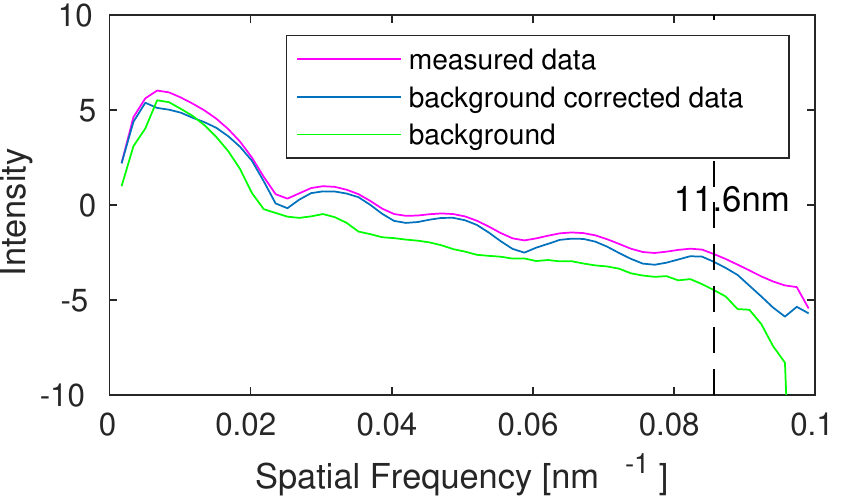}
    \label{fig:bg3k}}
  \caption{The \textit{left column} are results from the dataset
    $N_{1k}$, and the \textit{right column} are from
    $N_3k$. \protect\subref{fig:ori1k} and \protect\subref{fig:ori3k}
    were the cross-section plots of the EMC intensity at the final EMC
    iteration for $N_{1k}$ and $N_{3k}$, respectively.
    \protect\subref{fig:bgrm1k} and \protect\subref{fig:bgrm3k} were
    the corresponding background corrected data. The intensities and
    the background values were illustrated in
    \protect\subref{fig:bg1k} and \protect\subref{fig:bg3k}.  }
\label{fig:bgrm}
\end{figure*}

For given discrete shells $(s_u)_{u=1}^{U}$, let $S = (S_u)_{u=1}^{U}$
be the radial shells of a 3D Fourier intensity $\Wgrid$. The $u$th
shell is given by $S_{u} =\{s = (x,y,z); s_{u} \le \|s\| < s_{u+1}\}$,
where $s$ is a point (voxel) at position $(x,y,z)$, and $\|s\|$ is the
Euclidean norm.  The background noise in shell $u$�is then estimated
by
\begin{align}
B_u = \beta \min_{s\in S_u} (\Wgrid)_s,
\end{align}
and we chose $\beta = 0.5$ for our EMC reconstructions. We denote the
average shell of a 3D Fourier intensity by
\begin{align}
  D_u = |S_u|^{-1}\sum_{s\in S_u} (\Wgrid)_s,
\end{align}
where $|S_u|$ is the number of voxels in the $u$th shell.  Note that
$D_u$ can be also considered as the average power spectral density
(PSD).  The results of background subtraction showed larger contrasts
and clear boundaries, see Figure~\ref{fig:bgrm}.

The PSD profiles of $N_{1k}$ and $N_{3k}$ were similar, however, the
Fourier intensity for $N_{1k}$ was coarser in the high frequencies,
due to the lack of data to fill in the space, see
Figure~\subref*{fig:bg1k} and Figure~\subref*{fig:bg3k}.





\end{document}